\documentclass[12pt]{article}

\usepackage{amsmath,amssymb,amsthm,amscd,mathrsfs}
\usepackage{latexsym,amsmath,amssymb,graphicx}
\usepackage[fleqn]{mathtools}
\usepackage[all]{xy} 
\usepackage{multirow}
\usepackage{color}
\usepackage{hyperref}
\usepackage{epsf}
\usepackage{tikz}
\usetikzlibrary{shapes.geometric, arrows}

\usepackage{graphicx}

\makeatletter
\newcommand\opteq[1]{\mathrel{\mathpalette\opt@eq{#1}}}
\newcommand{\opt@eq}[2]{%
  \begingroup
  \sbox\z@{$#1#2$}%
  \sbox\tw@{\resizebox{!}{.5\ht\z@}{$\m@th#1($}}%
  \nonscript\hskip-\wd\tw@
  \mkern1mu
  \raisebox{-.35\ht\z@}[0pt][0pt]{\resizebox{!}{.5\ht\z@}{$\m@th#1($}}%
  \mkern-1mu
  {#2}%
  \mkern-1mu
  \raisebox{-.35\ht\z@}[0pt][0pt]{\resizebox{!}{.5\ht\z@}{$\m@th#1)$}}%
  \mkern1mu
  \nonscript\hskip-\wd\tw@
  \endgroup
}
\makeatother
\newcommand{\leoq}{\opteq{\leq}}

\usepackage{youngtab}
\xyoption{v2} \xyoption{2cell} \xyoption{dvips}
\CompileMatrices \numberwithin{equation}{section}
\newtheorem{prop}{Proposition}[section]

\newtheorem{lemm}[prop]{Lemma}
\newtheorem{coro}[prop]{Corollary}
\newtheorem{rema}[prop]{Remark}

\numberwithin{equation}{section}
\makeatletter
\newcommand{\subalign}[1]{%
  \vcenter{%
    \Let@ \restore@math@cr \default@tag
    \baselineskip\fontdimen10 \scriptfont\tw@
    \advance\baselineskip\fontdimen12 \scriptfont\tw@
    \lineskip\thr@@\fontdimen8 \scriptfont\thr@@
    \lineskiplimit\lineskip
    \ialign{\hfil$\m@th\scriptstyle##$&$\m@th\scriptstyle{}##$\crcr
      #1\crcr
    }%
  }
}
\makeatother
\newcommand{\be}{\begin{equation}}
\newcommand{\ee}{\end{equation}}
\newcommand{\IP}{\mathbb{P}}%{{\relax{\rm I\kern-.18em P}}}

\newcommand{\wgamma}{{\widetilde \gamma}}

\newcommand\IZ{\mathbb {Z}}

\newcommand{\IC}{\mathbb{C}}

\newcommand{\CU}{{\mathcal U}}
\newcommand{\ba}{\begin{array}}
\newcommand{\ea}{\end{array}}

\newcommand{\wH}{{\widetilde H}}

\newcommand{\CV}{{\mathcal V}}
\newcommand{\CX}{{\mathcal X}}

\newcommand{\CS}{{\mathcal S}}

\newcommand{\bal}{\begin{aligned}}
\newcommand{\eal}{\end{aligned}}

\newcommand{\wq}{{\widetilde {q}}}

\newcommand{\wSigma}{{\widetilde \Sigma}}

\newcommand{\longto}{\longrightarrow}

\newcommand{\wR}{{\widetilde R}}
\newcommand{\wCO}{{\widetilde{\mathcal O}}}
\newcommand{\ch}{{\mathrm{ch}}}

\newcommand{\CO}{{\mathcal O}}

\newcommand{\CE}{{\mathcal E}}
\newcommand{\CA}{{\mathcal A}}
\newcommand{\CH}{{\mathcal H}}
\newcommand{\CF}{{\mathcal F}}
\newcommand{\CG}{{\mathcal G}}

\newcommand{\CM}{{\mathcal M}}

\newcommand{\wCV}{{\widetilde {\mathcal V}}}
\newcommand{\sfC}{{\sf C}}
\newcommand{\sfA}{{\sf A}}

\newcommand{\CC}{{\mathcal C}}

\newcommand{\CT}{{\mathcal T}}

\newcommand{\sfB}{{\sf B}}

\newcommand{\us}{{\underline s}}

\newcommand{\ut}{{\underline t}}
\newcommand{\calP}{{\mathcal P}}

\newcommand{\wF}{{\widetilde F}}

\newcommand{\IA}{{\mathbb A}}

\newcommand{\uq}{{\underline q}}

\CompileMatrices \LaTeXdiagrams \UseAllTwocells
\newdimen\tableauside\tableauside=1.0ex
\newdimen\tableaurule\tableaurule=0.4pt
\newdimen\tableaustep
\def\phantomhrule#1{\hbox{\vbox to0pt{\hrule height\tableaurule width#1\vss}}}
\def\phantomvrule#1{\vbox{\hbox to0pt{\vrule width\tableaurule height#1\hss}}}
\def\sqr{\vbox{%
  \phantomhrule\tableaustep
  \hbox{\phantomvrule\tableaustep\kern\tableaustep\phantomvrule\tableaustep}%
  \hbox{\vbox{\phantomhrule\tableauside}\kern-\tableaurule}}}
\def\squares#1{\hbox{\count0=#1\noindent\loop\sqr
  \advance\count0 by-1 \ifnum\count0>0\repeat}}
\def\tableau#1{\vcenter{\offinterlineskip
  \tableaustep=\tableauside\advance\tableaustep by-\tableaurule
  \kern\normallineskip\hbox
    {\kern\normallineskip\vbox
      {\gettableau#1 0 }%
     \kern\normallineskip\kern\tableaurule}%
  \kern\normallineskip\kern\tableaurule}}
\def\gettableau#1 {\ifnum#1=0\let\next=\null\else
  \squares{#1}\let\next=\gettableau\fi\next}
\allowdisplaybreaks 
\begin{document}

\title{Twisted spectral correspondence and torus knots} 
\author{Wu-yen Chuang,
  Duiliu-Emanuel Diaconescu, Ron Donagi,\\ Satoshi Nawata, Tony Pantev}
\date{}
\maketitle

\begin{abstract} 
Cohomological invariants of twisted wild character varieties as constructed by Boalch and Yamakawa 
are derived from enumerative Calabi-Yau geometry and 
refined Chern-Simons invariants of torus knots.  Generalizing the untwisted case, the present approach is based on a spectral correspondence for meromorphic Higgs bundles with fixed conjugacy classes at the marked points. This construction is carried out for twisted wild character varieties associated to $(\ell,k\ell-1)$ torus knots, 
providing a colored generalization of existing results of Hausel, Mereb and Wong as well
as Shende, Treumann and Zaslow.
\end{abstract}

\tableofcontents

\section{Introduction}

This is the latest installment in a series of papers \cite{wallpairs,BPSPW,Par_ref,BPS_wild,Wild_deg} developing a string theoretic approach to the cohomology of character varieties that is an important problem in geometry, topology and number theory, as shown by the fairly extensive mathematical literature. 
There have been several approaches from different viewpoints to this question, by using arithmetic methods for character varieties \cite{HRV,HLRV,HLRVII,Arithmetic_wild,Zariski_closures,Hausel_slides,Hausel_private}, and by using arithmetic and/or motivic methods 
for Higgs bundles  \cite{Ind_vb_higgs,Counting_Higgs,Ind_obj,
Higgs_ind_proj_line,Mot_conn_higgs,Mot_chains,
y_genus_higgs}.  At the same time, the topology of wild character varieties has been related to polynomial invariants of legendrian knots in \cite{Fukaya_knots,Cluster_legendrian}. 

The current paper aims to extend the string theoretic approach to twisted wild character varieties. Mathematical framework on the subject has been established by Treumann, Shende and 
Zaslow \cite{Fukaya_knots} for character varieties on the disk, and Hausel, Mereb and Wong \cite{Arithmetic_wild,Hausel_slides,Hausel_private} for curves of arbitrary genus. Nevertheless, the string theoretic construction offers a new perspective on the subject, leading to new conjectural formulas, as well as a generalization to higher multiplicity. 

As in \cite{BPSPW,Par_ref,BPS_wild}, the relation between the topology of character varieties 
and BPS states in string theory is based on the $P=W$ conjecture 
of de Cataldo, Hausel and Migliorini \cite{hodgechar}, and a spectral construction for Higgs bundles. In particular, the spectral correspondence for twisted irregular Higgs bundles used in the 
present context is one of the main novelties of this paper. As explained in detail below, this result leads to new conjectural 
formulas for topological  invariants of twisted wild character varieties.

\subsection{Twisted meromorphic Higgs bundles}\label{twistedhiggs}

To begin with, let us provide a brief introduction to certain moduli spaces of meromorphic Higgs bundles on smooth projective curves, which constitute the main object of study in this paper. 

Let $C$ be a smooth projective curve over $\IC$ and
 $p_a\in C$, $1\leq a\leq m$ a set of 
 pairwise distinct marked points. Let $n_a\in \IZ$, $n_a\geq 1$ be integer multiplicities associated to the 
 marked points $p_a$ and let $D=\sum_{a=1}^m n_a p_a$. 
  The meromorphic Higgs bundles on $C$ we consider are 
pairs $(E,\Phi)$ with $E$ a vector bundle on $C$
 and $\Phi: E \to E\otimes K_C(D)$ a meromorphic Higgs field with polar divisor $D$. As usual $(E,\Phi)$ is (semi)stable if 
 $\mu(E')\leoq \mu(E)$ for any $\Phi$-invariant subbundle $E'\subset E$. Fixing the rank $N$ and degree $e$ of $E$, one obtains a quasi-projective moduli space of such objects. In the following it will be assumed that $(e,N)$ are coprime, in which case there are no strictly semistable objects and the moduli space is smooth. Furthermore, as shown in \cite{spectral_int}, it has a Poisson structure whose symplectic leaves are obtained by fixing the conjugacy class of the restriction $\Phi|_{D}$. The current study will be focused on symplectic leaves of special kind, where the local model for the Higgs field at each marked 
 point will be shown below. 
 
 Suppose $N=r\ell$ with $\ell,r\geq 1$ positive 
 integers. Let 
\be\label{eq:rJEformulas}
J_{\ell,r} = \left(\begin{array}{ccccc} 
0 & 0 & \cdots & 0 & 0 \\
{\bf 1}_r & 0 & \cdots & 0 & 0 \\
\vdots & \vdots & \vdots & \vdots& \vdots\\
0 & 0 & \cdots & {\bf 1}_r & 0\\
\end{array}\right),\qquad 
E_{\ell,r}= \left(\begin{array}{ccccc} 
0 & 0 & \cdots & 0 & {\bf 1}_r \\
0 & 0 & \cdots & 0 & 0 \\
\vdots & \vdots & \vdots & \vdots& \vdots\\
0 & 0 & \cdots & 0 & 0\\
\end{array}\right)~,
\ee
be $N\times N$ matrices, where all entries are $r\times r$ blocks. For any integer $n \geq 3$, let  
$$
Q_{n,\ell,r}(z)= {z^{1-n}\over 1-n} J_{\ell, r} + 
{z^{2-n}\over 2-n} E_{\ell,r}
$$
be a meromorphic $\mathfrak{gl}(N,\IC)$-valued function on the formal disk with coordinate $z$.

In this paper a twisted meromorphic Higgs bundle 
will be a Higgs bundle $(E,\Phi)$, $\Phi: E \to E\otimes K_C(D)$, $D = n \sum_{a=1}^m p_a$, satisfying the local condition:
\begin{itemize} 
\item[$(H.1)$] For each marked point $p_a$ there is a trivialization of $E$ on the $n$-th order infinitesimal 
neighborhood of $p_a$ which identifies $\Phi|_{np_a}$ with 
$dQ_{n,\ell,r}(z_a)$.
\end{itemize} 
Note that the word, ``twisted'', refers here to the fact that the fixed Laurent tail at any given marked point does not have coefficients in a Cartan subalgebra of $\mathfrak{gl}(N,\IC)$. In contrast, untwisted meromorphic Higgs bundles 
are defined similarly, except that the fixed Laurent tails do have coefficients in such a subalgebra. 

Now let $\CH_{n,\ell,r}$ denote the moduli space of stable rank $N$, degree $1$, twisted meromorphic Higgs bundles $(E,\Phi)$ as above.
This is a smooth holomorphic symplectic variety of complex dimension 
$$
d(g,m,n,\ell,r) = 2+ r^2(mn\ell(\ell-1)+2(g-1)\ell^2). 
$$
As in the untwisted case, these moduli spaces are related by the non-Abelian Hodge correspondence to certain moduli spaces $\CC_{n,\ell,r}$ of flat irregular $\mathfrak{gl}(N,\IC)$-connections on 
the curve $C$ with higher order poles at the marked points. This was proven by 
Sabbah \cite{Harmonic_metrics} and Biquard and
Boalch \cite{Wild_curves}. The higher dimensional 
case was 
proven by Mochizuki \cite{Wild_harmonic}.
The singular part of any such connection at the 
marked point $p_a$ is fixed and determined by the local data $Q_{m,\ell,r}(z_a)$ as shown in \S\ref{nonabHodge}. Furthermore the construction of Boalch and Yamakawa \cite{Twisted_wild} produces a  moduli space $\CS_{n,\ell,r}$ of twisted Stokes data associated to these connections, called twisted wild character varieties as in \cite{Twisted_wild}. These spaces will play a central role in this paper, and the details are provided in \S\ref{twistedStokes}. In particular, for $r=1$ they are related to irreducible $(\ell, (n-2)\ell-1)$ torus knots in the framework of \cite{Fukaya_knots}.

\subsection{Twisted spectral covers}\label{twistedcovers}

The spectral correspondence relates stable twisted meromorphic Higgs bundles on $C$ to stable pure dimension one sheaves on a holomorphic symplectic spectral 
surface $S$. Employing the approach of Kontsevich and Soibelman \cite[Sect 8.3]{structures}, the spectral surface is constructed in \S\ref{spectralsect} as the complement of an anticanonical divisor in a successive blow-up of the total space of the coefficient line bundle $M=K_C(n(p_1+\cdots+p_m))$. 
In order to explain the main features, it suffices to consider a single marked point $p\in C$ in which case $M=K_C(np)$. 
The generalization to an arbitrary number of marked points is immediate since the construction is local. 
Abusing notation, the total space of $M$ will be denoted again by $M$, and the distinction will be clear from the context. Let $z$ 
be a local coordinate on $C$ centered at $p$ and let $u$ 
be a linear vertical coordinate on $M$ in a neighborhood 
of the fiber $M_p$.  

The main observation is that a spectral cover for 
a twisted irregular Higgs bundle on $C$ with data $(n,\ell,1)$ 
is a compact irreducible curve $\Sigma$ in the total space of $M$ which has analytic type $z-u^\ell=0$ in the infinitesimal 
neighborhood of $M_p$. As shown in \S\ref{spectralsect}, one first constructs a semistable reduction 
of the higher order tangency point $z(z-u^\ell)=0$ by
$\ell-1$ successive blow-ups. This is followed by a second linear sequence of $(n-1)\ell+1$ blow-ups which is analogous 
to the spectral construction in \cite{BPS_wild}. The spectral surface $S$ is obtained by removing an anticanonical divisor 
supported on the strict transform of $M_p$ as well as all 
exceptional divisors except the last one in each chain. An important point is 
that there is a unique linear system whose generic elements are compact, 
irreducible and reduced divisors $\Sigma_{n\ell}$ on $S$, consisting of the 
strict transforms of the generic irreducible spectral covers $\Sigma$ on $M$. The numerical invariants of a pure dimension one sheaf $F$ on $S$ with compact support 
are $(r,c)\in \IZ_{\geq 1} \times \IZ$  where 
\[ 
\ch_1(F) = r [\Sigma_{n\ell}], \qquad c= \chi(F).
\]
The main claim of the twisted spectral correspondence  is:
\bigskip 

{\it There is an isomorphism of moduli stacks between the stack 
${\mathfrak H}_{n,\ell, r}$ 
of rank $N=r\ell$, degree $e$ semistable meromorphic Higgs bundles satisfying condition $(H.1)$ and the stack of semistable pure dimension one sheaves $F$ on $S$ with numerical invariants $(r,e-N(g-1))$. This isomorphism maps stable objects to stable objects.} 
\bigskip

The one-to-one correspondence between closed points is 
proven in \S\ref{spectralsect} by fairly technical 
local computations. The analogous correspondence for flat families is straightforward, but quite tedious, hence it will be omitted.

\subsection{Refined stable pairs and Gopakumar-Vafa expansion}\label{stpairsGV}

For the next step let $Y$ be the total space of $K_S$ which is 
identified to $S\times \IA^1$ by a choice of holomorphic symplectic form on $S$. Let
\[
Z^{\sf ref}_{PT}(Y;q,y,{\sf x}) = 1+\sum_{r\geq 1,\, c\in \IZ} PT^{\sf ref}_Y(r,c\,; y) q^c {\sf x}^r 
\]
be the generating function of refined stable pairs invariants 
of $Y$ with $\chi_1(F) = r[\Sigma_{n\ell}]$ and let 
\[ 
Z_{PT}(Y;q,{\sf x}) =1+ \sum_{r\geq 1,\, c\in \IZ} PT_Y(r,c) q^c {\sf x}^r 
\]
be its unrefined version. The refined Gopakumar-Vafa 
expansion conjectures the following formula 
\be\label{eq:refGVA}
{\rm ln}\, Z^{\sf ref}_{PT}(Y;q,y,{\sf x}) = -\sum_{s\geq 1} \sum_{r\geq 1} {{\sf x}^{sr}\over s}{y^{sre(g,m,n,\ell)} (qy^{-1})^{sd(g,m,n,\ell,r)/2} 
P_{n,\ell,r}(q^{-s}y^{-s}, y^s) \over (1-q^{-s}y^{-s})(1-q^sy^{-s})}~,
\ee
where $P_{n,\ell, r}(u,v)$ is the perverse Poincar\'e polynomial 
of the moduli space of stable Higgs bundles $\CH_{n,\ell,r}$
and
$$
e(g,m,n,\ell)={mn\ell(\ell-1)\over 2}+(g-1)\ell^2.
$$

The twisted wild version of the non-Abelian Hodge correspondence with $P=W$ conjecture \cite{hodgechar,Arithmetic_wild} identifies  $P_{n,\ell, r}(u,v)$ with the weighted 
Poincar\'e polynomial $WP_{n,\ell, r}(u,v)$ of the corresponding 
variety of Stokes data $\CS_{n,\ell,r}$, which is defined by 
\be\label{eq:weighted-Poincare}
WP_{n,\ell, r}(u,v) = \sum_{i,j} u^{i/2} (-v)^j{\rm dim}\, Gr^W_{i} H^j(\CS_{n,\ell,r}) ~,
\ee
where $Gr^W_{i}H^j(\CS_{n,\ell,r})$ are the successive quotients of the weight filtration on the cohomology of 
the character variety. By analogy with the character varieties without marked points treated in \cite{HRV}, 
one expects the weight filtration on cohomology to satisfy the condition $W_{2k}H^j(\CS_{n,\ell,r})= W_{2k+1}H^j(\CS_{n,\ell,r})$ for all $k,j$. Furthermore, 
as in \cite{hodgechar}, 
the wild variant of the $P=W$ conjecture is expected to identify $W_{2k}H^j(\CS_{n,\ell,r})$ with the degree $k$ 
subspace $P_kH^j(\CH_{n,\ell,r})$ of the perverse Leray filtration on the cohomology of the corresponding moduli space of Higgs bundles. This explains the factor $u^{i/2}$
in the right hand side of equation \eqref{eq:weighted-Poincare}. 

In conclusion, the refined Gopakumar-Vafa 
expansion yields conjectural explicit predictions for these polynomials provided the stable pairs theory of $Y$ 
can be explicitly computed. 

\subsection{Refined genus zero conjecture via torus knots} 

In this subsection $C$ will be the projective line and there will be a unique marked point $p$. These conditions  are essential 
for the conjecture derived below. Under such circumstances, 
there is a torus action ${\bf T} \times S\to S$ on the spectral surface $S$, which lifts easily to a torus action on $Y$ fixing the holomorphic three form. This localizes the theory to 
stable pairs with set theoretic support on a particular irreducible compact curve $\Sigma_{n\ell}\subset S$ preserved by the torus action. As shown in \S\ref{genzerosect}, 
this curve has a unique singular point of the form 
\[ 
v^\ell = w^{(n-2)\ell-1}.
\]
Reasoning by analogy to 
\cite[\S 4]{BPS_wild}, explicit predictions for refined stable pair invariants can be derived from 
the localization formalism of Nekrasov and Okounkov \cite{Membranes_Sheaves} using 
the refined colored generalization of the Oblomkov-Shende-Rasmussen conjecture \cite{OS,ORS}. Some details 
are spelled out in \S\ref{refCSpairs} for completeness. 

In order to write the resulting formulas, for any partition $\lambda$ let  $P_\lambda(s,t;{\sf z})$ denote the 
corresponding Macdonald symmetric functions, where ${\sf z} = (z_1, z_2, \ldots, )$ is an infinite set of formal variables. 
Then we define the principal specialization of Macdonald functions
\[
L_\lambda(q,y) = P_\lambda(t,s;\us)|_{s=qy,\ t=qy^{-1}}, \qquad \us= (s^{1/2}, s^{3/2}, \ldots)~, 
\]
and the framing factors
\be\label{framing}
g_\mu(q,y) = h_\mu(qy,qy^{-1}), \qquad h_\mu(s,t) = \prod_{\Box\in \mu} s^{a(\Box)} t^{-l(\Box)}.
\ee
Finally, for any pair of coprime positive integers $(p,q)$ let 
$W^{\sf ref}_{\lambda}(p,q; s,t)$ be the stable limit of the refined Chern-Simons invariant for the $(p,q)$ torus knot defined in \cite{refCS}. Alternative constructions based on  DAHA representations have been given in \cite{Cherednik:2011nr,cherednik2016daha,Ref_knots_Hilb}. 
Then the refined stable pair theory will be given by an expression of the form 
\be\label{eq:stpairsB} 
Z_{PT}^{\sf ref}(Y; q,y,{\sf x}) = 1 +\sum_{|\lambda|>0} 
L_{\lambda^t}(q,y) g_\lambda(q,y)^{-n\ell(\ell-1)+2\ell^2-1} W^{\sf ref}_{\lambda}(\ell, (n-2)\ell-1; qy, qy^{-1}) {\sf x}^{|\lambda|}.
\ee

Although no closed formula for $W^{\sf ref}_{\lambda}(\ell, (n-2)\ell-1; s,t)$ is known, refined Chern-Simons theory \cite{refCS}  provides a recursive algorithm which determines these invariants for arbitrary partitions in principle. This algorithm has been derived in \cite{Colored_HL} based on \cite{Super_evol} and is briefly summarized in \S\ref{refCSknots} for completeness. Implementing the algorithm of  \cite{Colored_HL}, one obtains explicit predictions for 
the weighted Poincar\'e polynomials $WP_{n,\ell,r}(u,v)$ 
for this particular case. Some examples are listed in Appendix 
\ref{examples}. For instance, the formula below has been conjectured in \cite{Super_evol} for the one-box partition, $\lambda = \Box$,
\be\label{1-box-torus}
W^{\sf ref}_{\Box}(\ell, (n-2)\ell-1;s,t)=\sum_{\mu \in \calP_{\ell} } \overline c_{\Box,\ell}^\mu(s,t) \gamma^\mu(s,t) h_\mu(s,t)^{2-n} P_\mu(s,t;\us) ~,
\ee
where the coefficients $\overline c_{\Box,\ell}^\mu(s,t)$ 
are defined through the identity of Macdonald functions
\[ 
P_{\Box}(s,t;{\sf z}^\ell) =\sum_{\mu \in \calP_{\ell}}
\overline c_{\Box,\ell}^\mu(s,t) P_{\mu}(s,t;{\sf z})~,
\]
and 
\[
\gamma^\mu(s,t)=\frac{1-s}{s^{1-|\mu|}-s}  \sum\limits_{(i,j)\in \mu} t^{1-i} s^{\mu_{i}-\mu_{1}+1-j} ~. 
\]
In fact, this yields explicit results for the polynomials $WP_{n,\ell,1}(u,v)$ 
with $\ell\in \{2,3\}$, as shown in Appendix \S\ref{refex}, Examples $(a)$ and $(b)$.

As supporting evidence for this conjecture, some localization computations of Poincar\'e polynomials of moduli spaces of spectral sheaves 
are carried out in \S\ref{locsect} for genus zero curves with one marked point.  As shown in detail in 
\S\ref{genzerosect} there is a torus action such that all 
fixed spectral sheaves are set theoretically supported on a 
fixed reduced irreducible curve $\Sigma \subset S$ with a unique marked 
point. Explicit results are then obtained for $n=4$, $\ell=2$, $r\geq 1$ (\S\ref{elliptic}), and for $n=5$, $\ell=2$, $r=2$ (\S\ref{genustwosect}). 

In \S\ref{elliptic}, the case that $\Sigma$ is a cuspidal elliptic curve is investigated. In this case, 
one can use the  Fourier-Mukai transform  \cite{Sheaves_genus_one} to prove that the Poincar\'e 
polynomial of the moduli space of spectral sheaves 
for $n=4$, $\ell=2$ is given by 
\[ 
P_{4,2,r}(v) = v^2+1 
\]
for any $r\geq 1$. As shown in Appendix \ref{refex},
this is in agreement with 
the $u=1$ specialization
of the results obtained from the formula \eqref{eq:stpairsB}
for $2\leq r \leq 3$. 

When $\Sigma$ is an 
arithmetic genus $2$ curve with a singular point of the form 
$v^2 = w^5$, the classification of the fixed loci for 
higher multiplicity $r\geq 2$ is a difficult problem involving 
fairly technical results on torsion free sheaves on singular plane curves. The explicit computations 
are carried out for $r=2$ in \S\ref{genustwosect}, and necessary technical results are proven in Appendices
\ref{torfreeapp} and \ref{spsheaves}. After a long computation, the final result
\[ 
P_{5,2,2}(v) = 2v^{10} + 4v^{8}+4v^{6}+3v^{4}+v^{2}+1
\]
also matches with the  $u=1$ specialization
of the result obtained from the formula \eqref{eq:stpairsB}.

\bigskip

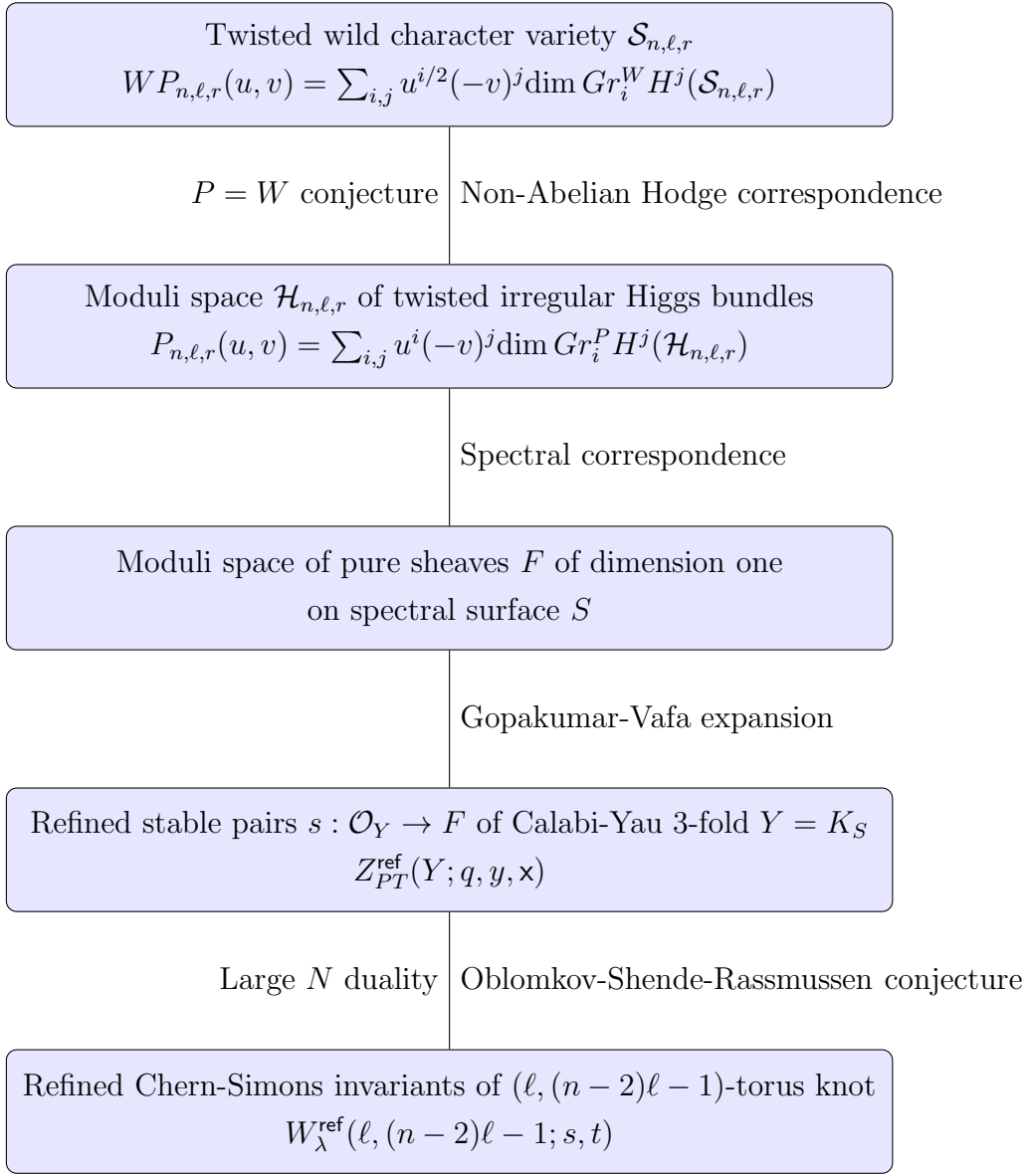
\begin{figure}\centering
% Define block styles
\tikzstyle{block} = [rectangle, draw, fill=blue!10, 
    text width=28em, text centered, rounded corners, minimum height=4em]
\tikzstyle{line} = [draw]
\begin{tikzpicture}[node distance = 3.5cm, auto]
    % Place nodes
    \node [block] (init) {Twisted wild character variety $\CS_{n,\ell,r}$ \\ ${WP_{n,\ell, r}(u,v) = \sum_{i,j} u^{i/2}(- v)^j{\rm dim}\, Gr^W_{i} H^j(\CS_{n,\ell,r})}$};
    \node [block, below of=init] (identify) {Moduli space $\CH_{n,\ell,r}$ of twisted irregular Higgs bundles  \\ ${P_{n,\ell,r}(u,v) = \sum_{i,j} u^i(- v)^j{\rm dim}\, Gr^P_iH^j(\CH_{n,\ell,r})}$};
    \node [block, below of=identify] (evaluate) {Moduli space of pure sheaves $F$ of dimension one \\ on spectral surface $S$};
    \node [block, below of=evaluate] (decide) {Refined stable pairs ${s:\CO_Y\to F}$ of Calabi-Yau 3-fold ${Y=K_S}$ \\ $Z^{\sf ref}_{PT}(Y;q,y,{\sf x})$};
    \node [block, below of=decide] (stop) {Refined Chern-Simons invariants of ${(\ell,(n-2)\ell-1)}$-torus knot  \\ ${W^{\sf ref}_{\lambda}(\ell, (n-2)\ell-1;s,t)}$};
    % Draw edges
    \path [line] (init) -- node[right] {Non-Abelian Hodge correspondence}node [left] {$P=W$ conjecture}(identify);
    \path [line] (identify) -- node {Spectral correspondence } (evaluate);
    \path [line] (evaluate) --node {Gopakumar-Vafa expansion } (decide) ;
    \path [line] (decide) -- node {Oblomkov-Shende-Rassmussen conjecture } node [left] { Large $N$ duality}(stop);
    \end{tikzpicture}
\caption{The broad map of the string theoretic approach to the cohomology groups of twisted wild character variety.}\label{fig:flow-chart}
\end{figure}

To conclude, the broad map of the string theoretic approach to the cohomology groups of a twisted wild character variety is depicted in Figure \ref{fig:flow-chart} by connecting all the above relationships.

\subsection{Unrefined all genus conjecture} 
For unrefined invariants, a higher genus generalization of the formula \eqref{eq:stpairsB} 
can be derived using degenerate Gromov-Witten theory 
by analogy with  \cite{Wild_deg}. In this approach the number $m$ of marked points can be arbitrary although the triples $(n,\ell,r)$ are the same for all marked points. Suppressing the details, the final formula 
is recorded below. 

For any partition $\rho$, let $S_\rho({\sf z})$ be the corresponding Schur function and we write its principal specialization
\[ 
S_\rho(\ut) = S_\rho(t^{1/2}, t^{3/2}, \ldots). 
\]
Let also 
\[
f_\rho(t) = \prod_{\Box \in \rho} 
t^{a(\Box)-l(\Box)} 
\]
be the unrefined specialization of the framing factor $h_\rho(s,t)$ in \eqref{framing}. 
Given two partitions $\lambda$ and $\mu \in \calP_{\ell|\lambda|}$, $\ell\geq 1$,  let 
the coefficients $c_\mu^\lambda$ be defined by the identity of Schur 
functions 
\[ 
S_\lambda({\sf z}^\ell) =\sum_{\mu \in \calP_{\ell|\lambda|}}
c_{\lambda,\ell}^\mu S_{\mu}({\sf z}). 
\] 
For any integers $\ell,r\geq 1$ and any partition
$\mu \in \calP_{r\ell}$ let 
\[ 
F_\mu(q) = f_\mu(q)^{-(n-1)+\ell^{-1}}\sum_{|\lambda|=r} f_\lambda(q)^{(n-1)\ell-1} S_{\lambda^t}(\uq) c_{\lambda,\ell}^\mu~.
\] 
Then the formula for unrefined stable pairs reads 
$$
Z_{PT}(Y; q,{\sf x}) = 1 + \sum_{r\geq 1} 
\sum_{\mu \in \calP_{r\ell}} {\sf x}^r F_\mu(q)^m S_{\mu^t}(\uq)^{2-2g-m} ,
$$
where $m$ is the number of marked points. 
For $g=0$ and $m=1$, using the identity $S_\mu(\uq) = f_\mu(q)^{-1} S_{\mu^t}(\uq)$, the above formula reduces to 
\[ 
\bal
Z_{PT}(Y; q,{\sf x})  & =  1 +\sum_{|\lambda|>0} 
\sum_{\mu\in \calP_{\ell|\lambda|}} {\sf x}^{|\lambda|} f_\lambda(q)^{(n-1)\ell-1} f_{\mu}(q)^{-(n-2)+\ell^{-1}}S_{\lambda^t}(\uq) S_\mu(\uq)c_{\lambda,\ell}^\mu\\
& = 1 +\sum_{|\lambda|>0} {\sf x}^{|\lambda|} f_\lambda(q)^{-n\ell(\ell-1)+2\ell^2-1} S_{\lambda^t}(\uq) 
W_\lambda(\ell, (n-2)\ell-1; q), \\
\eal
\]
where 
\be\label{eq:unrefinv} 
W_\lambda(\ell, (n-2)\ell-1; q) = f_\lambda(q)^{(n-2)\ell^2-\ell} \sum_{\mu\in \calP_{\ell|\lambda|}} c_{\lambda,\ell}^\mu f_\mu(q)^{-(n-2)+\ell^{-1}}
S_\mu(\uq)
\ee
is the large $N$ limit of the Rosso-Jones formula for $SU(N)$ quantum invariants
of the $(\ell, (n-2)\ell-1)$-torus knot. This is in agreement 
with the unrefined specialization of \eqref{eq:stpairsB}

To conclude, the unrefined specialization $y=1$ of the formula \eqref{eq:refGVA} 
yields predictions for the $E$-polynomials of twisted 
wild character varieties:
\be\label{eq:unrefGVA} 
{\rm ln}\,  Z_{PT}(Y; q,{\sf x})  = 
- \sum_{s\geq 1} \sum_{r\geq 1} 
{{\sf x}^{sr} \over s} {q^{sd(g,m,n,\ell,r)/2} E_{n,\ell,r}(q^{-s})
\over (1-q^s)(1-q^{-s})}
\ee
where the $E$-polynomials are the $v=1$ specializations of the weighted Poinca\'e polynomials \eqref{eq:weighted-Poincare}
$$
E_{n,\ell,r}(u) = \sum_{i,j} u^{i/2} (-1)^j  {\rm dim}\,Gr^W_{i} H^j(\CS_{n,\ell,r})~.
$$

In particular, for $r=1$ and arbitrary $g\geq 1$, $m\geq 1$, 
 one obtains 
$$ 
E_{n,\ell,1}(u) = q^{1-d(g,m,n,\ell,1)/2}
(1-q^{-1})^2\sum_{\mu \in \calP_{\ell}}  F_\mu(q)^m S_{\mu^t}(\uq)^{2-2g-m} \Big|_{q=u^{-1}}.
$$
Some explicit predictions for $r\geq 2$ via \eqref{eq:unrefGVA} are also listed in Appendix \ref{unrefex}. 

\subsection{The twisted formula of Hausel, Mereb and Wong} \label{sec:HMW}

$E$-polynomials of twisted wild character 
varieties were previously obtained by Hausel, Mereb and Wong \cite{Arithmetic_wild,Hausel_slides,Hausel_private} based on \cite{HLRV,HLRVII}. Their formula applies to character varieties related to $(\ell, k\ell+1)$ torus knots
in the framework of \cite{Fukaya_knots,Cluster_legendrian},
as opposed to the $(\ell, k\ell-1)$ knots considered here. 
However, the difference between these cases resides only in a 
different framing factor in the main formula, as explained below. 

To fix notation, recall that $m$ is the number of marked points. Two infinite sets of formal variables, 
${\sf x}_{(2a-1)}$, ${\sf x}_{(2a)}$, are assigned to the $a$-th marked point, for $1\leq a \leq m$. 
Then the generating function is defined by
\[ 
Z_{HMW}(z,w) = 1+\sum_{|\lambda|>0} \Omega_{\lambda}^{g,m,n}(z,w)
\prod_{i=1}^{2m}{\wH}_\lambda({\sf x}_{(i)}; z^2, w^2)~,
\]
where
$$
\Omega_\lambda^{g,m,n} = \prod_{\Box \in \lambda} {
  (-z^{2a(\Box)}w^{2l(\Box)})^{-mn-2g+2} (z^{2a(\Box)+1} -
  w^{2l(\Box)+1})^{2g} \over (z^{2a(\Box)+2}
  -w^{2l(\Box)})(z^{2a(\Box)} -w^{2l(\Box)+2})},
$$
for a Young diagram $\lambda$, and ${\wH}_\lambda({\sf x}; z^2,w^2)$ is the modified
  Macdonald function \cite{graded-rep,MDgeom} in the infinite set of variables ${\sf x} =
  (x_1, x_2, \ldots)$. 
  
Note that the only difference between the above expression for $\Omega_\lambda^{g,m,n}$ and the one used in \cite{Arithmetic_wild,Hausel_slides,Hausel_private}
is the factor $(-z^{2a(\Box)}w^{2l(\Box)})^{-mn-2g+2}$, which 
in loc. cit. reads $(-z^{2a(\Box)}w^{2l(\Box)})^{m(n-2)}$. Therefore, the above formula is in fact an analytic continuation to negative values of the exponent. 

Next let ${\mathbb H}_{g,m,n}(z,w; {\sf x}_{(1)}, \ldots, 
{\sf x}_{(m)})$ be defined by 
$$ {\rm
  ln}\, Z_{HMW}(z,w) = \sum_{k\geq 1} {(-1)^{(mn-1)|\mu|} {\mathbb H}_{g,m,n}(z^k,w^k;{\sf x}_{(1)}^k, \ldots, 
{\sf x}_{(m)}^k))\over
  (1-z^{2k})(w^{2k}-1) }.
$$
For each Young diagram $\rho$, let $h_\rho({\sf x})$ and $p_\rho({\sf x})$ be the complete and power sum symmetric functions respectively. 
Then the analytic continuation of the theorem of Hausel, Mereb and Wong states that 
$$
\bal
 E_{(g,m,n,\ell)}(u) =  u^{d(g,m,n,\ell,1)} \bigg\langle {\mathbb H}_{g,m,n}(u^{1/2},u^{-1/2};{\sf x}_{(1)}, \ldots, 
{\sf x}_{(m)})\, ,\, 
\bigotimes_{a=1}^m h_{(\ell-1,1)}({\sf x}_{(2a-1)})\otimes p_{(\ell)}
({\sf x}_{(2a)})\bigg\rangle \\
\eal
$$
where $\langle\ ,\ \rangle$ is the natural pairing on symmetric 
functions. 

This formula has been verified to be in agreement with \eqref{eq:unrefGVA} for $0\leq g \leq 5$, $1\leq m \leq 3$, 
$2\leq \ell \leq 3$ and $4\leq n \leq 6$. 

\bigskip

\

\noindent
{\bf Acknowledgments.} We owe special thanks to Philip Boalch and Tamas
Hausel for illuminating discussions and comments on the manuscript. We would also like to thank Davesh Maulik, Alex Soibelman
and Yan Soibelman for discussions. Wu-yen Chuang was supported by NCTS grant and
MOST grant 105-2115-M-002-015-MY2. The work of
Duiliu-Emanuel Diaconescu was partially supported by NSF grant
DMS-1501612.  During the preparation of this work Ron Donagi was
supported in part by NSF grant DMS 1603526 and by Simons HMS
Collaboration grant \# 390287 and Tony Pantev was supported in part by
NSF grant DMS 1601438 and by Simons HMS
Collaboration grant grant \# 347070. Satoshi Nawata would like to thank Keio University, Simons Center for Geometry and Physics, Rutgers University, Aspen Center for Physics, Banff International Research Station and Max-Planck Institute for Mathematics for warm hospitality where a part of work has been carried out.

\section{Meromorphic Higgs bundles and twisted wild character varieties}\label{nonabHodge}  

The goal of this section is to explain the relation between the twisted meromorphic Higgs bundles introduced in \S\ref{twistedhiggs} and twisted wild character varieties in two-step process. Applying the non-Abelian Hodge correspondence proven by Sabbah \cite{Harmonic_metrics}, Biquard and
Boalch \cite{Wild_curves}, one first obtains  a moduli space of flat connections with fixed singular parts at the marked points up to gauge transformations. 
Next, employing the construction of Boalch and Yamakawa
\cite{Twisted_wild}, one obtains a moduli space of 
Stokes data associated to the resulting flat connections, called in loc. cit. a twisted wild character variety.

\subsection{From Higgs bundles to connections} 

Recall that in the present framework, $C$ is a smooth projective curve over $\IC$ and
 $p_a\in C$, $1\leq a\leq m$ is a set of 
 pairwise distinct marked points on $C$. As in \S\ref{twistedhiggs}, given a triple  
 $(n,\ell,r)$ of positive integers with $n\geq 3$, one 
 has a moduli space of stable rank $N=r\ell$, degree $1$ 
 meromorphic Higgs bundles $(E,\Phi:E \to E\otimes K_C(D))$ on $C$. By construction, the restriction of the Higgs field 
 $\Phi|_{np_a}$ is equivalent to the fixed Laurent tail $dQ_{n,\ell,r}(z_a)$ by conjugation, where 
$$
Q_{n,\ell,r}(z)= {z^{1-n}\over 1-n} J_{\ell, r} + 
{z^{2-n}\over 2-n} E_{\ell,r}~,
$$
and the matrices $J_{\ell,r}, E_{\ell,r}$ are as in \eqref{eq:rJEformulas}.

% \[
%J_{\ell,r} = \left(\begin{array}{ccccc} 
%0 & 0 & \cdots & 0 & 0 \\
%{\bf 1}_r & 0 & \cdots & 0 & 0 \\
%\vdots & \vdots & \vdots & \vdots& \vdots\\
%0 & 0 & \cdots & {\bf 1}_r & 0\\
%\end{array}\right),\qquad 
%E_{\ell,r}= \left(\begin{array}{ccccc} 
%0 & 0 & \cdots & 0 & {\bf 1}_r \\
%0 & 0 & \cdots & 0 & 0 \\
%\vdots & \vdots & \vdots & \vdots& \vdots\\
%0 & 0 & \cdots & 0 & 0\\
%\end{array}\right).
%\]

In order to determine the local form of the flat connections related to such Higgs bundles by the non-Abelian Hodge correspondence, one has to diagonalize
the Laurent tail of the Higgs field on a suitable cyclic local cover. Hence, let $\Delta$ be a formal disk with coordinate $z$ and $\rho:{\widetilde \Delta}\to \Delta$ be the $\ell:1$ cover $z=w^\ell$. 
The main observation is that the 
pull-back of the local meromorphic one-form 
\[ 
\rho^*dQ_{n,\ell, r} = \ell w^{\ell-1} \left(w^{-\ell n}J_{\ell,r} + w^{\ell(1-n)}E_{\ell,r}\right) dw~,
\]
via $\rho$ is diagonalizable. 
More precisely, there exists a meromorphic $GL(N,\IC)$-valued function $g(w)$ such that 
\[ 
\rho^*dQ_{n,\ell, r} = \ell w^{\ell(1-n)}\left(g(w) 
\Lambda g(w)^{-1} \right) dw ~,
\]
where $\Lambda$ is the block-diagonal matrix 
\[
\Lambda  = \left(\begin{array}{ccccc} 
{\bf 1}_r & 0 & \cdots & 0 & 0 \\
 0 & \zeta_\ell {\bf 1}_r & \cdots & 0 & 0 \\
\vdots & \vdots & \vdots & \vdots& \vdots\\
0 & 0 & \cdots & 0 & \zeta_\ell^{\ell-1} {\bf 1}_r\\
\end{array}\right)
\]
with $\ell$-th root of unity $\zeta_\ell= e^{2\pi\sqrt{-1}/\ell}$. 
Explicitly, $g(w)$ can be written in the block form 
\[ 
g(w) = \big(\underbrace{g_0, \ldots, g_0}_{r}, \
\ldots, \underbrace{g_{\ell-1}, \ldots, g_{\ell-1}}_r
\big) 
\]
where for each $0\leq j \leq \ell-1$, $g_j$ is the block matrix 
\[ 
g_j = \left(\begin{array}{c} 
{\bf 1}_r  \\
 \zeta_\ell^{-j} w^{-1}{\bf 1}_r \\
\vdots \\
\zeta_\ell^{-j(\ell-1)} w^{1-\ell} {\bf 1}_r\\
\end{array}\right)~.
\]
The local model for the corresponding irregular connections is obtained 
by applying the gauge transformation 
$g(w)$ to the diagonal $GL(N,\IC)$-connection one-form
\be\label{eq:diagconnection}
\Gamma_{n,\ell,r} = -2 \ell w^{\ell(1-n)} \Lambda dw
\ee
on ${\widetilde \Delta}$. 
Suppressing the details of the computation, this yields 
the one-form connection $\rho^*A_{n,\ell,r}$ where 
\be\label{eq:connectionA}
A_{n, \ell, r} = -2 dQ_{n,\ell,r} - \ell^{-1} R {dz\over z} 
\ee
is a $\mathfrak{gl}(N,\IC)$-valued one-form on $\Delta$. 
The residue $R$ is the diagonal block-matrix 
\[ 
R = \left(\begin{array}{ccccc} 
0 & 0 & \cdots & 0 & 0 \\
 0 &  {\bf 1}_r & \cdots & 0 & 0 \\
\vdots & \vdots & \vdots & \vdots& \vdots\\
0 & 0 & \cdots & 0 & (\ell-1) {\bf 1}_r\\
\end{array}\right)~.
\]
In conclusion, the non-Abelian Hodge correspondence relates 
the moduli space $\CH_{n,\ell,r}$ with the following moduli space of irregular connections on $C$. 

For any triple
$(n,\ell, r)$ let 
$\CC_{n,\ell, r}$ denote the moduli space of 
pairs $(V,A)$ where $V$ is a rank $N=r\ell$, degree $0$ vector
bundle on $C$ and $A$ a meromorphic $\mathfrak{gl}(N,\IC)$-connection on $V$ subject to the following conditions: 
\begin{itemize}
\item[$(C.1)$] $A$ has a pole of order $n$ at each marked point $p_a$ and there exists a local trivialization of $V$ at $p_a$ which identifies it to 
$A_{n,\ell,r}$  up to holomorphic terms. 
\item[$(C.2)$] $A$ has a simple pole at $\infty$ with 
residue ${\bf 1}_N/2i\pi N$. 
\item[$(C.3)$] Any proper nontrivial sub-bundle $V'\subset V$ preserved by $A$ has degree ${\rm deg}(V')<0$. 
\end{itemize} 

\noindent 
Note that the extra marked point and condition $(C.2)$ are needed because the Higgs bundles on the other side of this correspondence have degree $1$ rather than 0. 
This is necessary in order to rule out strictly semistable objects which would render the moduli space singular. 

The variety of Stokes data associated to the type of  irregular connections satisfying conditions $(C.1)$-$(C.3)$ above is obtained from the construction of Boalch and Yamakawa \cite{Twisted_wild}. Using quasi-hamiltonian reduction, this construction generalizes 
previous results of Boalch \cite{Quasi_hamiltonian,Braiding_stokes} for untwisted wild 
character varieties. 
As a first step one has to determine the Stokes rays and associated Stokes groups 
for such connections. This is carried out below.

\subsection{Twisted Stokes data}\label{twistedStokes} 
The formal flat sections for the connection $d-\Gamma_{n,\ell,r}$ on ${\widetilde \Delta}$ are 
\be\label{eq:flatsections}
\psi_{a} = {\rm exp}\left({2\ell\over (n-1)\ell-1}\zeta_\ell^a w^{1+\ell(1-n)}\right)
\ee
with $0\leq a\leq \ell-1$, each of them having multiplicity $r$. 
For each pair of flat sections $(\psi_{a}, \psi_{b})$, $a\neq b$,  the associated anti-Stokes rays are the  directions 
of steepest descent as $\rho \to 0$ of the ratio 
\[ 
\psi_{a}\psi_{b}^{-1} = {\rm exp}\left( {2\ell\over (n-1)\ell-1}
(\zeta_\ell^a-\zeta_\ell^b) w^{1+\ell(1-n)} \right)
\]
with $w= \rho e^{2\pi \sqrt{-1} \phi}$, $\rho>0$.
In other words, the anti-Stokes rays are given by the directions $\phi$ for which the exponent 
\[
(\zeta_\ell^a-\zeta_\ell^b) e^{2\pi \sqrt{-1} (1+\ell(1-n))\phi}
\]
is real and negative. 
As in \cite[Definition 7.4]{Braiding_stokes},  
the Stokes group ${\mathbb S}{\rm to}_\phi$ associated to each such direction $\phi$ is the subgroup of $GL(N,\IC)$ consisting of all matrices $S_\phi$ of the form 
$$
S_\phi = {\bf 1}_N + (\sigma_{b,a}) ~,
$$
where $\sigma_{b,a}$, $0\leq a,b \leq \ell-1$, are $r\times r$ blocks such that $\phi$ is an anti-Stokes 
direction for $(a,b)$. This is a unipotent subgroup of $GL(N,\IC)$. 
As shown in \cite[Appendix A]{Twisted_wild}, the Stokes groups associated to the irregular singular point $z=0$ for connections of the local form \eqref{eq:connectionA}
are identified with the Stokes groups ${\mathbb S}{\rm to}_\phi$ for the connections of the local form \eqref{eq:diagconnection} with $\phi$ in an 
interval $[\beta,\ \beta+1/\ell)$. The choice of $\beta$ reflects the choice of a base point on $S^1$ and affects 
the Stokes groups only by conjugation in $GL(N, \IC)$. 
This is inconsequential for the construction of wild character varieties.

Furthermore, for untwisted connections, the set of all anti-Stokes directions 
is naturally divided into \emph{complete half-periods}
\cite[Lemma 7.13]{Braiding_stokes}. The subgroups ${\mathbb S}{\rm to}_\phi$ with $\phi$ in any given complete half-period directly span the unipotent radical 
of a parabolic subgroup of $GL(N, \IC)$ with Levi subgroup $H$, which is the centralizer of the leading singular term of the connection \eqref{eq:diagconnection}. Hence $H$ is the subgroup of 
invertible block diagonal matrices 
\[
\left(\begin{array}{ccccc} 
\ast & 0 & \cdots & 0 & 0 \\
0& \ast & \cdots & 0 & 0 \\
\vdots & \vdots & \vdots & \vdots& \vdots\\
0 & 0 & \cdots & 0 & \ast\\
\end{array}\right)
\]
with arbitrary $r\times r$ blocks on the diagonal.

For twisted connections\footnote{We are very grateful to Philip Boalch for very helpful explanations on this point.}, one will also have 
\emph{incomplete half-periods} since the range of values of $\phi$ is truncated to lie in the interval $[\beta,\ \beta+1/\ell)$. The Stokes subgroups corresponding to an incomplete  half-period span a smaller unipotent subgroup of $GL(N,\IC)$. The details for $\ell \geq 6$ are displayed below. The case $\ell=2$ was considered in 
\cite[Example 6.2]{Twisted_wild} while the cases $3\leq \ell\leq 5$ will be left to the reader since the computations are very similar the ones below.

For the sake of simplicity, let us set $\kappa = (n-1)\ell-1$. 
Then the 
anti-Stokes rays for a pair $(a,b)$, $0\leq a,b\leq \ell-1$, $a\neq b$, are of the form 
\[
\phi_j(a,b)= {a+b\over 2\kappa\ell} + {j\over 2\kappa} - {1\over 4\kappa} 
\]
with $j \in \IZ$ subject to the selection rule 
\[
a>b \ \Leftrightarrow \ j \ {\rm even}, \qquad {a<b}\ \Leftrightarrow\ j\ {\rm odd}. 
\]
Let
\[
\beta = {1\over 4\kappa} - \epsilon, \qquad {1\over \kappa\ell} < \epsilon < {3\over 2\kappa\ell}~,
\]
and note that $1\leq a+b \leq 2\ell-3$ since  $0\leq a,b\leq \ell-1$, $a\neq b$. 
Then there are $2n-3$ complete half-periods in the 
interval $[\beta,\ \beta+1/\ell)$ consisting of the following sets of rays 
\be\label{eq:compleven}
\{\phi_j(a,b)\,|\, \ell-2\leq a+b \leq 2\ell-3,\ a>b\} 
\cup 
\{\phi_{j+1}(a,b)\,|\, 1\leq a+b \leq \ell-3,\ a<b\}
\ee
for $0\leq j \leq 2n-4$, $j$ even, respectively
 \be\label{eq:complodd}
\{\phi_j(a,b)\,|\, \ell-2\leq a+b \leq 2\ell-3,\ a<b\} 
\cup 
\{\phi_{j+1}(a,b)\,|\, 1\leq a+b \leq \ell-3,\ a>b\}
\ee
for $0\leq j \leq 2n-4$, $j$ odd. One is then left with the  incomplete half-period 
\be\label{eq:inchalfperiod} 
\{\phi_{2n-3}(a,b)\,|\, \ell-2\leq a+b \leq 2\ell-5,\ a<b\} 
\cup 
\{\phi_{2n-2}(a,b)\,|\, 1\leq a+b \leq \ell-5,\ a>b\}.
\ee
The unipotent subgroup $U_j$ associated to a complete half period of the form \eqref{eq:compleven} consists of Stokes matrices 
${\bf 1}_N + (\sigma_{b,a})$
supported on 
\[
\{ 1\leq a+b \leq \ell-3,\ a<b\} \cup \{ \ell-2 \leq a+b \leq 2\ell-3,\ a>b\}.
\]
The unipotent subgroup $U_j$ associated to a complete half period of the form \eqref{eq:complodd} consists of Stokes matrices ${\bf 1}_N + (\sigma_{b,a})$ supported on 
\[
\{ 1\leq a+b \leq \ell-3,\ a>b\} \cup \{ \ell-2 \leq a+b \leq 2\ell-3,\ a<b\}.
\]
For illustration, in the case of $\ell=6$, this yields block matrices of the form 
\[ 
j\ {\rm even:}\quad 
\left(\begin{array}{cccccc}
{\bf 1}_r & 0 & 0 & 0 & \ast & \ast \\
\ast & {\bf 1}_r & 0 & \ast & \ast & \ast \\ 
\ast & \ast & {\bf 1}_r & \ast & \ast & \ast \\ 
\ast &  0 & 0 & {\bf 1}_r & \ast & \ast \\
0 & 0 & 0 & 0 & {\bf 1}_r & \ast \\
0 & 0 & 0 & 0 & 0  & {\bf 1}_r\\
\end{array} \right)\quad\qquad
j\ {\rm odd:}\quad 
\left(\begin{array}{cccccc}
{\bf 1}_r & \ast & \ast & \ast & 0 & 0 \\
0 & {\bf 1}_r & \ast & 0 & 0 & 0 \\ 
0 & 0 & {\bf 1}_r & 0 & 0 & 0 \\ 
0 & \ast & \ast & {\bf 1}_r & 0 & 0 \\
\ast & \ast & \ast & \ast & {\bf 1}_r  & 0\\
\ast & \ast & \ast & \ast & \ast  & {\bf 1}_r\\
\end{array} \right)~.
\]
The unipotent subgroup $U_{2n-3}$ associated to the incomplete half-period \eqref{eq:inchalfperiod} consists of Stokes matrices supported on 
\[
\{ 1\leq a+b \leq \ell-5,\ a>b\} \cup \{ \ell-2 \leq a+b \leq 2\ell-5,\ a<b\}.
\]
Again, for $\ell=6$, these are the block matrices of the form 
\[ 
\left(\begin{array}{cccccc}
{\bf 1}_r & \ast & 0 & 0 & 0 & 0 \\
0 & {\bf 1}_r & 0 & 0 & 0 & 0 \\ 
0 & 0 & {\bf 1}_r & 0 & 0 & 0 \\ 
0 & \ast & \ast & {\bf 1}_r & 0 & 0 \\
\ast & \ast & \ast & \ast & {\bf 1}_r  & 0\\
\ast & \ast & \ast & 0 & 0  & {\bf 1}_r\\
\end{array} \right)~.
\]

To conclude this section, we also note that up to conjugacy, the formal 
monodromy of connections of the form \eqref{eq:connectionA} at the 
singular point $z=0$  is given by the block clock matrix 
\be\label{eq:formalmon}
{\sf M} =  \left(\begin{array}{ccccc} 
0 & {\bf 1}_r & \cdots & 0 & 0 \\
\vdots & \vdots & \vdots & \vdots& \vdots\\
0 & 0 & \cdots & 0 & {\bf 1}_r\\
{\bf 1}_r & 0 & \cdots & 0 & 0\\
\end{array}\right).
\ee
This follows easily from the effect of a rotation $w\mapsto \zeta_\ell w$ on the space of the formal flat sections
\eqref{eq:flatsections}.

\subsection{Twisted wild character varieties}

This section concludes the construction of the twisted 
wild character varieties following \cite{Twisted_wild}. 
Recall that $H\subset GL(N,\IC)$ is the  subgroup of 
invertible block diagonal matrices 
\[
\left(\begin{array}{ccccc} 
\ast & 0 & \cdots & 0 & 0 \\
0& \ast & \cdots & 0 & 0 \\
\vdots & \vdots & \vdots & \vdots& \vdots\\
0 & 0 & \cdots & 0 & \ast\\
\end{array}\right)
\]
with arbitrary $r\times r$ blocks on the diagonal. 
Let also $H(\partial), H({\overline \partial}) \subset GL(N,\IC)$ be the subsets of matrices of block form 
\[ 
h = \left(\begin{array}{ccccc} 
0 & \ast & \cdots & 0 & 0 \\
\vdots & \vdots & \vdots & \vdots& \vdots\\
0 & 0 & \cdots & 0 & \ast\\
\ast & 0 & \cdots & 0 & 0\\
\end{array}\right),\qquad 
{\overline h} = 
\left(\begin{array}{ccccc} 
0 & 0 & \cdots & 0 & \ast \\
\ast & 0 & \cdots & 0 & 0 \\
\vdots & \vdots & \vdots & \vdots& \vdots\\
0 & 0 & \cdots & \ast & 0\\
\end{array}\right)~,
\]
respectively. 
Note that there is an $H\times H$ action on 
$H(\partial)$ sending $(k_1, k_2) \times h \mapsto 
k_1 h k_2^{-1}$, and similarly an $H\times H$ action 
on $H({\overline \partial})$ sending 
$(k_1, k_2) \times h \mapsto 
k_2 h k_1^{-1}$. Furthermore, the map $h \mapsto h^{-1}$ is an $H$-equivariant isomorphism $H({\partial}) {\buildrel \sim \over \longto} H({\overline \partial})$.
Note also that any $h\in H(\partial)$ determines
twisted $H$-conjugacy classes $\CC_h = \{ k h k^{-1}\, |\, k\in H\}$, $ \CC_{h^{-1}} = \{ k h k^{-1}\, |\, k\in H\}$ in $H({\partial})$ and 
$H({\overline \partial})$ respectively. 
In particular, this holds for the formal monodromy $\sf M$ 
in \eqref{eq:formalmon}, which is an element of $H({\partial})$. 

Using \cite[Theorems 21, 24 and Corollary 22]{Twisted_wild}, 
the variety of twisted Stokes data associated to the flat connections parameterized by $\CC_{n,\ell, r}$ is constructed as follows. Let   
\[
\CA =\left( GL(N,\IC)\times GL(N,\IC)\right)^{g} \times 
 GL(N,\IC)^{m}\times H(\partial)^m \times 
 \Big(\prod_{0\leq j \leq 2n-3} U_j\Big)^{m}~,
\]
where the factors in the product $\prod_{0\leq j\leq 2n-3} U_j$ are ordered such that $j$ decreases from left to right. 
Elements of $\CA$ will be denoted by 
\[
{\alpha}= ({\sf A}_i, {\sf B}_i, {\sf C}_1,\ldots, {\sf C}_m, h_1,
\ldots, h_m, S_{1,j}, \ldots, S_{m,j}),
\]
where  $1\leq i \leq g$, $0\leq j \leq 2n-3$.  
For each $1\leq i \leq g$ let $(\sfA_i,\sfB_i)=\sfA_i\sfB_i\sfA_i^{-1} \sfB_i^{-1}$. 
We define $\mu: \CA \to GL(N,\IC) \times H({\overline \partial})^m$ as
\[ 
\bal
\mu(\alpha) = \bigg(({\sf A_1},{\sf B_1})\cdots ({\sf A_g},{\sf B_g}) & {\sf C}_1^{-1} 
h_1 \mathop{\prod}_{0\leq j \leq 2n-3} S_{1,j}\,  
\sfC_1
\cdots \\
&  {\sf C}_m^{-1} 
h_m 
\mathop{\prod}_{0\leq j \leq 2n-3} S_{m,j}\,
  \sfC_m\, ,\,  h_1^{-1}, \ldots, h_m^{-1}\bigg)~,\\
\eal
\]
where all products $\mathop{\prod}_{0\leq j \leq 2n-3}$ are ordered in such a way that $j$ decreases from left to right. 
Note that there is a natural action $GL(N,\IC)\times H^m\times \CA\to\CA$ 
given by 
\[ 
\bal
(g,k_1, \ldots, k_m) \times \alpha \mapsto 
\bigg(g\sfA_i g^{-1}, g\sfB_i g^{-1}, 
 k_a\sfC_ag^{-1}, k_ah_ak_a^{-1},
k_aS_{a,j}k_a^{-1}
\bigg) 
\eal 
\]
where $1\leq i \leq g$, $1\leq a \leq m$, and 
$0\leq j \leq2n-3$. 
The map $\mu$ is equivariant with respect to the action $GL(N, \IC) \times H^m \times 
GL(N,\IC)\times H({\overline\partial})^m\to 
GL(N,\IC)\times H({\overline\partial})^m$, 
\[
(g,k_a) \times ({\sf C}, {\overline h}_a) \mapsto (g {\sf C} g^{-1}, k_a {\overline h}_a k_a^{-1})
\]
on the target. 
Then the variety of Stokes data for the flat connections 
parameterized by  $\CC_{n,\ell,r}$ is the quotient 
\[ 
\CS_{n,\ell,r} = \mu^{-1}(e^{-2\pi\sqrt{-1}/N}{\bf 1}_N,\CC_{{\sf M}^{-1}}, \ldots, \CC_{{\sf M}^{-1}})/GL(N,\IC)\times H^m~,
\]
 where ${\sf M}$ is the formal monodromy 
 \eqref{eq:formalmon}. According to \cite[Thm. 1]{Twisted_wild}, this is a smooth quasi-projective variety.

In conclusion, the above construction associates a moduli space $\CH_{n,\ell,r}$ of  twisted meromorphic Higgs bundle to a twisted wild character variety $\CS_{n,\ell,r}$. Then the 
 twisted wild variant of the $P=W$ conjecture of de Cataldo, Hausel and Migliorini \cite{hodgechar} implies that the perverse Poincar\'e polynomial of $\CH_{n,\ell,r}$ is identical to the weighted Poincar\'e polynomial of $\CS_{n,\ell,r}$. The main point of the present paper is that explicit conjectural formulas for 
 the former can be derived from string theory and enumerative geometry. As the first step, we will build a spectral construction for twisted meromorphic Higgs bundles  in the next section.

\section{Spectral correspondence}\label{spectralsect}

In this section $C$ will be an arbitrary smooth projective curve, $p\in C$ a point in $C$ and $n \geq 3$, $\ell\geq 2$ fixed  integers. Let $M = K_C(D)$ where $D=np$. Abusing notation, the total space of $M$ will be also denoted by $M$ and the canonical projection to $C$ will be denoted by $\pi:M \to C$. 
Let $f_0\subset M$ be the fiber of $M$ over $p$, let $C_0\subset M$ be the zero section, and let $q_0$ be 
their transverse intersection point.

\subsection{The holomorphic symplectic surface}\label{gencurve}
Let $(\CU,z,u)$ be an affine coordinate chart on $M$ 
such that $(\CU\cap C_0, z|_{C_0})$ is an affine chart 
on $C_0$ centered at $q_0$, which is canonically identified with a chart on $C$ centered at $p$. 
The vertical coordinate $u$ is defined in such a way that 
\be\label{eq:loctautsect}
y\big|_\CU = u {dz\over z^n} 
\ee
where $y\in H^0(M,\pi^*M)$ is the tautological section. 
Suppose that $\Sigma_0\subset M$ is a compact irreducible reduced divisor in the linear system 
$|\ell C_0|$ such that $\Sigma_0$ intersects $f_0$ at $q_0$ 
with multiplicity $\ell$, and $\Sigma_0$ is smooth at $q_0$. 
Thus, $\Sigma_0$ is isomorphic to the divisor 
$$
z-u^\ell = 0~,
$$
in the infinitesimal neighborhood of $f_0$ in $M$. 
The symplectic surface constructed below is independent of $\Sigma_0$, but it will be helpful to keep track of the 
strict transforms of such a divisor through the sequence of 
blow-ups. 

The symplectic surface is obtained by a sequence of $n\ell$ successive blow-ups of $M$ starting with a blow-up 
at $q_0$. The surface obtained at the $i$-th blow-up will be denoted by $M_i$ while the $i$-th strict transform of $\Sigma_0$ will be denoted by $\Sigma_i$. 
The first $\ell-1$ successive blow-ups are given in affine coordinates by 
\be\label{eq:blowupA}
z_i = z_{i+1} u_i, \qquad u_{i}=u_{i+1}, \qquad 
0 \leq i \leq \ell-2,
\ee
where $z_0=z$ and $u_0=u$. 
The strict transforms of $\Sigma_0$ are locally given by 
$$
u_i^{\ell-i} = z_i, \qquad 1\leq i \leq \ell-1~.
$$
The $\ell$-th blow-up is given by 
$$
u_{\ell-1} = u_{\ell} z_{\ell-1}~, \qquad z_{\ell}=z_{\ell-1}~,
$$
and the strict transform $\Sigma_\ell$ is locally given by 
$$
u_\ell =1~.
$$
Here we mean by ``locally'' the infinitesimal neighborhood 
of the most recent exceptional divisor. 
Next one blows-up the transverse intersection point 
between $\Sigma_\ell$ and the exceptional divisor, 
$u_\ell-1=0$, $z_{\ell}=0$. Therefore, in affine coordinates the blow-up map is 
$$
u_{\ell}-1= z_{\ell}u_{\ell+1},\qquad z_{\ell+1}=z_\ell,
$$
while the strict transform of $\Sigma_\ell$ is given by $u_{\ell+1}=0$. 

\begin{figure}
%\hspace{30pt}
\vspace{-10pt}
\setlength{\unitlength}{1mm}
\begin{picture}(200,100)
\thicklines
\put(10,5){\line(1,0){125}} 
\put(15,2){\line(-1,2){10}}
\multiput(7,28)(0,3){3}{\huge .}
\put(4,43){\line(1,2){10}}
\put(16,55){\line(-1,2){12}}
\put(6,71){\line(0,1){25}}
\put(7,65){\line(2,1){25}}
\put(27,77){\line(2,-1){25}}
\multiput(60,72)(3,0){3}{\huge .}
\put(78,64){\line(2,1){25}}
{\color{blue}\qbezier(107,83)(95,80)(115,65)}
%{\color{blue}\qbezier(120.5,5)(122,22)(130,45)}
\put(105,65){$\Sigma_{n\ell}$}
\put(65,0){$C_{0,{n\ell}}$}
%\put(119,2){$\sigma_{n\ell}$}
\put(9,13){$\Xi_1$}
\put(5,43){$\Xi_{\ell-1}$}
\put(15,53){$\Xi_\ell$}
\put(16,67){$\Xi_{\ell+1}$}
\put(37,72){$\Xi_{\ell+2}$}
\put(91,75){$\Xi_{n\ell}$}
\put(83,63){$q_{n\ell}$}
\put(1,84){$f_{n\ell}$}
\end{picture}
\caption{Schematic description for $n\ell$ successive blow-ups $M_{n\ell}\to M$.}
\label{Blowup} 
\end{figure}
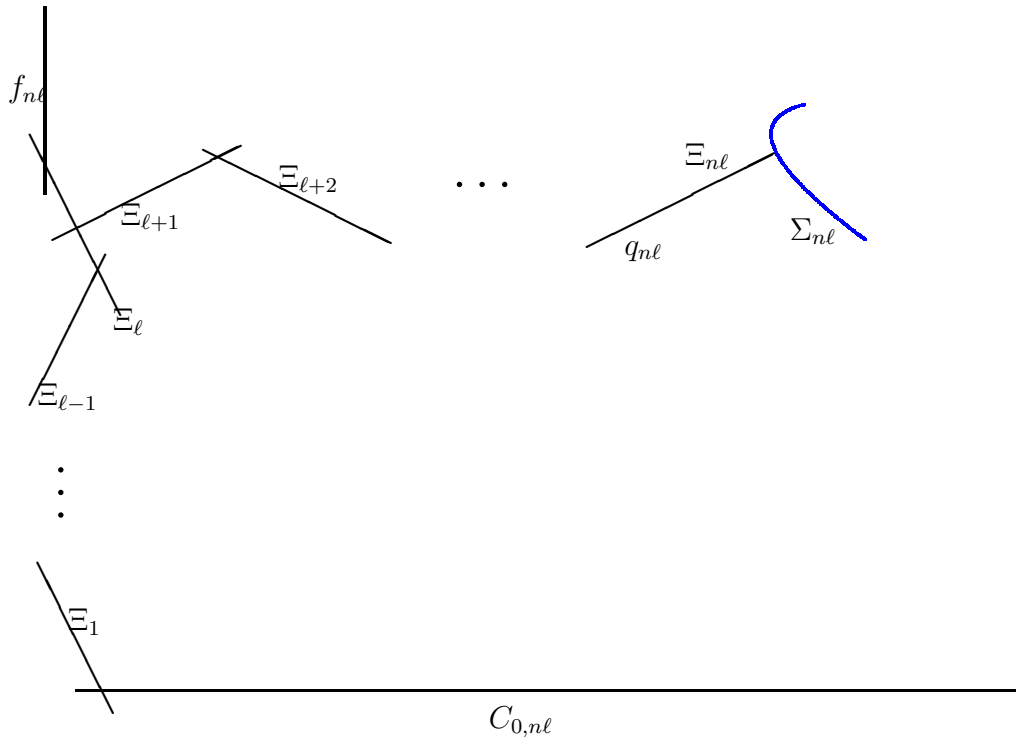

The construction is then concluded by a sequence of $(n-1)\ell-1$ successive blow-ups. Starting with a blow-up at $u_{\ell+1}=z_{\ell+1}=0$, one recursively blows-up the 
transverse intersection point of the strict transform of $\Sigma_{\ell+1}$ with the exceptional divisor obtained 
from the previous blow-up. In affine coordinates, this sequence is given by 
$$ 
u_{i} = z_{i+1} u_{i+1},\qquad z_i=z_{i+1} \qquad \ell+1 \leq i \leq n\ell-1.
$$
The $i$-th strict transform of $\Sigma_0$ is given 
by 
$$
u_{i} =0, \qquad  \ell+1 \leq i \leq n\ell.
$$
Note that the natural projection $\rho_{n\ell}: M_{n\ell} \to M$ 
maps $\Sigma_{n\ell}$ isomorphically to $\Sigma_0$. 
The resulting configuration of exceptional curves, 
and the strict transform $\Sigma_{n\ell}$ are 
shown in Figure \ref{Blowup}. The strict transforms of $f_0, \Sigma_0$ and the zero section $C_0$ in $M$ are denoted by $f_{n\ell}, \Sigma_{n\ell}$ and $C_{0,n\ell}$, respectively. 
The self-intersection number of $\Sigma_{n\ell}$ in $M_{n\ell}$ is 
\be\label{eq:selfintA}
(\Sigma_{n\ell})^2 = mn\ell(\ell-1)+2(g-1)\ell^2.
\ee
The canonical class of the surface $M_{n\ell}$ is given by 
\begin{align}
K_{M_{n\ell}} = & -nf_{n\ell}-(n-1)\Xi_1 - 2(n-1)\Xi_2 -\cdots - \ell(n-1)\Xi_\ell \cr
& -(\ell(n-1)-1)\Xi_{\ell+1} - \cdots  
- \Xi_{n\ell-1}. \nonumber
\end{align}
Here, the last exceptional divisor $\Xi_{n\ell}$ has multiplicity 0 in the above formula. Therefore, the complement 
\[
S= M_{n\ell} \setminus\left(f_{n\ell}\cup \bigcup_{i=1}^{n\ell-1}\Xi_i\right)
\]
is a holomorphic symplectic surface. The strict transform $\Sigma_0$ is entirely contained in $S$
while the divisor $\Xi_{n\ell}$ restricts to an affine line in $S$. In fact, by construction, any compactly supported divisor on $S$ must be linearly equivalent to a multiple $r\Sigma_{n\ell}$ for some integer $r\geq 1$.

\subsection{From sheaves on $S$ to Higgs bundles on $C$} 

Let $F$ be a pure dimension one sheaf on $S$ with proper support. Hence ${\rm det}(F) = \CO_S(r \Sigma_{n\ell})$ with $r\geq 1$, and the set theoretic support of $F$ intersects $\Xi_{n\ell}$ at finitely many points. The direct image $\pi_{S*}(F)$ is a rank $N=r\ell$ locally free sheaf $E$ on $C$. Since the projection $\pi_S: S \to C$ factors through the blow-up map $\rho_{n\ell}: M_{n\ell}\to M$, there is a natural Higgs field $\Phi: E \to E \otimes M$ induced by multiplication by the tautological section 
$y \in H^0(M, \pi^* M)$. Let $z$ be a local coordinate on $C$ centered at $p$. Given an arbitrary local holomorphic frame for $E$ 
at $p$, the
 restriction $\Phi|_{D}$ to the $n$-th infinitesimal 
neighborhood $D=np$ is written as 
\[ 
\Phi \big|_D = \varphi_D {dz\over z^n} ~,
\]
where $\varphi_D$ takes the value in the ring $\CM_{N}(\CO_D)$ of $N\times N$ matrices with coefficients in $\CO_D$
\[
\varphi_D = \varphi|_D \in \CM_{N}(\CO_D)\simeq \CM_{N}(\IC[z]/(z^n)), \qquad N= r\ell. 
\]
Note that the equivalence class of $\varphi_D$ with respect to conjugation in 
$GL(N,\IC[z]/(z^n))$ does not depend on the choice of a local holomorphic frame for $E$ at $p$. 
Then the main result proved in this subsection is 
\bigskip

{\it 
Let 
\be\label{eq:lochiggsA}
\varphi_0(\ell,r) = J_{\ell,r}+ z E_{\ell,r} \in \CM_{N}(\IC[z]/(z^n))
\ee
where $J_{\ell,r}, E_{\ell,r}$ are the $N\times N$  matrices defined in \eqref{eq:rJEformulas}. Then $\varphi_D$ is equivalent to $\varphi_0(\ell,r)$ by conjugation in
$GL(N,\IC[z]/(z^n))$. }

\bigskip

This is proven by a (somewhat lengthly) local computation. To simplify the notation, the affine local coordinates $(z_{n\ell}, u_{n\ell})$ on $M_{n\ell}$ will be denoted by 
\[
z_{n\ell}=t, \qquad u_{n\ell} =s.
\]
Then the local equation of $\Sigma$ is $s=0$. The blow-up map $\rho_{n\ell}: M_{n\ell} \to M$ is locally given by 
\be\label{eq:blowupG}
z = t^\ell(1+t^{(n-1)\ell}s)^{\ell-1}, \qquad 
u = t (1+t^{(n-1)\ell}s).
\ee
The space of local sections of $F$ in this affine coordinate chart is a $\IC[t,s]$-module $\Gamma_F$
which is isomorphic to $\IC[t]^{\oplus r}$ as a 
$\IC[t]$-module. Multiplication by $s$ determines an 
endomorphism of $\IC[t]$-modules $\eta_s : \IC[t]^{\oplus r}\to \IC[t]^{\oplus r}$.
Using the blow-up equations 
\eqref{eq:blowupG}, $\IC[t]$ has a structure of $\IC[z]$-module, which is compatible with the multiplication by $s$. 
The space of local holomorphic sections of $E = \pi_{S*}(F)$ is isomorphic to $\Gamma_F$ with the 
induced $\IC[z]$-module structure. The Higgs field is determined by multiplication by the tautological section $y$, which is related to the local coordinate $u$ by 
\eqref{eq:loctautsect}.  
 Since only the infinitesimal structure is of interest, it suffices to work with the local completion ${\widehat \Gamma}_F\simeq \IC[[t]]^{\oplus r}$ of $\Gamma_F$ at 
$t=0$. Again, ${\widehat \Gamma}_F$ has a $\IC[[t]][s]$-module structure, as well as a $\IC[[z]][u]$-module structure. The main goal is to make the latter explicit. 

\medskip
{\bf Step 1}. Any compactly supported divisor on $S$ in the linear system $r\Sigma$ is locally defined by an equation of the form 
\be\label{eq:locdet}
s^r + \sum_{i=0}^{r-1} f_i(t) s^i=0~,
\ee
where $f_i(t)$, $1\leq i \leq r-1$ are polynomial functions of $t$. Since the determinant of $F$ is such a divisor by assumption, ${\widehat \Gamma_F}$ will be annihilated by a polynomial as in the left hand side of \eqref{eq:locdet}.  This implies that the 
endomorphism 
\be\label{eq:invend}
{\bf 1}+t^{(n-1)\ell}\eta_s: \IC[[t]]^{\oplus r} \to 
\IC[[t]]^{\oplus r}
\ee
is invertible. In particular, this implies that 
${\widehat \Gamma_F}$ is a torsion free $\IC[[z]]$-module, hence it must be isomorphic to a free $\IC[[z]]$-module. 

\medskip
{\bf Step 2.} Next one has to construct an explicit isomorphism 
\[ 
\xi : \IC[[z]]^{\oplus N} {\buildrel \sim \over 
\longto} {\widehat \Gamma}_F
\]
of $\IC[[z]]$-modules.
For this purpose it will be convenient to use the canonical isomorphism
 $$\IC[[z]]^{\oplus N} \simeq \IC[[z]]^{\oplus \ell} 
 \otimes_{\IC[[z]]} \IC[[z]]^{\oplus r}.$$
  Let $e_i\otimes f_a$, $1\leq i \leq \ell$, $1\leq a\leq r$ be the canonical generators of the tensor product. 
Then let $\xi: \IC[[z]]^{\oplus \ell} 
 \otimes_{\IC[[z]]} \IC[[z]]^{\oplus r}\to {\widehat \Gamma}_F$ be the $\IC[[z]]$-module morphism defined by 
\be\label{eq:zmorphismA}
\xi( e_i\otimes f_a ) = t^{i-1} h_a, \qquad 
\xi( z\, e_i\otimes f_a ) = t^{\ell+i-1}( {\bf 1}+ 
t^{(n-1)\ell} \eta_s)^{\ell-1}(h_a)~,
\ee
where $h_a$, $1\leq a \leq r$ are the standard generators of ${\widehat \Gamma}_F\simeq \IC[[t]]^{\oplus r}$ as a $\IC[[t]]$-module. Since ${\widehat \Gamma}_F$ is a locally free $\IC[[z]]$-module by construction, in order to prove that $\xi$ is an isomorphism it suffices to prove that the induced $\IC$-linear map  
\[ 
{\bar \xi} : \IC^{\oplus \ell}\otimes_\IC \IC^{\oplus r} \to {\widehat \Gamma}_F/ z{\widehat \Gamma}_F 
\] 
is an isomorphism of vector spaces. 

Since $z \equiv 0$ mod $t^\ell$, there is a natural map 
\be\label{eq:ztmap}
{\widehat \Gamma}_F/ z{\widehat \Gamma}_F \to 
{\widehat \Gamma}_F/ t^\ell\, {\widehat \Gamma}_F. 
\ee
Given the explicit form of $\xi$ in \eqref{eq:zmorphismA}, it follows that ${\bar \xi}$ is an isomorphism if and only if the map \eqref{eq:ztmap} 
is an isomorphism. However this is immediate since the endomorphism \eqref{eq:invend} is invertible. Therefore, ${\bar \xi}$ is indeed an isomorphism. 

\medskip
{\bf Step 3}. Next, multiplication by $t$  
induces a $\IC[[z]]$-module endomorphism of 
$\IC[[z]]^{\oplus \ell} \otimes_{\IC[[z]]} \IC[[z]]^{\oplus r}$ via $\xi$. Let  
$\phi_t \in \CM_{N}( \IC[[z]])$ be the corresponding matrix with respect to the standard generators $e_i\otimes f_a$. Similarly, let 
$\phi_u \in \CM_{N}( \IC[[z]])$ be the matrix 
associated to 
multiplication by $u$. 
Then the first equation in \eqref{eq:blowupG} implies the congruence 
\[
\phi_t  \equiv \varphi_{0}(\ell,r)\ \ {\rm mod}\ z^n.
\]
Abusing notation, $\varphi_{0}(\ell,r)=J_{\ell,r}+zE_{\ell,r}$ is regarded here as an element of $\CM_N(\IC[[z]])$ rather than 
$\CM_N(\IC[z]/(z^n))$ as in \eqref{eq:lochiggsA}. The distinction will be clear from the context.

Now let ${\sf S}\in \CM_{r}(\IC[[t]])$ be the matrix 
associated to the morphism $\eta_s$ with respect to the standard generators $h_a$. Then ${\sf S}$ has an expansion
\[ 
{\sf S} = \sum_{k\geq 0} {\sf S}_k t^k 
\]
with ${\sf S}_k \in \CM_{r}(\IC)$ complex $r\times r$ matrices. Then the second equation in \eqref{eq:blowupG} implies the congruence 
\[
\phi_u \equiv \varphi_{0}(\ell,r) + z^{n-1} \sum_{k=1}^{\ell-1} 
{\sf S}_{k-1} J_{\ell,r}^k \quad {\rm mod}\ z^n
\]
where $J_{\ell,r}$ is as in \eqref{eq:rJEformulas}. 
The sum in the right hand side can be easily computed as 
\[ 
\sum_{k=1}^{\ell-1} 
{\sf S}_{k-1} J_{\ell,r}^k = \left(\begin{array}{cccccc} 
0 & 0 & \cdots & 0 & 0 & 0 \\
{\sf S}_0 & 0 & \cdots & 0 & 0 & 0 \\
{\sf S}_1 & {\sf S}_0 & \cdots  & 0 & 0 & 0 \\
\vdots & \vdots & \vdots & \vdots & \vdots & \vdots \\
{\sf S}_{\ell-2} & {\sf S}_{\ell-3}  & \cdots &  {\sf S}_1 & {\sf S}_0 & 0\\
\end{array}\right).
\]

\medskip
{\bf Step 4}. To conclude the proof, let 
\[ 
g = {\bf 1}_{N} + z^{n-1} 
\left(\begin{array}{cccccc} 
(\ell-1) {\sf S}_0 & 0 & \cdots & 0 & 0 & 0 \\
(\ell-2){\sf S}_1 & (\ell-2){\sf S}_0 & \cdots & 0 & 0 & 0 \\
(\ell -3) {\sf S}_2& (\ell-3) {\sf S}_1 & \cdots  & 0 & 0 & 0 \\
\vdots & \vdots & \vdots & \vdots & \vdots & \vdots \\
{\sf S}_{\ell-2} & {\sf S}_{\ell-3}  & \cdots &  {\sf S}_1 & {\sf S} _0 & 0\\
0 & 0  & \cdots & 0 & 0 & 0\\
\end{array}\right) \in GL(N, \IC[z]/(z^n)).
\]
Then a straightforward computation shows that 
\[
g \phi_u g^{-1} \equiv \varphi_{0}(\ell,r) \quad {\rm mod}\ z^n.
\]

\subsection{The inverse direction} 

Let $(E,\Phi)$, $\Phi: E \to E\otimes M$ 
 be a stable rank $N=r\ell$ irregular Higgs bundle on $C$. 
Suppose that there exists a local trivialization of $E$ at $p$ such that $\Phi\big|_D$ is equivalent to 
 $\varphi_0(\ell,r){dz/z^n}$ by conjugation in $GL(N, \IC[z]/(z^n))$
as in the main statement proved in the previous subsection. 
In the following it will be more convenient to work with a 
different representative of this conjugacy class, 
\be\label{eq:lochiggsB} 
{\widetilde\varphi}_0(\ell,r) = \left(\begin{array}{ccccc} 
J_{\ell,1}+zE_{\ell,1} & 0 & \cdots & 0 & 0 \\
0 & J_{\ell,1}+zE_{\ell,1} & \cdots & 0 & 0 \\
\vdots & \vdots & \vdots & \vdots& \vdots\\
0 & 0 & \cdots & 0 & J_{\ell,1}+zE_{\ell,1}\\
\end{array}\right),
\ee
where the $\ell\times \ell$ matrices $J_{\ell,1}, E_{\ell,1} $ are the $r=1$ specializations of \eqref{eq:rJEformulas}.
Note that ${\widetilde\varphi}_0(\ell,r)$ and $\varphi_0(\ell,r)$ in \eqref{eq:lochiggsA}  are related by conjugation in $GL(N,\IC)$. 
Proceeding  by analogy with 
 \cite[\S 3.3]{BPS_wild}, one constructs a stable pure dimension 
 one sheaf $F$ on $S$ such that $\pi_{S*}(F)$ is isomorphic to the Higgs bundle $(E\otimes M^{-1}, 
 \Phi\otimes {\bf 1}_{M^{-1}})$. This is carried out in detail below. 

\medskip
{\bf Step 1.} Let $y$ be the tautological section of $\pi^*M$ and consider the monad complex
\[
\pi^* (E\otimes M^{-1}) \xrightarrow{y - \pi^*\Phi} \pi^*E 
\]
on $M$. This is an injective morphism and 
its cokernel, $Q_0$ is a pure dimension one sheaf on $M$. 
Taking direct images, one obtains a Higgs bundle  $(\pi_*Q_0, \pi_*(y \otimes {\bf 1}_{Q_0}))$ on $C$ which 
is isomorphic to $(E,\Phi)$. Let $\Gamma_0\subset M$ denote the spectral curve given by ${\rm det}(y -\pi^*\Phi)=0$. 
This is a compact divisor in $M$ which contains the scheme theoretic support of $Q$ as a closed subscheme. Let $q_0$ be the transverse intersection point between the fiber $M_p$ and the zero section of $M$. Then the following holds:
\bigskip 

{\it (S.1) The set theoretic intersection of $\Gamma_0$ 
with the fiber $M_p$ consists of the single point $q_0$. Moreover, there is an isomorphism 
\be\label{eq:pointisomA} 
Q_0 \otimes \CO_{q_0} \simeq \CO_{q_0}^{\oplus r}.
\ee
}
\bigskip

Let $(u,z)$ be an affine local coordinate chart on $M$, where $z$ is a local affine coordinate on $C$ centered at $p$, and $u$ is a vertical local coordinate such that 
\[
y|_{\CU} = u {dz\over z^n}. 
\]
Let $\CU\simeq {\rm Spec}\, \IC[[z]][u]$ be the infinitesimal neighborhood of the fiber 
$z=0$ in $M$. For any sheaf $F$ on $M$ let $\Gamma(\CU,F)$ denote the $\IC[[z]][u]$-module 
consisting of sections of the local completion of $F$ at 
$z=0$. This will be loosely referred to as the restriction 
of $F$ to $\CU$ in the following. 
Using the monad construction, the space of local sections $\Gamma(
\CU, Q_0)$ is isomorphic 
to the cokernel of a morphism 
\[ 
u {\bf 1}_{N}-\phi_0:\IC[[z]][u]^{\oplus N} \to \IC[[z]][u]^{\oplus N} 
\]
of $\IC[u,z]$-modules. By assumption, here 
\[
\phi_0 \equiv {\widetilde \varphi}_0(\ell,r) \quad {\rm mod}\ z^n,
\]
where ${\widetilde \varphi}_0(\ell,r)$ is the block matrix 
defined in equation \eqref{eq:lochiggsB}. 
Note also that the local equation of $\Gamma_0$ in $\CU$ is 
\[
{\rm det}(u {\bf 1}_{N}-\phi_0) = 0. 
\]
The left hand side of this equation satisfies  
\[
{\rm det}(u {\bf 1}_{N}-\phi_0) \equiv 
(u^\ell-z)^r \quad {\rm mod}\ z^n. 
\]
In particular, the set theoretic intersection of $\Gamma_0$ 
with the fiber $M_p$ consists of the single point $q_0$ given by  $u=z=0$. 

To conclude the local analysis, note there is 
also an isomorphism 
\[ 
Q_0\otimes \CO_{q_0} \simeq {\rm Coker}(f_{0,q_0}) \otimes \CO_{q_0}
\]
where $f_{0,q} : \IC^{N} \to \IC^{N}$
is the linear map
\[ 
\left(\begin{array}{ccccc} 
J_{\ell,1} & 0 & \cdots & 0 & 0 \\
0 & J_{\ell,1} & \cdots & 0 & 0 \\
\vdots & \vdots & \vdots & \vdots& \vdots\\
0 & 0 & \cdots &  0 & J_{\ell,1}\\
\end{array}\right). 
\]
This implies the isomorphism \eqref{eq:pointisomA}. 
  
\medskip
{\bf Step 2}. Let $\rho_1: M_1 \to M$ be the first blow-up in the construction of the spectral surface $S$, and let $\Xi_1$ be the exceptional divisor. Let $q_1$ be the center of the next blow-up in the sequence, which in this case is the transverse intersection point of $\Xi_1$ and the strict transform of the fiber 
$M_p$. Then 
there is a surjective morphism 
\[
\rho_1^* Q_0 \twoheadrightarrow \rho_1^*(Q_0\otimes \CO_q) \simeq \CO_{\Xi_1}^{\oplus r}.
\]
Let $Q_1$ be its kernel. Note that $\rho_1^*Q_0$ is pure of dimension one since  $\rho_{1*}\rho_1^* Q_0 \simeq Q_0$ by the projection formula. Therefore, if $T \subset \rho_1^*Q_0$ is a zero-dimensional subsheaf, then $\rho_{1*}T\subset Q_0$ is also zero-dimensional, leading to a contradiction. 
In particular, $Q_1$ is also pure of dimension one. Let $\Gamma_1$ be its determinant.
Then the following statement will be proven below by local computations:
\bigskip 

 {\it (S.2) The set theoretic intersection of $\Gamma_1$ with $\Xi_1$ consists of the single point $q_1$ and there is an isomorphism 
\[ 
Q_1\otimes \CO_{q_1} \simeq \CO_{q_1}^{\oplus r}~.
\]
Moreover, $\Gamma_1= \rho_1^{-1}(\Gamma_0)-r\Xi_1$ is the strict transform of $\Gamma_0$.}
\bigskip 

Let $(u_1,z_1)$ be the affine coordinate chart on 
$M_1$ such that the blow-up map is locally given by 
\[ 
z = z_1u_1, \qquad u = u_1 
\]
as in \eqref{eq:blowupA}. Let $\CU_1\simeq {\rm Spec}\, \IC[[z_1]][u_1]$ be an infinitesimal neighborhood of the strict transform of the fiber $z=0$.  Then the space of local sections $\Gamma(
\CU_1, \rho_1^*Q_0)$ is isomorphic 
to the cokernel of the morphism of $\IC[u,z]$-modules
\[ 
\rho_1^*\phi_0:\IC[[z_1]][u_1]^{\oplus N} \to \IC[[z_1]][u_1]^{\oplus N} 
\]
which satisfies 
\[ 
\rho_1^* \phi_0 \equiv u_1 {\bf 1}_{N} -
\left(\begin{array}{ccccc} 
J_{\ell,1} + z_1u_1E_{\ell,1} & 0 & \cdots & 0 & 0 \\
0 & J_{\ell,1}+ z_1u_1E_{\ell,1} & \cdots & 0 & 0 \\
\vdots & \vdots & \vdots & \vdots& \vdots\\
0 & 0 & \cdots &   0 & J_{\ell,1}+ z_1u_1E_{\ell,1} \\
\end{array}\right)\quad {\rm mod}\ (z_1u_1)^n. 
\]
In order to find an explicit local presentation for $Q_1$ let ${\sf U}_{1}$ be the $\ell\times \ell$ matrix 
\[
{\sf U}_{1} = \left(\begin{array}{ccccc} 
u_1 & 0 & 0 & \cdots & 0 \\
0 & 1 &  0 & \cdots & 0 \\
\vdots & \vdots & \vdots & \cdots  & \vdots \\ 
0 & 0 & 0 & \cdots & 1 \\
\end{array}
\right)~,
\]
and let $\gamma_1$ be the 
block diagonal matrix 
\[
\gamma_1=\left(\begin{array}{ccccc} 
{\sf U}_{1} & 0 & \cdots & 0 & 0 \\
0 & {\sf U}_{1} & \cdots & 0 & 0 \\
\vdots & \vdots & \vdots & \vdots& \vdots\\
0 & 0 & \cdots &   0 & {\sf U}_{1} \\
\end{array}\right),
\]
with $r$ identical blocks on the diagonal. 
Let also $\widetilde{{\sf U}}_1$ be the $\ell\times \ell$ matrix 
\[ 
\widetilde{{\sf U}}_{1} = \left(\begin{array}{ccccc} 
1 & 0 & 0 & \cdots & 0 \\
0 & u_1 &  0 & \cdots & 0 \\
\vdots & \vdots & \vdots & \cdots  & \vdots \\ 
0 & 0 & 0 & \cdots & u_1 \\
\end{array}
\right)~,
\]
and $\wgamma_1$ be the block matrix 
\[ 
\wgamma_1=\left(\begin{array}{ccccc} 
\widetilde{{\sf U}}_{1} & 0 & \cdots & 0 & 0 \\
0 & \widetilde{{\sf U}}_{1} & \cdots & 0 & 0 \\
\vdots & \vdots & \vdots & \vdots& \vdots\\
0 & 0 & \cdots &   0 & \widetilde{{\sf U}}_{1} \\
\end{array}\right),
\]
consisting of $r$ diagonal blocks. 
Then, there is the matrix identity 
\be\label{eq:matrixidA}
u_1{\bf 1}_{N}-
\left(
\begin{array}{ccccc} 
J_{\ell,1} + z_1u_1E_{\ell,1} & 0 & \cdots & 0 & 0 \\
0 & J_{\ell,1}+ z_1u_1E_{\ell,1} & \cdots & 0 & 0 \\
\vdots & \vdots & \vdots & \vdots& \vdots\\
0 & 0 & \cdots &   0 & J_{\ell,1}+z_1u_1E_{\ell,1} \\
\end{array}\right) = \gamma_1 (\wgamma_1-\varphi_1)~,
\ee
where 
\[
\varphi_1 = 
\left(\begin{array}{ccccc} 
 J_{\ell,1} +z_1E_{\ell,1} & 0 & \cdots & 0 & 0 \\
0 & J_{\ell,1}+z_1E_{\ell,1} & \cdots & 0 & 0 \\
\vdots & \vdots & \vdots & \vdots& \vdots\\
0 & 0 & \cdots &   0 & J_{\ell,1} + z_1E_{\ell,1} \\
\end{array}\right).
\]
This implies that there is a commutative diagram 
\[ 
\xymatrix{ 
\IC[[z_1]][u_1]^{\oplus N} \ar[r]^-{\phi_1}
\ar[d]_-{\bf 1} &  \IC[[z_1]][u_1]^{\oplus N} 
\ar[d]^-{\gamma_1} \\
\IC[[z_1]][u_1]^{\oplus N} \ar[r]^-{\rho_1^*\phi_0}\ar[d]_-{0} &  \IC[[z_1]][u_1]^{\oplus N} \ar[d]^-{\jmath_1} \\
0 \ar[r] & (\IC[[z_1]][u_1]/(u_1))^{\oplus r} ~,\\}
\]
where $\jmath_1$ is the canonical projection and the difference $\delta_1=\phi_1-\varphi_1$ satisfies 
the congruence relations
\[ 
(\delta_1)_{i,j} \equiv \begin{cases} z_1^nu_1^{n-1}, & {\rm if}\ i \equiv 1\ {\rm mod}\ \ell  \\
z_1^nu_1^{n}, & {\rm  otherwise}
\end{cases}
\]
for $1\leq i,j \leq N$. 
Note further that $\gamma_1$ is injective and $\IC[[z_1]][u_1]/(u_1)$ is isomorphic to the space of local sections $\Gamma(\CU_1, \CO_{\Xi_1})$. 
Therefore, one obtains an isomorphism of $\IC[[z_1]][u_1]$-modules 
\be\label{eq:strictA}
\Gamma(\CU_1, Q_1)\simeq 
{\rm Coker}(\phi_1).
\ee

Now let $\Gamma_1$ be the determinant of $Q_1$. 
Then, the isomorphism \eqref{eq:strictA} implies that $\Gamma_1$ is locally given by 
\[ 
{\rm det}(\wgamma_1-\phi_1) =0
\]
in $\CU_1$. In particular, its scheme theoretic intersection with $\Xi_1$ in this particular chart is given by 
\[
u_1=0,\qquad {\rm det}(\wgamma_1-\phi_1) =0.
\]
However, $\delta_1|_{u_1=0}=0$ since $n\geq 3$, 
hence 
\begin{align}\nonumber
& {\rm det}(\wgamma_1-\phi_1)|_{u_1=0} = \cr
&  {\rm det} \left(\begin{array}{ccccc} 
 I_{1,1}-J_{\ell,1} -z_1E_{\ell,1} & 0 & \cdots & 0 & 0 \\
0 & I_{1,1}-J_{\ell,1} -z_1E_{\ell,1} & \cdots & 0 & 0 \\
\vdots & \vdots & \vdots & \vdots& \vdots\\
0 & 0 & \cdots &   0 & I_{1,1}-J_{\ell,1} -z_1E_{\ell,1}\\
\end{array}\right) = (-1)^{r\ell} z_1^r~,
\end{align}
where $I_{1,1}=\widetilde{{\sf U}}_{1}|_{u_1=0} $ is the $\ell\times\ell$ matrix with 1 at (1,1) entry and 0 at the other entries.
Moreover, one can easily check that there are no other intersection points between $\Gamma_1$ and $\Xi_1$. Precisely speaking, one has to check that the point at $\infty$ on $\Xi_1$ in the coordinate chart $(z_1,u_1)$ is not an intersection point. 
To this end let $(z_1',u_1')$ be the second standard 
chart on the blow-up, with transition functions 
\[ 
z_1' = z_1^{-1}, \qquad u_1' = u_1z_1. 
\]
Then the local equation of $\Gamma_1$ in this chart is 
\[
(z_1')^r\, {\rm det}(\wgamma_1-\phi_1)\bigg|_{\substack{z_1= (z_1')^{-1}\\ u_1=z_1'u_1'}} =0. 
\]
The left hand side of this equation is equal to $(-1)^{r\ell}$ when restricted on $\Xi_1=\{ u_1'=0\}$. Thus, there are no further intersection points. 

Consequently, the scheme theoretic intersection 
of $\Gamma_1$ and $\Xi_1$ is the closed subscheme of $\CU_1$ given by $u_1=0$, $z_1^{r-1}=0$. Moreover, using the matrix identity \eqref{eq:matrixidA} it also follows that 
\[
\Gamma_1 = \rho_1^{-1}(\Gamma_0) -r\Xi_1
\]
is the strict transform of $\Gamma_0$. 

To conclude, let $q_1$ be the closed point defined by $z_1=u_1=0$. 
Then, the isomorphism \eqref{eq:strictA} implies that 
$Q_1\otimes \CO_{q_1} \simeq \CO_{q_1}^{\oplus r}$. 

\medskip
{\bf Step 3}. Proceeding recursively, one constructs a sequence of pure dimension one sheaves $Q_i$ on the successive blow-ups $M_i$, $1\leq i \leq n\ell$ such that
\bigskip

{\it (S.3) The set theoretic intersection between determinant $\Gamma_i$ of $Q_i$ and the 
$i$-th exceptional divisor consists only of the $(i+1)$-th blow-up center $q_i$ for $1\leq i \leq n\ell-1$. Moreover, there is an isomorphism 
$$
Q_i \otimes \CO_{q_i} \simeq \CO_{q_i}^{\oplus r} 
$$
for all $1\leq i \leq n\ell-1$. 
}
\bigskip 

This is proven by local computations analogous to those at Step 2, and the details are omitted. 

Note that the statement $(S.3)$ does not hold for the last blow-up in the sequence. Namely the determinant $\Gamma_{n\ell}$ generically will intersect the 
exceptional divisor $\Xi_{n, \ell}$ at finitely many arbitrary points such that the total intersection multiplicity is $r$. At the same time, the $\Gamma_{n\ell}$ will not intersect any of the strict transforms of the previous exceptional divisors, $\Xi_i$, $1\leq i \leq n\ell-1$. 

Moreover, by construction, the irregular Higgs bundle obtained by pushing $Q_{n\ell}$ forward to $C$ is isomorphic to $(E\otimes M^{-1}, \Phi\otimes {\bf 1}_{M^{-1}})$. 

\subsection{Isomorphism of moduli stacks} 

The constructions carried in the previous subsections yield an isomorphism of moduli stacks of stable objects. Equivalence of stability is completely analogous to 
\cite[\S3.2, 3.3]{BPS_wild} while the extension to flat families does not pose any significant problems. Therefore, the details will be omitted. 

This concludes the spectral construction for twisted irregular Higgs bundles. As explained in \S\ref{stpairsGV}, this is 
a central element of the string theoretic approach to the cohomology groups of twisted wild character varieties. Another important step is the connection to refined Chern-Simons invariants  of 
torus knots, which is reviewed in the next section.

\section{Refined stable pairs via torus knots}\label{refCSpairs}

As discussed in \cite{BPSPW,Par_ref,BPS_wild}, through the spectral construction, perverse Betti numbers of  moduli spaces of stable 
irregular Higgs bundles can be interpreted as D2-D0 BPS degeneracies in string theory on the canonical bundle $Y=K_S$ of the spectral surface $S$. The mathematical theory of D2-D0 BPS states has been constructed as stable pairs on the three-fold $Y$ \cite{stabpairsI}. Thus, we shall show an explicit conjectural formula  for refined stable pair invariants on the three-fold $Y$ by using string theoretic framework. The formulation is parallel to the case of torus links so that we refer to \cite[\S 4]{BPS_wild} for more detail. 
In this section, we assume that $C$ is the projective line with a unique marked point $p$, \textit{i.e.} $g=0$ and $m=1$. As shown in \S\ref{genzerosect}, in this case there is a rank one torus action ${\bf T}\times S\to S$ on the spectral surface $S$ which preserves  a unique compact reduced spectral 
curve $\Sigma_{n\ell}$. 

A stable pair on $Y$ consists of a compactly
supported pure dimension one sheaf $F$ on $Y$ equipped with a
generically surjective section $s:\CO_{Y}\to F$.  
In general, one can define refined stable pair invariants using the motivic theory of Kontsevich and Soibelman \cite{wallcrossing}. 
In the present case, one can also employ the approach of Nekrasov and Okounkov \cite{Membranes_Sheaves} using the torus 
action ${\bf T}\times Y \to Y$ which is the unique lift of the natural $\IC^\times$ action ${\bf T}\times S\to S$ fixing the holomorphic three-form on $Y$. In fact the two constructions are equivalent, as shown in \cite{HRV_proof,Non_arch_sheaves}.

 It turns out that the above torus action 
 localizes the theory on stable pairs $(F,s)$ 
 where $F$ is set theoretically supported on 
 the torus invariant spectral curve 
 $\Sigma_{n\ell}\subset S$. There is a disjoint union of fixed loci $\mathcal{P}_\mu$ labelled by Young diagram $\mu$. Each such diagram corresponds to a torus invariant scheme theoretic thickening of the reduced curve $\Sigma_{n\ell}$ in $Y$. Using the formalism of  \cite{Membranes_Sheaves}, one is then led to conjecture that the refined stable pair partition function takes the form
$$
Z^{\sf ref}_{PT}(Y;q,y,{\sf x}) = 
\sum_{\lambda} {V}^{(n,\ell)}_{\lambda}(q,y) 
L_{\lambda^t}(q,y)  {\sf x}^{|\lambda|} ~,
$$
where 
$$L_\lambda(q,y) = P_\lambda(t,s;\us)|_{s=qy,\ t=qy^{-1}}~, \qquad  \us= (s^{1/2}, s^{3/2}, \ldots)$$
is one leg refined vertex along a preferred direction. Here ${V}^{(n,\ell)}_{\lambda}(q,y)$
is a vertex factor associated to the unique singular point of the curve $\Sigma_{n\ell}$. 
 
Since it is hard to perform the direct localization computation of ${V}^{(n,\ell)}_{\lambda}(q,y)$, a more effective strategy is to derive a conjectural formula for  
${V}^{(n,\ell)}_{\lambda}(q,y)$ via the Oblomkov-Shende-Rasmussen conjecture \cite{OS,ORS}. More specifically, one has to employ their refined colored generalization derived in 
\cite{DHS} from large N duality. 
The main observation is that $\Sigma_{n\ell}$ 
is a rational curve in $S$ with a unique singular point of the form 
\[ 
v^\ell = w^{(n-2)\ell-1}.
\] 
Then, the intersection of the curve $\Sigma_{n\ell}$ with a small three-sphere centered at $(0,0)\in \IC^2$ creates the 
$(\ell, (n-2)\ell-1)$-torus knot.
Hence, the refined colored variant of \cite{OS,ORS} tells us that explicit conjectural formulas for 
${V}^{(n,\ell)}_{\lambda}(q,y)$ can be derived in terms of the  refined Chern-Simons invariants of the $(\ell, (n-2)\ell-1)$-torus knot. To this end, we review  the construction of the refined Chern-Simons invariants below. We also note 
that the unrefined specialization of this claim has been proven by Maulik \cite{Homfly_pairs}.

\subsection{Refined Chern-Simons invariants of torus knots}\label{refCSknots} 

First let us review the refined invariants  \cite{refCS} of the $(p,q)$ torus knot in $S^3$. Aganagic and Shakirov proposed a refinement of $SU(N)$ Chern-Simons theory with level $k$.
Given a $(p,q)$ torus knot in $S^3$, the 
  refined Chern-Simons theory of level $k$ associates a Hilbert space $\mathcal{H}_{N,k}$ to the boundary torus of the knot by decomposing $S^3$ into two solid tori. The Hilbert space $\mathcal{H}_{N,k}$  is spanned by a basis $\{| P_\lambda \rangle \}$ of Macdonald polynomials labeled by Young diagrams $\lambda$ inscribed in a $k\times N$ rectangle, which correspond to the unknot colored by $\lambda$ in the solid torus. The mapping class group $SL(2,\mathbb{Z})$ of the boundary torus acts on the Hilbert space $\mathcal{H}_{N,k}$, which will realize a torus knot in the solid torus. This action can be understood by introducing a knot operator $\mathcal{O}_\lambda^{p,q}$ where 
$$
\mathcal{O}_\lambda^{p,q}|\varnothing\rangle\in \mathcal{H}_{N,k}
$$
is a vector associated to  the solid torus with the $(p,q)$ torus knot colored by $\lambda$ inside. In particular, $\mathcal{O}_\lambda^{1,0}$ creates the unknot colored by $\lambda$  as $\mathcal{O}_\lambda^{1,0} |\varnothing\rangle=| P_\lambda \rangle$, and its action on the basis is 
$$
\mathcal{O}_\lambda^{1,0}| P_\mu \rangle=\sum_{\rho\,\in\calP_{|\lambda|}} N_{\lambda\mu}^\rho| P_\rho \rangle
$$
where $N_{\lambda\mu}^\rho$ are the $(s,t)$-Littlewood-Richardson coefficients associated to Macdonald polynomials: $P_\lambda P_\mu= \sum_{\rho\, \vdash |\lambda|+|\mu|}N_{\lambda\mu}^\rho P_\rho $. 
Then, the action of $SL(2,\mathbb{Z})$ is given by
$$
\mathcal{O}^{p,q}_{\lambda}=K\mathcal{O}^{1,0}_{\lambda}K^{-1}~,\qquad K=\begin{pmatrix}p&q\\\ast&\ast\end{pmatrix}\in SL(2,\mathbb{Z})~.
$$
This action can be explicitly realized via the modular $S$ and $T$ matrices associated to Macdonald polynomials \cite{cherednik1995macdonald,kirillov1996inner,
etingof1996representation,refCS}:
\begin{equation}
\label{S and T}
S_{\lambda\mu}=P_{\lambda}(t^{\rho}s^{\mu})P_{\mu}(t^{\rho})~,\qquad T_{\lambda\mu}=\delta_{\lambda\mu}s^{\frac{1}{2}\sum_{i}\lambda_i(\lambda_i-1)}t^{\sum_{i}\lambda_i(i-1)}~,
\end{equation}
where $\rho$ is the Weyl vector of $\mathfrak{sl}(N)$. Here the parameters $(s,t)$ are identified by
$$
s=\exp\left(\frac{2\pi i }{k+\beta N} \right)~, \qquad t=\exp\left(\frac{2\pi i \beta}{k+\beta N} \right)~,
$$
and $s=t$ is the unrefined limit.
Since the three-sphere can be obtained by identifying the boundaries of two solid tori by the modular $S$-transformation, the refined Chern-Simons invariants of the $(p,q)$ torus knot in $S^3$ are then defined as
\begin{equation}
\label{CS evaluation}
\mathcal{W}^{p,q}_{\lambda,N}(s,t):=\langle \varnothing|S\mathcal{O}^{p,q}_{\lambda}|\varnothing \rangle~.
\end{equation}
by inserting the extra $S$-matrix. In fact, one can also define the refined invariants by using the $PSL(2,\mathbb{Z})$ action on DAHA  \cite{Cherednik:2011nr,cherednik2016daha}. In \cite{Ref_knots_Hilb} it is proven that there exists a unique stable limit $\mathcal{W}^{p,q}_\lambda(a,s,t)$ for large $k$ and $N$ such that
$$
\mathcal{W}^{p,q}_{\lambda,N}(s,t)=\mathcal{W}^{p,q}_\lambda(a=t^N,s,t)~.
$$
An illuminating example is the refined Chern-Simons invariant of the unknot that is expressed as
$$
P_\lambda^{\,\star} :=\mathcal{W}_\lambda^{1,0}(a,s,t)= \prod_{\Box\in \lambda} \frac{a^{\frac12}s^{\frac{a'(\Box)}{2}}t^{-\frac{l'(\Box)}{2}} -
(a^{\frac12}s^{\frac{a'(\Box)}{2}}t^{-\frac{l'(\Box)}{2}} )^{-1}} {s^{\frac{a(\Box)}{2}}t^{\frac{l(\Box)+1}{2}} - (s^{\frac{a(\Box)}{2}}t^{\frac{l(\Box)+1}{2}})^{-1}} ~,
$$
where $a(\Box)$ and $l(\Box)$ are the arm and leg length of a box $\Box$ in the Young diagram $\lambda$, and $a'(\Box)$ and $l'(\Box)$ are the  dual arm and leg length. The $a=t^N$ specialization reduces to the Macdonald dimension 
$$
P_\lambda^{\,\star}(a=t^N,s,t) =P_\lambda(t^\rho)~.
$$

Defining ``$\Gamma$-factors'' as in \cite{Colored_HL}
$$
\Gamma^{p,q}_{\lambda|\mu,\nu}:=\langle P_\mu| \mathcal{O}^{p,q}_{\lambda}|P_\nu\rangle~,
$$
the refined invariants of the $(p,q)$ torus knot can then  be written as
\begin{equation}\label{refine-knot}
\mathcal{W}_{\lambda}^{p,q}(a,s,t):=\sum_{\mu \,\in\calP_{p|\lambda|}} \Gamma_{\lambda|\mu,\varnothing}^{p,q} P_\mu^{\,\star}~.
\end{equation}
When $(p,q)=(1,0)$, the $\Gamma$-factors are indeed equal to  the $(s,t)$-Littlewood-Richardson coefficients 
\be\label{initial}
\Gamma^{(1,0)}_{\lambda|\mu,\nu}=N^\mu_{\lambda\nu}~.
\ee
In fact, in the case of an $(\ell,m\ell)$ torus link considered in \cite{BPS_wild}, it is easy to determine the $\Gamma$-factors
\begin{equation}\label{refine-link}
\mathcal{W}_{\lambda_1,\ldots,\lambda_\ell}^{\ell, m\ell}=\sum_{\mu\,\in\calP_{|\lambda_1|+\cdots +|\lambda_\ell|}}N^{\mu}_{\lambda_1,\ldots,\lambda_\ell}  (T_{\mu\mu})^m P_\mu^{\,\star}~,
\end{equation}
which can be interpreted as the deformation of the formula
$$
P_{\lambda_1}\cdots P_{\lambda_\ell}=\sum_{\mu\,\in\calP_{|\lambda_1|+\cdots +|\lambda_\ell|}}N^{\mu}_{\lambda_1,\ldots,\lambda_\ell}  P_\mu~,
$$
by the braiding operation with the $T$-matrix in \eqref{S and T}. The equation \eqref{refine-link} is the refined generalization \cite{Super_evol} of the Rosso-Jones formula. 

However, it turns out that the refined generalization of the Rosso-Jones formula \eqref{eq:unrefinv}  cannot be applied to  the case of the $(p,q)$ torus knot ($(p,q)$ are coprime) because the formula is not symmetric under the exchange of $(p,q)\leftrightarrow (q,p)$ \cite{Super_evol}. This can be seen in \eqref{1-box-torus} where the extra factor $\gamma^\mu(s,t)$ is inserted. Therefore, in general, one  has to determine the $\Gamma$-factors by the following recursion relation \cite{Colored_HL}
\begin{align}\label{recursion}
T_{\mu\mu} \Gamma^{(p,q)}_{\lambda|\mu,\nu} = \Gamma^{(p,q+p)}_{\lambda|\mu,\nu} T_{\nu\nu}~, \qquad
\sum\limits_{\rho} S_{\mu\rho} \Gamma^{(p,q)}_{\lambda|\rho,\nu} = \sum\limits_{\sigma} \Gamma^{(p,-q)}_{\lambda|\mu,\sigma} S_{\sigma\nu} ~,
\end{align}
where the initial condition is given by \eqref{initial}.

There is an important property of refined Chern-Simons invariants. The formula \eqref{refine-knot} with \eqref{recursion} is  valid even for negative integers $p$ and $q$. In fact, the change from $(p,q)$ to $(p,-q)$ is equivalent to the orientation reverse of the corresponding torus knot which amounts to
$$
\mathcal{W}^{p,-q}_{\lambda}(a,s,t)=\mathcal{W}^{p,q}_{\lambda}(a^{-1},s^{-1},t^{-1})~.
$$
In particular, the case of $n\equiv 1\ \textrm{mod} \ m$ is related to that  of $n\equiv -1\ \textrm{mod} \ m$ via
$$
\mathcal{W}^{\ell,\ell k-1}_{\lambda}(a,s,t)=\mathcal{W}^{\ell,-\ell k+1}_{\lambda}(a^{-1},s^{-1},t^{-1})~.
$$
It is expected that this property would be related to analytic continuation discussed in \S\ref{sec:HMW}. 

For our purpose, we introduce $SU(\infty)$ refined Chern-Simons invariants
$$W^{\sf ref}_\lambda(p,q;s,t):=a^{-\#}\, \mathcal{W}^{p,q}_\lambda(a,s,t)\Big|_{a=0}~.$$
Here $\#$ is the minimum $a$-degree of the rational function $\mathcal{W}^{p,q}_\lambda(a,s,t)$ so that ${W}^{p,q}_\lambda(s,t)$ is called \emph{bottom row} of $\mathcal{W}^{p,q}_\lambda(a,s,t)$ in \cite{Gorsky:2013jxa}.

\subsection{The formula for refined stable pairs}

Based on the refined colored generalization \cite{DSV,DHS} of the 
conjecture of \cite{OS,ORS}, the  vertex ${V}^{(n,\ell)}_{\lambda}(q,y)$ 
is expected to be related to $W^{\sf ref}_{\lambda}(\ell, (n-2)\ell-1;s,t)$ by a change of variables $s=qy,\ t=qy^{-1}$, up to a normalization factor
\be\label{eq:largeNrelA}
{V}^{(n,\ell)}_{\lambda}(q,y)= w^{(n,\ell)}_{\lambda}(s,t)
W^{\sf ref}_{\lambda}(\ell, (n-2)\ell-1;s,t)\Big|_{s=qy,\ t=qy^{-1}} ~,
\ee
where $w^{(n,\ell)}_{\lambda}(q,y)$ is a monomial in $(q,y)$. With the unrefined limit and the direction computations, we fix this normalization factor as 
$$
w^{(n,\ell)}_{\lambda}(s,t)= h_\lambda(s,t)^{-n\ell(\ell-1)+2\ell^2-1}~, \qquad h_\lambda(s,t) = \prod_{\Box\in \lambda} s^{a(\Box)} t^{-l(\Box)}.
$$
In conclusion, we conjecture that the refined stable pair theory of $Y$ be given by an expression of the form 
\be \label{eq:stpairsB2} 
Z_{PT}^{\sf ref}(Y; q,y,{\sf x}) = 1 +\sum_{|\lambda|>0} 
L_{\lambda^t}(q,y) g_\lambda(q,y)^{-n\ell(\ell-1)+2\ell^2-1} W^{\sf ref}_{\lambda}(\ell, (n-2)\ell-1; qy, qy^{-1}) {\sf x}^{|\lambda|}~,
\ee
with $g_\lambda(q,y) = h_\lambda(qy,qy^{-1})$.

This formula of refined stable pair invariants on $Y$ admits refined Gopakumar-Vafa expansion. Indeed, 
as explained in detail in \cite{BPSPW}, the connection to moduli spaces of twisted irregular Higgs bundles can be seen in the refined Gopakumar-Vafa expansion. From the viewpoint of string theory, the spectral correspondence  identifies moduli spaces of  
D2-D0 BPS bound states on $Y$ with moduli spaces of stable 
twisted irregular Higgs bundles. These spaces are expected to be independent of the D0-brane charges $c$, and the moduli spaces are therefore denoted by  $\CH_{n,\ell,r}$ in \S\ref{twistedcovers}. Moreover, on general grounds, the space of BPS states in string theory is expected to have an $SU(2)\times  SU(2)$ Lefschetz action \cite{HST}. As explained in \cite{BPSPW}, for the type of local geometries considered here, only an $SU(2)\times U(1)$ subgroup is manifest, which can be expressed by  the perverse degree $i\geq 1$ and  the cohomological degree $j\geq 0$ of  $Gr^P_iH^j(\CH_{n,\ell,r})$. Consequently, we denote the perverse Poincar\'e polynomial of the moduli space by
\[ 
P_{n,\ell,r}(u,v) = \sum_{j,k} u^i(-v)^j{\rm dim}\, Gr^P_iH^j(\CH_{n,\ell,r})~. 
\]
All in all, the relation between geometry of $\CH_{n,\ell,r}$ and refined stable pair on $Y$ based on the spectral correspondence can be conjecturally summarized as 
\be\label{eq:refGVA2}
{\rm ln}\, Z^{\sf ref}_{PT}(Y;q,y,{\sf x}) = -\sum_{s\geq 1} \sum_{r\geq 1} {{\sf x}^{sr}\over s}{y^{sre(0,1,n,\ell)} (qy^{-1})^{sd(0,1,n,\ell,r)/2} 
P_{n,\ell,r}(q^{-s}y^{-s}, y^s) \over (1-q^{-s}y^{-s})(1-q^sy^{-s})}
\ee
where $d(g,m,n,\ell,r) = 2+ r^2(mn\ell(\ell-1)+2(g-1)\ell^2) $ is the dimension of $\CH_{n,\ell,r}$ and 
\[
e(g,m,n,\ell)={mn\ell(\ell-1)\over 2}+(g-1)\ell^2~.
\]

\section{Poincar\'e polynomials via localization}\label{locsect} 

The goal in this section is to test the refined conjecture \eqref{eq:refGVA2} against localization computations of 
 Poincar\'e  polynomials of several moduli spaces of spectral sheaves. This will first require a few more details on the spectral construction in genus zero. 

\subsection{Genus zero base curves}\label{genzerosect}
Let $({\sf X}_1,{\sf X}_2)$ be homogeneous coordinates on $C=\IP^1$ such that $p$ is 
given by ${\sf X}_1=0$.  Then  $M=K_C(np)\simeq \CO_C(kp)$ with $k=n-2$ is a toric surface with homogeneous 
coordinates $({\sf X}_1, {\sf X}_2,{\sf Y})$ of degrees $(1,1,k)$. 
In this case for each $\ell\geq 1$ let $\Sigma_0\subset M$ be the divisor
\be\label{eq:singcurveA} 
{\sf Y}^\ell = {\sf X}_1{\sf X}_2^{k\ell-1}. 
\ee
This curve is the unique spectral cover in the linear system 
$\ell C_0$ which is preserved by the torus action 
$\IC^\times \times M \to M$ scaling the homogeneous coordinates $({\sf X}_1,{\sf X}_2,{\sf Y})$ with weights
$(\ell, 0, 1)$. 

Let $(u,z)$ be affine coordinates on the open 
subset $\CU=\{{\sf X}_2\neq 0\}$ with 
\[
u= {\sf X}_2^{-k}{\sf Y}, \qquad z = {\sf X}_2^{-1}{\sf X}_1.
\]
Let $(v,w)$ be affine coordinates on the open subset $\CV=\{{\sf X}_1\neq 0\}$ with 
\[ 
v = {\sf X}_1^{-k}{\sf Y}, \qquad w = {\sf X}_1^{-1}{\sf X}_2.
\]
The local equations of $\Sigma_0$ in the two coordinate charts are respectively 
\[ 
u^\ell=z \ \ \textrm{on}\ \ \CU~, \qquad\qquad v^\ell = w^{k\ell-1} \ \ \textrm{on}\ \ \CV~.
\]
Hence $\Sigma_0$ is a reduced irreducible curve with a 
unique singular point $\sigma_0$ at ${\sf Y}=0,\ {\sf X}_2=0$. Moreover, the reduced inverse image of $p$ in $\Sigma$ is the point $q_0\in \Sigma_0$ determined by ${\sf Y}=0$, ${\sf X}_1=0$. In local coordinates the torus action reads  respectively
\be\label{torus-action}
\begin{array}{ll}
\zeta \times (z,u) \mapsto (\zeta^\ell z, \zeta u)&  \textrm{on} \ \  \CU~,\cr
\zeta \times (v,w) \mapsto (\zeta^{1-k\ell} v, \zeta^{-\ell} w)&  \textrm{on} \ \ \CV ~.
\end{array}\ee

Note that $\Sigma_0$ is the scheme theoretic 
image of the map $\IP^1\to M$ given in homogeneous 
coordinates by 
\[
{\sf X}_1 = {\sf T}_1^{\ell}, \qquad {\sf X}_2= {\sf T}_2^{\ell}, \qquad 
{\sf Y} = {\sf T}_1{\sf T}_2^{k\ell-1}.
\]
The resulting map $\nu : \IP^1 \to \Sigma_0$ is the normalization of $\Sigma_0$. 
The torus
 action lifts to the normalization such that $({\sf T}_1,{\sf T}_2)$ have weights $(1,0)$. Then the 
the map $\nu$ is equivariant.

\begin{figure}
%\hspace{30pt}
\vspace{-10pt}
\setlength{\unitlength}{1mm}
\begin{picture}(200,100)
\thicklines
\put(10,5){\line(1,0){125}} 
\put(15,2){\line(-1,2){10}}
\multiput(7,28)(0,3){3}{\huge .}
\put(4,43){\line(1,2){10}}
\put(16,55){\line(-1,2){12}}
\put(6,71){\line(0,1){25}}
\put(7,65){\line(2,1){25}}
\put(27,77){\line(2,-1){25}}
\multiput(60,72)(3,0){3}{\huge .}
\put(78,64){\line(2,1){25}}
{\color{blue}\qbezier(82,73)(120,40)(122,5)}
{\color{blue}\qbezier(120.5,5)(122,22)(140,25)}
\put(108,45){$\Sigma_{n\ell}$}
\put(65,0){$C_{0,{n\ell}}$}
\put(119,2){$\sigma_{n\ell}$}
\put(9,13){$\Xi_1$}
\put(5,43){$\Xi_{\ell-1}$}
\put(15,53){$\Xi_\ell$}
\put(16,67){$\Xi_{\ell+1}$}
\put(37,72){$\Xi_{\ell+2}$}
\put(96,70){$\Xi_{n\ell}$}
\put(83,63){$q_{n\ell}$}
\put(1,84){$f_{n\ell}$}
\end{picture}
\caption{The torus action preserves $\Sigma_{n\ell}$ in the spectral surface $S$.}
\label{BlowupB} 
\end{figure}
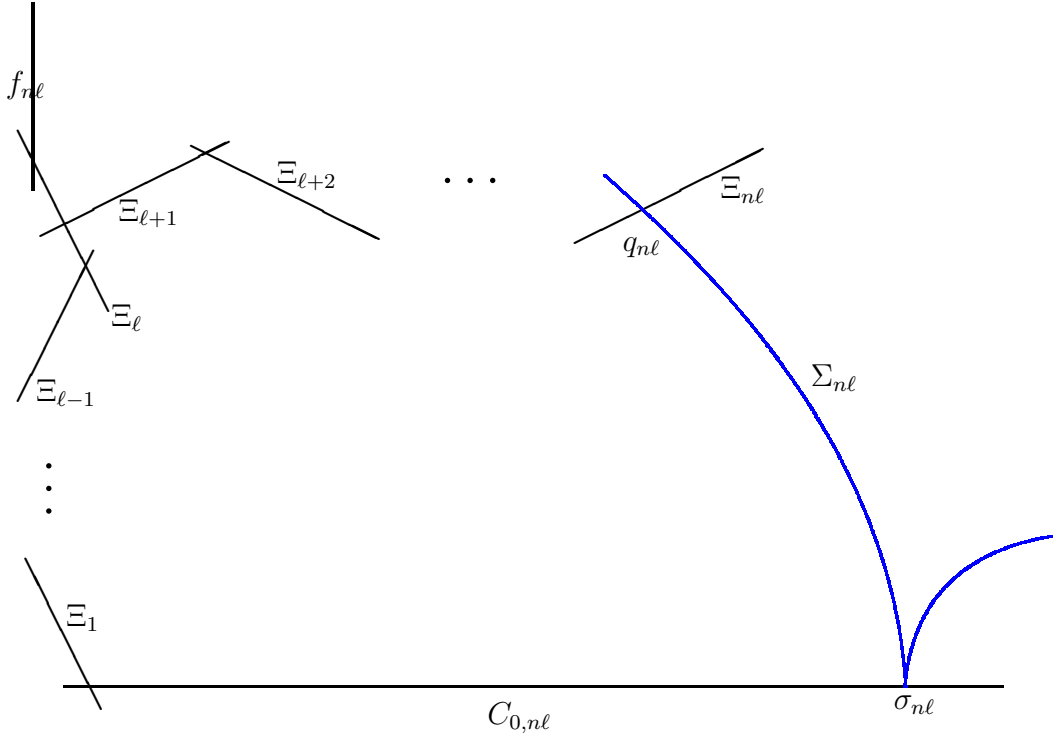

The torus action \eqref{torus-action} on $M$ lifts to a torus action on 
each surface $M_i$ in the sequence of blow-ups constructed in \S\ref{gencurve}, which preserves 
the exceptional divisors, and the strict transforms $\Sigma_i$. In particular, there is a torus action 
on the holomorphic symplectic surface $S$ preserving the strict transform $\Sigma_{n\ell}$, which is isomorphic to $\Sigma$. This configuration  is 
shown in Figure \ref{BlowupB}. 

Furthermore there is a canonical isomorphism 
$\rho_{n\ell}^{-1}(\CV)\simeq \CV$ where $\CV=\{{\sf X}_1\neq 0\}\subset M$ is the domain of the local coordinates $(v,w)$ on $M$. Therefore, $(v,w)$ can be used as local 
coordinates on $M_{n\ell}$ as well, and the distinction will be clear from the context. The local equation of $\Sigma_{n\ell}$ in $ \rho_{n\ell}^{-1}(\CV)$ is $$v^\ell =w^{k\ell-1}~,$$ and the singular point $\sigma_{n\ell}\in \Sigma_{n\ell}$ is at $v=w=0$.

The self-intersection of $\Sigma_{n\ell}$ in $S$ is the  $g=0$, $m=1$  specialization of \eqref{eq:selfintA}
$$
(\Sigma_{n\ell})^2 = (n-2)\ell^2 - n\ell~,
$$
while 
$$
\chi(\CO_{\Sigma_{n\ell}}) = -{1\over 2}\Sigma_{n\ell}(\Sigma_{n\ell}+K_S) = 
-{1\over 2}((n-2)\ell^2 -n\ell ) = \chi(\CO_{\Sigma_0})
$$
since $K_S=0$. Note also that $\Sigma_{n\ell}$ is Gorenstein with dualizing 
line bundle 
$$
\omega_{\Sigma_{n\ell}} \simeq \CO_S(\Sigma_{n\ell})\otimes \CO_{\Sigma_{n\ell}}
\simeq \CO_{\Sigma_{n\ell}}\big(((n-2)\ell^2-n\ell)q_{n\ell}\big)~,
$$
where $q_{n\ell}$ is the transverse intersection point of $\Sigma_{n\ell}$ with $\Xi_{n\ell}$.

The Poincar\'e polynomial of the moduli space of spectral sheaves will be computed by localization with respect to (the lift of) the torus action \eqref{torus-action}. 
This action extends to a torus action on the spectral surface $S$ with fixed points $S^{\bf T}=\{\sigma_{n\ell}, q_{n\ell}\}$. 
Furthermore, any compactly supported pure dimension one sheaf $F$ on $S$ fixed under the torus action up to isomorphism must be set theoretically supported on $\Sigma_{n\ell}$. Therefore, classifying such objects requires 
 detailed understanding of the local geometry of $\Sigma_{n\ell}$ and its infinitesimal neighborhood in $S$.
For ease of exposition, the curve $\Sigma_{n\ell}$, the singular point $\sigma_{n\ell}$ and the point at infinity $q_{n\ell}$ will be denoted in the following by $\Sigma$, $\sigma$ and $q$, respectively.   

\subsection{Gluing construction}\label{gluingsect} 
The main observation is that $\Sigma$ can be intrinsically constructed by gluing two affine schemes 
\[ 
\CU = {\rm Spec}(S)~, \qquad \CV = {\rm Spec}(R)~,
\]
where 
\[
S= \IC[s], \qquad R = 
\IC[t^\ell, t^{k\ell-1}].
\]
Using the canonical ring inclusion $R \subset \IC[t]$, any ideal in $\IC[t]$ yields by intersection an ideal in $R$. Let $q\in \CU$ be the closed point defined by $(s) \subset S$ and 
$\sigma\in \CV$ be the closed point defined by $R\cap (t)$. 
The open subschemes $\CU\setminus \{q\}$ and 
$\CV\setminus \{\sigma\}$ are identified using the transition function $s=t^{-1}$. 

The resulting curve 
$\Sigma$ admits a normalization 
$\nu : \wSigma \to \Sigma$ with $\wSigma=\IP^1$, which can be also explicitly given in terms of local data. Let $\wCV = {\rm Spec}(\wR)$ with $\wR = \IC[t]$. Then $\wSigma$ 
is obtained in the standard way 
by gluing $\CU$ and $\wCV$ via the transition function $s=t^{-1}$. The projection $\nu: \wSigma \to \Sigma$
 restricts to the identity on $\CU$ while its restriction to $\CV$ is determined by the natural  injective morphism $R\hookrightarrow \wR$. 

Let $\wCO_\Sigma = \nu_*\CO_\wSigma$. This
is a rank one torsion free sheaf on $\Sigma$. 
Moreover, for any integer $d\in \IZ$, let 
$\wCO_\Sigma(dq) =\wCO_\Sigma \otimes \CO_\Sigma(dq)$. Then,
$\wCO_\Sigma(dq) \simeq 
\nu_*\CO_{\wSigma}(d\wq)$, where $\wq$ is the point 
$t=0$ in $\wSigma$. 
Then for any $d$ there is a canonical exact sequence 
\be\label{eq:normseqA}
0\to \CO_\Sigma(dq) \to \wCO_\Sigma(dq) \to \CT_\Sigma\to 0
\ee
where $\CT_\Sigma\simeq \wCO_\Sigma/\CO_\Sigma$ is a zero-dimensional sheaf with set theoretic support $\{\sigma\}$. Note that each of these sheaves has a natural torus 
equivariant structure. For ease of exposition, the tensor product $R\otimes F$ between a torus representation and 
an equivariant sheaf $F$ will be denoted below by $RF$. 

Since $\Sigma \subset S$ is a divisor in a smooth surface, it is Gorenstein and its dualizing line bundle is given by the adjunction formula 
$$
\omega_\Sigma \simeq \omega_S \otimes \CO_\Sigma(\Sigma). 
$$
Since $S$ is holomorphic symplectic, $\omega_S$ is isomorphic to the trivial line bundle. As an equivariant line bundle, there is an isomorphism 
$$
\omega_S \simeq T^{\kappa} \CO_S
$$
where $\kappa = (k+1)\ell-1$. Moreover, the line bundle 
$\CO_S(\Sigma)$ must carry a linearization $T^{-\ell(k\ell-1)}$ where $\ell(k\ell-1)$ is the weight of the local defining equation of $\Sigma$. 
Then, equivariantly, $\omega_\Sigma$ is isomorphic to 
$$
\omega_\Sigma \simeq T^{\kappa - \ell(k\ell-1)} \CO_\Sigma(2eq) = T^{-2e-1}\CO_\Sigma(2eq)
$$
where $e=\Sigma^2/2=k\ell(\ell-1)/2 - \ell$.

Some useful computations of cohomology and extension groups are recorded below. The proof reduces to straightforward computations by \v Cech cohomology and will be omitted. 

\medskip
 
\noindent $\bullet$ For any $d\in \IZ$, there are isomorphisms 
$$ 
H^i(\Sigma, \wCO_\Sigma(dq)) \simeq H^i(\Sigma, \CO_{\wSigma}(d\wq))
$$
with $0\leq i \leq 1$. In particular, 
$$
 \chi(\wCO_\Sigma(dq)) = d+1
$$
 for any $d\in \IZ$.
Using the exact sequence 
\eqref{eq:normseqA}, this further
yields 
$$ 
\chi(\CO_\Sigma(dq)) = d+1 -\chi(\CT_\Sigma).
$$

\medskip

\noindent $\bullet$ For any $d_1, d_2\in \IZ$ 
$$ 
{\rm Hom}_\Sigma(\wCO_\Sigma(d_1q), \wCO_\Sigma(d_2q)) 
\simeq H^0(\wSigma, \CO_{\wSigma}((d_2-d_1)\wq))).
$$

\medskip

\noindent $\bullet$ For any two compactly supported torus equivariant sheaves $E_1,E_2$ on $S$ the equivariant 
$K$-theoretic Euler characteristic is defined by 
$$
\CX_S(E_1,E_2) = \sum_{i=0}^2 (-1)^i {\rm Ext}^i_S(E_1,E_2)
$$
where the right hand side belongs to the representation ring of ${\bf T}$. 

For any two equivariant torsion free sheaves $E_1,E_2$ on $\Sigma$, extended 
by zero to $S$, Lemma \ref{extlemma} yields 
$$
\CX_S(E_1,E_2) = \CX_\Sigma(E_1,E_2) + T^{-\kappa}{\rm Ext}_\Sigma^0(E_2,E_1)^\vee - 
T^{- \ell(k\ell-1)}{\rm Ext}^0_\Sigma(E_1,E_2)~,
$$
where 
\[
\CX_\Sigma(E_1,E_2)= {\rm Ext}^0_\Sigma(E_1,E_2) - 
{\rm Ext}^1_\Sigma(E_1,E_2).
\]

\subsection{Cuspidal elliptic curves}\label{elliptic} 

Suppose $k=\ell=2$ i.e. $\Sigma$ has a local singularity of the form $v^2=w^3$. In this case
\[ 
\omega_\Sigma \simeq T^{-1}\CO_\Sigma.
\]
Then one can prove the following result.
\begin{lemm}\label{ellipticstab}
Suppose $F$ is a stable pure dimension one sheaf on $S$ with set theoretic support $\Sigma$. Then $F$ is scheme theoretically supported on $\Sigma$.
\end{lemm}

{\it Proof.}
Let $\zeta\in H^0(S, \CO_\Sigma(\Sigma))$ be a defining section of $\Sigma$, as above. Then there is a morphism 
\[ 
{\bf 1}\otimes \zeta: F \to F \otimes\CO_\Sigma(\Sigma) \simeq F.
\]
Since $F$ is stable this morphism must be either identically zero or an isomorphism. Moreover, it 
cannot be an isomorphism because 
${\bf 1}\otimes \zeta$ has a nontrivial kernel, as $F$ 
is scheme theoretically supported on some thickening of $\Sigma$. Thus, ${\bf 1}\otimes \zeta$ must be identically zero, proving the claim. 

\hfill $\Box$

Since all stable pure dimension one sheaves on $S$ are in this case scheme theoretically supported on $\Sigma$,
one can use the results of \cite{Sheaves_genus_one} 
to determine the torus fixed locus in the moduli space. 
Let $\Phi: D^b(\Sigma) \to D^b(\Sigma)$ be the Fourier-Mukai transform determined by the 
ideal sheaf of the diagonal in $\Sigma\times\Sigma$. 
Then, according to 
\cite[Prop. 1.13]{Sheaves_genus_one}, any stable rank $r$ torsion-free sheaf $F$ on $\Sigma$ with $\chi(F)=1$ is $WIT_1$ with respect to $\Phi$ (i.e. its Fourier-Mukai transform is a complex consisting of a single sheaf of degree $1$.)
 Let $\Phi^1(F)$ denote its image, which is a rank $(r-1)$ torsion free sheaf 
on $\Sigma$ with $\chi(\Phi^1(F))=1$. 
Moreover, \cite[Thm. 1.20]{Sheaves_genus_one} proves 
that $F$ is stable if and only if  $\Phi^1(F)$ is stable. Proceeding recursively one obtains an 
isomorphism 
\[
\varphi: \CM_\Sigma(1,1) \simeq \CM_\Sigma(r,1)
\]
of moduli stacks of stable torsion-free sheaves on $\Sigma$, as in \cite[Cor. 1.25]{Sheaves_genus_one}. 
This isomorphism is naturally equivariant
with 
respect to the torus action so that it yields an isomorphism of fixed loci. This implies that for $r\geq 1$ 
there are only two fixed points in $\CM_\Sigma(r,1)$, 
namely $F_1=\varphi(\CO_\Sigma(q))$ and 
$F_2=\varphi(\wCO_\Sigma)$. Moreover, since the Fourier-Mukai transform is a derived equivalence, it preserves extension groups. 
Therefore, one can easily compute the tangent spaces 
to the coarse moduli space at $[F_1]$, $[F_2]$, obtaining 
\[ 
\CT_{[F_1]} =\CT_{[F_2]} = T+T^{-6}. 
\]
This shows that the Poincar\'e polynomial of the coarse moduli space will be $P_{4,2,r}(v)=1+v^2$ for any $r\geq 1$.

\subsection{Singular genus two curves}\label{genustwosect} 

Let $k=3$ and $\ell=2$. Then the curve $\Sigma$ has a 
local singularity $v^2 = w^5$. In this case the normalization map reads locally $v=t^5$, $w=t^2$. Hence  
\[
R = \langle 1, t^2, t^4, t^5,t^6, \ldots \rangle 
\]
as a vector space over $\IC$, and $\wR/R \simeq \langle t,t^3\rangle$. Thus, $\CT_\Sigma$ is isomorphic to $T \CO_Z$ as an equivariant sheaf, where 
$Z\subset \Sigma$ is the closed subscheme of $\Sigma$ defined by
the ideal $(v,w^2)$. 

Let $\CE$ be the rank one torsion free sheaf on $\Sigma$ such that 
\[
\Gamma(\CU, \CE) =\IC[s], \qquad \Gamma(\CV,\CE) = 
\langle 1,t^2,t^3, \ldots \rangle.
\]
Then, any rank one sheaf fixed by the torus action must be isomorphic to one of the followings
\be\label{eq:shlist}
T^a\CO_\Sigma(dq),\qquad T^a\wCO_\Sigma(dq), \qquad T^a\CE(dq) ~,
\ee
for some $d\in \IZ$, where $a\in \IZ$ encodes the choice of a linearization. 

The main goal of this section is to classify all stable pure dimension one 
sheaves $F$ on $S$ with $\ch_1(F)= 2\Sigma$ and 
$\chi(F)=1$ which are fixed by the torus action up to isomorphism. Any such sheaf is set theoretically supported 
on $\Sigma$ and fits in an extension 
\be\label{eq:Tinvext}
0 \to F_1\to F \to F_2 \to 0
\ee
on $S$, where $F_1,F_2$ are isomorphic to one of the 
sheaves listed in \eqref{eq:shlist}. Using scaling gauge transformations, one can choose the linearization $a=0$ for $F_1$ without loss of generality. Thus, 
there is only one discrete parameter $a\in \IZ$ 
encoding the linearization of $F_2$. 
Furthermore, an extension class associated to 
\eqref{eq:Tinvext} must belong to the 
weight zero part of ${\rm Ext}^1_S(F_2,F_1)$. 
This yields nine possible cases, each such case splitting into several subcases depending on the values of $a$
for $F_2$. 
In all these cases,
stability requires $\chi(F_1)\leq 0$, $\chi(F_2)\geq 1$, 
and that the extension class be nontrivial. Moreover, in order to avoid redundancy, in all cases where $F$ is scheme theoretically supported on $\Sigma$, it will be assumed by 
construction that $F_1$ is a maximal torsion free subsheaf 
as in Lemma \ref{TFc}. 

Lemmas \ref{extlemma}, \ref{stabext}, \ref{nonlocfree}, \ref{TFd} yield a complete 
 classification of all such fixed points, although the argument is rather involved. One has to first prove some structure results for the extension group ${\rm Ext}^1_S(F_2,F_1)$.

\begin{lemm}\label{nonlocfree}
Suppose $F_1,F_2$ are isomorphic to either 
$\wCO(dq)$ or $\CE(dq)$, $d\in \IZ$ up to linearization, 
and 
$$
\chi(F_1)+\chi(F_2)=1, \qquad \chi(F_1)\leq 0.
$$
Then there is an isomorphism 
$$
{\rm Ext}_S^1(F_2,F_1) \simeq {\rm Ext}_\Sigma^1(F_2,F_1).
$$
\end{lemm}

{\it Proof}. 
As shown in Lemma \ref{extlemma}, there is an exact sequence 
\[
\bal
0 & \to {\rm Ext}^1_\Sigma (F_2,F_1) \to 
{\rm Ext}^1_S(F_2,F_1) \to  {\rm Ext}^0_\Sigma(F_2, F_1(\Sigma))\\
& \to {\rm Ext}^2_\Sigma(F_2,F_1) \to 
{\rm Ext}^2_S(F_2,F_1) 
 \to {\rm Ext}^1_\Sigma(F_2,F_1(\Sigma)) \\
 &\to {\rm Ext}^3_\Sigma(F_2,F_1)\to 0~. 
 \eal
\]
Moreover,
\[
{\rm Ext}_S^2(F_2,F_1)\simeq {\rm Ext}_S^0(F_1,F_2)^\vee
\]
by Serre duality on $S$, 
and 
\[ 
 {\rm Ext}_S^0(F_2,F_1)\simeq {\rm Ext}_\Sigma^0(F_2,F_1)
 \]
 since $F_1, F_2$ are scheme theoretically supported on $\Sigma$. 
The idea  is to prove that the first map from the left is an isomorphism while the second is identically zero by dimension counting. This is clearly the case if ${\rm Ext}^0_\Sigma(F_2, F_1(\Sigma))=0$. 
Suppose \[
{\rm Ext}^0_\Sigma(F_2, F_1(\Sigma))\simeq
{\rm Ext}^0_\Sigma(F_2,F_1(2q))
\]
is nonzero. This implies $\chi(F_1)+2 \geq \chi(F_2)$, hence $\chi(F_1) \geq -1/2$ since $\chi(F_2)=1-\chi(F_1)$. Since $\chi(F_1)\leq 0$ by assumption, this implies 
\[
\chi(F_1)=0, \qquad \chi(F_2)=1~.
\]
This yields four cases: 
\begin{align}
(a)&\quad F_1\simeq \wCO_\Sigma(-q)~,\quad F_2\simeq \wCO_\Sigma~,\cr
(b)&\quad F_1\simeq \wCO_\Sigma(-q)~,\quad F_2\simeq \CE(q)~,\cr
(c)&\quad F_1\simeq \CE~,\quad F_2\simeq \wCO_\Sigma~,\cr
(d)&\quad F_1\simeq \CE~,\quad F_2\simeq \CE(q)~.\nonumber
\end{align}
Using the Ext group computations in Appendix 
\ref{extgenustwo} it is then straightforward to 
prove the claim by dimension counting. 
 For example, in case $(a)$, computations in \eqref{Ext-results} show
\[
 {\rm Ext}^0_\Sigma(F_2, F_1(\Sigma))\simeq
{\rm Ext}^0_\Sigma(F_2,F_1(2q))={\rm Ext}^0_\Sigma(\wCO_\Sigma,\wCO_\Sigma(q)) = 1+T 
 \]
 is two-dimensional. 
 At the same time, one can read off from \eqref{Ext-results} 
\begin{align}\nonumber
& {\rm Ext}^1_\Sigma(F_2, F_1(\Sigma))\simeq {\rm Ext}^1_\Sigma(\wCO_\Sigma,\wCO_\Sigma(q))= 
 T^{-5}(1+T+T^2+T^3)\cr
& {\rm Ext}^3_\Sigma(F_2, F_1)=  {\rm Ext}^3_\Sigma(\wCO_\Sigma,\wCO_\Sigma(-q))= 
 T^{-15}(1+T+T^2+T^3)
 \end{align}
 are both four-dimensional. Therefore, the last morphism in the above exact sequence must be an isomorphism. 
Next, 
\[ 
 {\rm Ext}^2_\Sigma(F_2, F_1)= {\rm Ext}^2_\Sigma(\wCO_\Sigma,\wCO_\Sigma(-q))= 
 T^{-10}(1+T+T^2+T^3)
 \]
 is four-dimensional and, by Serre duality on $S$, 
 \[ 
 {\rm Ext}^2_S(F_2, F_1)=  {\rm Ext}_S^2(\wCO_\Sigma, \wCO_\Sigma(-q))\simeq 
  T^{-7} {\rm Ext}^0_S(\wCO_\Sigma(-q), \wCO_\Sigma)^\vee = T^{-7}(1+T)
  \]
  is two-dimensional. 
This implies that the first map must be an isomorphism 
by dimension counting. 
In conclusion $F$ must indeed be scheme theoretically supported on $\Sigma$. 

The remaining three cases are entirely analogous. 

\hfill $\Box$ 

Next suppose that at least one of $F_1, F_2$ is locally free on $\Sigma$. Then 
Lemma \ref{extlemma} yields a short exact sequence 
\[
0  \to {\rm Ext}^1_\Sigma (F_2,F_1) \to 
{\rm Ext}^1_S(F_2,F_1) \to  {\rm Ext}^0_\Sigma(F_2, F_1(\Sigma))\to 0~,
\]
where
\[
{\rm Ext}^1_\Sigma (F_2,F_1)=\sum_{w\in 
W_1(F_1,F_2)} m_wT^{w}, \qquad 
{\rm Ext}^0_\Sigma(F_2, F_1(\Sigma)) = 
\sum_{w\in W_0(F_1,F_2)} n_w T^w.
\]
Then the main observation is 
\begin{lemm}\label{nooverlap}
There is no overlap, 
$W_0(F_1,F_2) \cap W_1(F_1,F_2) = \varnothing$. 
\end{lemm}

{\it Proof}. Since at least one of  $F_1, F_2$ is locally free 
by assumption, 
Serre duality on $\Sigma$ yields an isomorphism 
\[ 
{\rm Ext}^1_\Sigma (F_2,F_1)\simeq T^3 {\rm Ext}^0_\Sigma(F_1, F_2(2q))^\vee.
\]
Moreover, 
\[ 
{\rm Ext}^0_\Sigma(F_2, F_1(\Sigma)) \simeq T^{-10} 
{\rm Ext}^0_\Sigma(F_2, F_1(2q)).
\]
Suppose that $F_1$ is locally free and then there is a natural pairing 
\[ 
{\rm Ext}^0_\Sigma(F_2, F_1(2q)) \times 
{\rm Ext}^0_\Sigma(F_1, F_2(2q)) \to {\rm Ext}^0_\Sigma(F_1, F_1(4q))
\]
given by $(f_2, f_1) \mapsto (f_2\otimes {\bf 1}_{\CO_\Sigma(2q)}\circ f_1)$. Since $F_1$ is invertible 
\[ 
{\rm Ext}_\Sigma^0(F_1, F_1(4q)) \simeq H^0(\CO_\Sigma(4q)) = 
1+T^2+T^4.
\]
Therefore, if $f_1, f_2$ are homogeneous elements of weights $\alpha_1, \alpha_2$, it follows that 
$\alpha_1+\alpha_2\in \{0,2,4\}$. 
At the same time, they correspond to elements of weights 
\[ 
w_1 = 3-a-\alpha_1, \qquad w_2 = -10-a +\alpha_2~,
\]
yielding
\[ 
w_2 - w_1 = -13 +\alpha_1+\alpha_2 \neq 0~. 
\]
If $F_2$ is locally free, the proof is analogous. 

\hfill $\Box$

In conclusion, if at least one of $F_1,F_2$
is locally free, there is an equivariant splitting 
$$
{\rm Ext}^1_S(F_2,F_1) \simeq 
{\rm Ext}^1_\Sigma (F_2,F_1)\oplus 
{\rm Ext}^0_\Sigma(F_2, F_1(\Sigma)).
$$
 The 
first term parametrizes extensions $F$ 
scheme theoretically supported on $\Sigma$, while 
the second parametrizes extensions $F$ 
with scheme theoretic support on $2\Sigma$. 
If both $F_1, F_2$ are not locally free, and $\chi(F_1)=1-\chi(F_2)\leq 0$, there is an isomorphism 
$$
{\rm Ext}^1_S(F_2,F_1) \simeq 
{\rm Ext}^1_\Sigma (F_2,F_1).
$$
In particular, in this case, all extensions $F$ are 
scheme theoretically supported on $\Sigma$. 
As stated above, whenever $F$ is scheme theoretically supported on $\Sigma$, it will be assumed by construction that $F_1$ is a maximal 
subsheaf as in Lemma \ref{TFc} by construction. Therefore, 
according to Lemma \ref{TFd}, one has
\be\label{eq:homboundA}
{\rm dim}\, {\rm Hom}_\Sigma(F_1, F_2) \leq 3.
\ee
This yields strong conditions on $\chi(F_1)$, confining it to a finite set of values. 
If $F$ is  scheme theoretically supported on $2\Sigma$, 
one has 
\be\label{eq:homboundB}
{\rm dim}\, {\rm Hom}_\Sigma(F_2, F_1(2q)) \geq 1~,
\ee
which yields again a finite set of possible values for 
$\chi(F_1)$. 

Now let 
\[ 
{\rm Ext}^1_S(F_2,F_1) \simeq \sum_{w\in W(F_1, F_2)} m_w T^w 
\]
be the character decomposition of the extension group,
where the sum is finite and $m_w\geq 0$ are multiplicities assigned to each character.  Then the allowed values of the linearization of $F_2$ are
$a\in \{-w\, |\, w\in W(F_1,F_2)\}$. 
Given a fixed pair $(F_1,F_2)$ for each $a=-w$ there is an $m_w$-dimensional family of equivariant 
extensions. Furthermore   Lemma \ref{stabext} 
provides a necessary and sufficient criterion for stability. 
An important issue to keep in  mind is that  different extensions of the form \eqref{eq:Tinvext} can in general 
yield isomorphic sheaves $F$. All such isomorphisms must 
be identified in order to obtain an accurate classification 
of the fixed points.  
Finally,  in  each case, the 
equivariant tangent space 
$\CT_{[F]}\simeq {\rm Ext}^1_S(F,F)$ to the moduli 
space is computed using Lemma \ref{extlemma}. 

To summarize, the classification algorithm is as follows.

\medskip
{\bf Step 1.} Pick a pair $(F_1,F_2)$ from the list \eqref{eq:shlist} such that $\chi(F_1) = 1-\chi(F_2) \leq 0$. 
Compute the character decomposition of ${\rm Ext}^1_S(F_2,F_1)$. Let $w\in W(F_1,F_2)$ be a weight occurring in this decomposition. If $w$  belongs to 
${\rm Ext}^1_\Sigma(F_2,F_1)$, the condition 
\eqref{eq:homboundA} yields a finite set of invariants 
$(d_1,d_2)$. If $w$ belongs to 
${\rm Ext}^0_\Sigma(F_2,F_1(\Sigma))$, the condition 
\eqref{eq:homboundB} yields a finite set of invariants 
$(d_1,d_2)$. Therefore, for a fixed $w$, one obtains 
a finite set of $m_w$-dimensional families of torus 
invariant extensions. This step is straightforward, but 
very tedious, hence the details will be omitted. 

\medskip
{\bf Step 2.} For all the resulting extensions 
one has to check stability using Lemma \ref{stabext}. 
This requires the extension class to be nontrivial, and 
forbids the existence of nontrivial subsheaves
$F_2'\subset F_2$ with $\chi(F_2')\geq 1$ such that 
the extension class corresponding to $F$ is in the kernel 
of the natural map 
\[
{\rm Ext}^1_\Sigma(F_2,F_1) \to {\rm Ext}^1_\Sigma(F_2', F_1). 
\]
Moreover, as observed below Lemma \ref{invHN}, 
it suffices to consider subsheaves $F_2'\subset F_2$ which are themselves fixed by the torus action. Thus, 
$F_2'$ must be isomorphic to one of the rank one sheaves on $\Sigma$ listed in \eqref{eq:shlist}.

Similarly, if $F$ is scheme theoretically supported on $\Sigma$, one also has to check that $F_1$ 
is a maximal subsheaf as in Lemma \ref{TFc}. This forbids the existence of nontrivial subsheaves
$F_2'\subset F_2$ with $\chi(F_2')>\chi(F_1)$ such that 
the extension class corresponding to $F$ is in the kernel 
of the natural map 
\[
{\rm Ext}^1_\Sigma(F_2,F_1) \to {\rm Ext}^1_\Sigma(F_2', F_1). 
\]
This is again straightforward though fairly tedious.

\bigskip

For illustration, let us first consider the case 
$F_1 = \CE_\Sigma(-q)$, $F_2 = T^a\CE_\Sigma(2q)$. 
Using the computations in Appendix \ref{extgenustwo}, 
the local to global spectral sequence yields an exact sequence 
$$
0\to {\rm H}^1({\mathcal Ext}^0_\Sigma (\CE_\Sigma(2q), \CE_\Sigma(-q)))
\to {\rm Ext}^1_S(\CE_\Sigma(2q), \CE_\Sigma(-q)) 
\to{\rm H}^0({\mathcal Ext}^1_\Sigma (\CE_\Sigma(2q), \CE_\Sigma(-q))) \to 0
$$
where 
\[ 
{\rm H}^1({\mathcal Ext}^0_\Sigma (\CE_\Sigma(2q), \CE_\Sigma(-q))) = T^{-2} + T^{-1} + T, 
\qquad 
{\rm H}^0({\mathcal Ext}^1_\Sigma (\CE_\Sigma(2q), \CE_\Sigma(-q))) = T^{-5} + T^{-2}.
\] 
Note also that there is a unique injective morphism $f: \CO_\Sigma(2q) \to \CE_\Sigma(2q)$ given locally by 
\[ 
1\mapsto 1, \qquad s^{-2}\mapsto s^{-2}.
\]
Moreover, the local to global spectral sequence 
yields an isomorphism 
\[ 
{\rm H}^1({\mathcal Ext}^0_\Sigma(\CO_\Sigma(2q), \CE_\Sigma(-q))) {\buildrel \sim \over \longto} 
{\rm Ext}^1_\Sigma(\CO_\Sigma(2q), \CE_\Sigma(-q)) ~,
\]
where 
\[
 {\rm H}^1({\mathcal Ext}^0_\Sigma(\CO_\Sigma(2q), \CE_\Sigma(-q))) 
= T^{-2}+T^{-1}+T. 
\]
Since the local to global spectral sequence is functorial, the pull-back
$f^*: {\rm Ext}^1_S(\CE_\Sigma(2q), \CE_\Sigma(-q)) \to 
{\rm Ext}^1_\Sigma(\CO_\Sigma(2q), \CE_\Sigma(-q)) $ 
restricts to an isomorphism 
\[
{\rm H}^1({\mathcal Ext}^0_\Sigma (\CE_\Sigma(2q), \CE_\Sigma(-q))) 
{\buildrel \sim \over \longto} {\rm H}^1({\mathcal Ext}^0_\Sigma(\CO_\Sigma(2q), \CE_\Sigma(-q))). 
\]
This yields an equivariant splitting 
$$
\begin{aligned} 
 {\rm Ext}^1_S(\CE_\Sigma(2q), \CE_\Sigma(-q))  & \simeq 
{\rm H}^1({\mathcal Ext}^0_\Sigma (\CE_\Sigma(2q), \CE_\Sigma(-q))) \oplus {\rm H}^0({\mathcal Ext}^1_\Sigma (\CE_\Sigma(2q), \CE_\Sigma(-q)))\\ 
& \simeq {\rm H}^1({\mathcal Ext}^0_\Sigma (\CO_\Sigma(2q), \CE_\Sigma(-q))) \oplus {\rm H}^0({\mathcal Ext}^1_\Sigma (\CE_\Sigma(2q), \CE_\Sigma(-q)))\\ 
& = \left(T^{-2}+T^{-1}+T\right) + \left(T^{-5}+T^{-2}\right)
\end{aligned}
$$
Let $\wp_1, \wp_2$ respectively denote the projections onto the two 
summands. 

Next, for torus fixed extensions, one must have $a\in \{-5,-2,-1,1\}$. Moreover, a necessary condition for stability is $\wp_1(\epsilon)\neq 0$ for an extension class $\epsilon \in  {\rm Ext}^1_S(\CE_\Sigma(2q), \CE_\Sigma(-q))$. Otherwise $\CO_\Sigma(2q)$ is a destabilizing subsheaf of $F$. This 
rules out $a=-5$. 

Similarly, there is a unique injection $\iota: \wCO_\Sigma\hookrightarrow \CE_\Sigma(2q)$, 
\[ 
(1,t)\mapsto (t^2,t^3), \qquad 1\mapsto s^{-2}. 
\]
At the same time, there is an exact sequence 
$$
0\to {\rm H}^1({\mathcal Ext}^0_\Sigma 
(\wCO_\Sigma, \CE_\Sigma(-q))) 
\to {\rm Ext}^1_\Sigma(\wCO_\Sigma, \CE_\Sigma(-q)) 
\to {\rm H}^0({\mathcal Ext}^1_\Sigma (\wCO_\Sigma, \CE_\Sigma(-q))) \to 0~,
$$
where 
\[
{\rm H}^1({\mathcal Ext}^0_\Sigma 
(\wCO_\Sigma, \CE_\Sigma(-q)))  = 1+T, \qquad 
{\rm H}^0({\mathcal Ext}^1_\Sigma (\wCO_\Sigma, \CE_\Sigma(-q))) = T^{-3}+T^{-2}.
\]
This 
destabilizes $a=1$, leaving $a=-2$, in which there is two-parameter family of extensions, and 
$a=-1$ where there is one-parameter family. 
However, we have 
\[
{\rm Ext}^0_\Sigma(\CE_\Sigma, \CE_\Sigma(2q)) 
= 1+ T^2. 
\]
The two generators $f_i: \CE_\Sigma\to \CE_\Sigma(2q)$ 
are locally given by 
\begin{align}
f_1&:\quad (1,t^3) \mapsto (1,t^3), \qquad 1\mapsto 1 ~,\cr
f_2&: \quad (1,t^3)\mapsto (t^2, t^5), \qquad 1\mapsto s^{-2}~,\nonumber
\end{align}
respectively. 
At the same time, there is an exact sequence 
$$
0\to {\rm H}^1({\mathcal Ext}^0_\Sigma 
(\CE_\Sigma, \CE_\Sigma(-q))) 
\to {\rm Ext}^1_\Sigma(\CE_\Sigma, \CE_\Sigma(-q)) 
\to {\rm H}^0({\mathcal Ext}^1_\Sigma (\CE_\Sigma, \CE_\Sigma(-q))) \to 0~,
$$
where 
\[ 
 {\rm H}^1({\mathcal Ext}^0_\Sigma 
(\CE_\Sigma, \CE_\Sigma(-q))) =T~,
\]
and 
\[
{\rm H}^0({\mathcal Ext}^1_\Sigma (\CE_\Sigma, \CE_\Sigma(-q)))  = T^{-5}+T^{-2}~.
\]
For $a=-1$ one has $f_1^*\epsilon=0$. Hence 
$f_1$ lifts to an injection $\CE\hookrightarrow F$,
contradicting the maximality of $\CE_\Sigma(-q)\subset F$. 
For $a=-2$ one has $f_2^*\epsilon=0$, leading again 
to a similar contradiction. In conclusion this case is entirely ruled out. 

\bigskip 

As a second example, we consider the case  $F_1=\CO_\Sigma$, $F_2=T^a\wCO_\Sigma(q)$. Since $F_1$ is locally free 
on $\Sigma$, Lemma \ref{extlemma} yields an exact sequence 
\[
0  \to {\rm Ext}^1_\Sigma (\wCO_\Sigma(q),\CO_\Sigma) \to 
{\rm Ext}^1_S  (\wCO_\Sigma(q),\CO_\Sigma)\to  
{\rm Ext}^0_\Sigma (\wCO_\Sigma(q),\CO_\Sigma(\Sigma))\to 0.
\]
Since $\CO_\Sigma(\Sigma) \simeq T^{-10}\CO_\Sigma(2q)$, one has ${\rm Ext}^0_\Sigma (\wCO_\Sigma(q),\CO_\Sigma(\Sigma))=0$. Moreover, using Serre duality, we have 
\[
{\rm Ext}^1_\Sigma (\wCO_\Sigma(q),\CO_\Sigma)
\simeq {\rm Ext}^0_\Sigma (\CO_\Sigma, \omega_\Sigma\otimes \wCO_\Sigma(q))^\vee.
\]
Because of $\omega_\Sigma \simeq T^{-3}\CO_\Sigma(2q)$ as an equivariant line bundle, one obtains
\[
{\rm Ext}^1_\Sigma (\wCO_\Sigma(q),\CO_\Sigma)
\simeq {\rm Ext}^0_\Sigma (\CO_\Sigma, \omega_\Sigma\otimes \wCO_\Sigma(q))^\vee
= 1+T+T^2+T^3~.
\]
Therefore, $a\in \{0,1,2,3\}$ and all extensions are scheme theoretically supported on $\Sigma$. 

The next observation is that cases $a\in \{0,2,3\}$ are unstable. There are two injective morphisms 
$f_i: \wCO_\Sigma \to \wCO_\Sigma(q)$ given locally by 
\begin{align}
f_1&: (1,t) \mapsto (1,t) \qquad \ 1\mapsto 1 ~,\cr
f_2&: (1,t) \mapsto (t,t^2) \qquad 1\mapsto s^{-1}~. \nonumber
\end{align}
Moreover, by Serre duality, 
\[ 
{\rm Ext}^1_\Sigma(\wCO_\Sigma, \CO_\Sigma) \simeq {\rm Ext}^0_\Sigma (\CO_\Sigma, \omega_\Sigma\otimes \wCO_\Sigma)^\vee
= T+T^2+T^3. 
\]
Thus, given an extension class $\epsilon\in{\rm Ext}^1_\Sigma (\wCO_\Sigma(q),\CO_\Sigma)$, one has $f_1^*\epsilon=0$ for $a=0$ while one obtains $f_2^*\epsilon=0$ for 
$a=3$. In both cases
the injection $\wCO_\Sigma\hookrightarrow \wCO_\Sigma(q)$ 
lifts to $F$, destabilizing it. The case $a=2$ is similarly 
destabilized by the morphism $f:\CE_\Sigma(q)\to \wCO_\Sigma(q)$ given locally by 
\[
(1,t^3) \mapsto (1,t^3), \qquad 1\mapsto 1. 
\]

Similarly, one can check that the remaining case, $a=1$, 
is stable by testing against all possible equivariant rank one subsheaves as in Lemma \ref{invHN}. However, this case violates the maximality condition on $F_1$. Let $f:\wCO_\Sigma(-q) \to \wCO_\Sigma(q)$ be locally given by 
\[ 
(1,t)\mapsto (1,t) \qquad 1\mapsto 1. 
\]
Then for $a=1$ one has $f^*\epsilon=0$, and hence $\wCO_\Sigma(-q)$ is a subobject of $F$. 

\bigskip

Proceeding in complete analogy, one is left with the following 
cases. Let us define the Morse index of a fixed point as the dimension 
of the positive weight summand of the tangent space at 
a fixed point. This corresponds to counting Betti numbers for 
compactly supported cohomology. 
\medskip

\noindent{\bf Morse index 0} 

\noindent$(0.1)\ F_1 \simeq \wCO_\Sigma(-q),\ F_2 \simeq T^{a} \wCO_\Sigma, 
\ {\rm Ext}^1_S (F_2,F_1) \simeq {\rm Ext}^1_\Sigma (F_2,F_1) \simeq T^{-a}T^{-5}(T^{-1}+T^{-2}+T^{-3}+T^{-4}),\ a=-2.$
\[
\CT_{[F]}=2T^{-2}+3T^{-3}+3T^{-4}+2T^{-5}~.
\]

\noindent$(0.2)\ F_1 \simeq \wCO_\Sigma(-q),\ F_2 \simeq T^{a} \CE_\Sigma(q), 
\ {\rm Ext}^1_S (F_2,F_1) \simeq {\rm Ext}^1_\Sigma (F_2,F_1) \simeq T^{-a}(T^{-5}+T^{-4}+T^{-1}),\ a=-1.$
 \[
\CT_{[F]}=2T^{-2}+3T^{-3}+3T^{-4}+2T^{-5}~.
\]

\noindent{\bf Morse index 1} 

\noindent$(1.1)\ F_1 \simeq \wCO_\Sigma(-q),\ F_2 \simeq T^{a} \wCO_\Sigma, 
\ {\rm Ext}^1_S (F_2,F_1) \simeq {\rm Ext}^1_\Sigma (F_2,F_1) \simeq T^{-a}T^{-5}(T^{-1}+T^{-2}+T^{-3}+T^{-4}),\ a=-3.$
\[
\CT_{[F]}=
T+T^{-1}+2T^{-2}+T^{-3}+T^{-4}+2T^{-5}+T^{-6}+T^{-8}.
\]

\noindent$(1.2)\ F_1 \simeq \CE_\Sigma,\ F_2 \simeq T^{a} \CE_\Sigma(q), 
\ {\rm Ext}^1_S (F_2,F_1) \simeq {\rm Ext}^1_\Sigma (F_2,F_1) \simeq T^{-a}(T^{-5}(1+T^3)+T),\ a=1.$
\[
\CT_{[F]}=
T+T^{-1}+2T^{-2}+T^{-3}+T^{-4}+2T^{-5}+T^{-6}+T^{-8}.
\]

\noindent{\bf Morse index 2}

\noindent$(2.1)\ F_1 \simeq \CO_\Sigma(q),\ F_2 \simeq T^{a} \wCO_\Sigma,\ {\rm Ext}^1_S (F_2,F_1) \simeq {\rm Ext}^1_\Sigma (F_2,F_1) \simeq T^{3-a}(1+T^{-1}),\ a=3.$
\[
\CT_{[F]} =  T^3+T+T^{-1}+
T^{-2}+T^{-3}+T^{-4}+T^{-5}+T^{-6}+T^{-8}+T^{-10}.
\]

\noindent$(2.2)\ F_1 \simeq \wCO_\Sigma(-q),\ F_2 \simeq T^{a} \CO_\Sigma(2q), 
\ {\rm Ext}^1_\Sigma (F_2,F_1) \simeq T^{3-a}(T^{-4}+T^{-5}),\ a=-1.$
\[
\CT_{[F]} =  T^3+T+T^{-1}+
T^{-2}+T^{-3}+T^{-4}+T^{-5}+T^{-6}+T^{-8}+T^{-10}.
\]

\noindent$(2.3)\ F_1 \simeq \wCO_\Sigma(-q),\ F_2 \simeq T^{a} \CE_\Sigma(q), 
\ {\rm Ext}^1_S (F_2,F_1) \simeq {\rm Ext}^1_\Sigma (F_2,F_1) \simeq T^{-a}(T^{-5}+T^{-4}+T^{-1}),\ a=-4.$
\[
\CT_{[F]} =  
T^3+T+T^{-1}+T^{-2}+T^{-3}+T^{-4}+
T^{-5}+T^{-6}+T^{-8}+T^{-10}.
\]

\noindent$(2.4)\ F_1 \simeq \wCO_\Sigma(-q),\ F_2 \simeq T^{a} \wCO_\Sigma, 
\ {\rm Ext}^1_S (F_2,F_1) \simeq {\rm Ext}^1_\Sigma (F_2,F_1) \simeq T^{-a}T^{-5}(T^{-1}+T^{-2}+T^{-3}+T^{-4}),\ a=-4.$
\[
\CT_{[F]} = 
T^2+T+T^{-1}+2T^{-2}+2T^{-5}+T^{-6}+T^{-8}+T^{-9}.
\]

\noindent{\bf Morse index 3} 

\noindent$(3.1)\ F_1 \simeq \wCO_\Sigma(-q),\ F_2 \simeq T^{a} \wCO_\Sigma, 
\ {\rm Ext}^1_S (F_2,F_1) \simeq {\rm Ext}^1_\Sigma (F_2,F_1) \simeq T^{-a}(T^{-1}+T^{-2}+T^{-3}+T^{-4}),\ a=-5.$
\[
\CT_{[F]} = T^3+T^2+T+T^{-2}+T^{-3}+
T^{-4}+T^{-5}+T^{-8}+T^{-9}+T^{-10}.
\]

\noindent$(3.2)\ F_1 \simeq \CO_\Sigma(q),\ F_2 \simeq T^{a} \wCO_\Sigma,\ {\rm Ext}^1_S (F_2,F_1) \simeq {\rm Ext}^1_\Sigma (F_2,F_1) \simeq T^{3-a}(1+T^{-1}),\ a=2.$
\[
\CT_{[F]} = 
T^3+2T+T^{-2}+T^{-3}+T^{-4}+T^{-5}+2T^{-8}+T^{-10}.
\]

\noindent$(3.3)\ F_1 \simeq \wCO_\Sigma(-q),\ F_2 \simeq T^{a} \CO_\Sigma(2q), 
\ {\rm Ext}^1_\Sigma (F_2,F_1) \simeq T^{3-a}(T^{-4}+T^{-5}),\ a=-2.$
\[
\CT_{[F]}=
T^3+2T+T^{-2}+T^{-3}+T^{-4}+T^{-5}+2T^{-8}+T^{-10}. 
\]

\noindent$(3.4)\ F_1 \simeq \CE_\Sigma,\ F_2 \simeq T^{a} \CE_\Sigma(q), 
\ {\rm Ext}^1_S (F_2,F_1) \simeq {\rm Ext}^1_\Sigma (F_2,F_1) \simeq T^{-a}(T^{-5}+T^{-2}+T),\ a=-2.$
\[
\CT_{[F]} = 
T^3+2T+T^{-2}+T^{-3}+T^{-4}+T^{-5}+2T^{-8}+T^{-10}.
\]

\noindent$(3.5)\ F_1 \simeq \CO_\Sigma(q),\ F_2 \simeq T^{a} \CE_\Sigma(q), 
\ {\rm Ext}^1_S (F_2,F_1) \simeq {\rm Ext}^1_\Sigma (F_2,F_1) + {\rm Ext}^0_\Sigma (F_2,F_1(\Sigma)) \simeq T^{3-a}(1+T^{-2})+T^{-8-a},\ a=3$
\[
\CT_{[F]} = T^4+2T+2T^{-2}+2T^{-5}+2T^{-8}+T^{-11}.
\]

\noindent$(3.6)\ F_1 \simeq \wCO_\Sigma(-q),\ F_2 \simeq T^{a} \CE_\Sigma(q), 
\ {\rm Ext}^1_S (F_2,F_1) \simeq {\rm Ext}^1_\Sigma (F_2,F_1) \simeq T^{-a}(T^{-5}+T^{-4}+T^{-1}),\ a=-5.$
\[
\CT_{[F]} =  T^4+2T+2T^{-2}+2T^{-5}+2T^{-8}+T^{-11}.
\]

\noindent$(3.7)\ F_1 \simeq \CE_\Sigma,\ F_2 \simeq T^{a} \wCO_\Sigma, 
\ {\rm Ext}^1_S (F_2,F_1) \simeq {\rm Ext}^1_\Sigma (F_2,F_1) \simeq T^{-a}(T^{-3}+T^{-2}+T),\ a=-3$.
\[
\CT_{[F]} =  T^4+2T+2T^{-2}+2T^{-5}+2T^{-8}+T^{-11}.
\]

\noindent$(3.8)\ F_1 \simeq \CE_\Sigma,\ F_2 \simeq T^{a} \CO_\Sigma(2q), 
\ {\rm Ext}^1_S (F_2,F_1) \simeq {\rm Ext}^1_\Sigma (F_2,F_1) \simeq T^{3-a}(T^{-2}+T^{-4}),\ a=1.$
\[
\CT_{[F]} =  T^4+2T+2T^{-2}+2T^{-5}+2T^{-8}+T^{-11}.
\]

\noindent{\bf Morse index 4}

\noindent$(4.1)\ F_1 \simeq \CO_\Sigma(q),\ F_2 \simeq T^{a} \CO_\Sigma(2q),\ 
{\rm Ext}^1_S (F_2,F_1) \simeq {\rm Ext}^1_\Sigma (F_2,F_1)+{\rm Ext}^0_\Sigma (F_2,F_1(\Sigma)) \simeq T^{3-a}(1+T^{-2})+T^{-10-a},\ a=3\ $
\[
\CT_{[F]} =  T^6+T^3+2T+T^{-2}+T^{-5}+2T^{-8}+T^{-10}+T^{-13}.
\]

\noindent$(4.2)\ F_1 \simeq \CO_\Sigma(q),\ F_2 \simeq T^{a} \CE_\Sigma(q), 
\ {\rm Ext}^1_S (F_2,F_1) \simeq {\rm Ext}^1_\Sigma (F_2,F_1) + {\rm Ext}^0_\Sigma (F_2,F_1(\Sigma)) \simeq T^{3-a}(1+T^{-2})+T^{-8-a},\ a=1$
\[
\CT_{[F]} =
T^3+2T^2+T+T^{-2}+T^{-5}+T^{-8}+2T^{-9}+T^{-10}.
\]

\noindent$(4.3)\ F_1 \simeq \CO_\Sigma(q),\ F_2 \simeq T^{a} \CE_\Sigma(q), 
\ {\rm Ext}^1_S (F_2,F_1) \simeq {\rm Ext}^1_\Sigma (F_2,F_1) + {\rm Ext}^0_\Sigma (F_2,F_1(\Sigma)) \simeq T^{3-a}(1+T^{-2})+T^{-8-a},\ a=-8$
\[
\CT_{[F]} = 
T^{11}+T^9+T^3+T+T^{-2}+T^{-5}+T^{-10}+T^{-16}+
T^{-18}.
\]

\noindent$(4.4)\ F_1 \simeq \CE_\Sigma,\ F_2 \simeq T^{a} \CE_\Sigma(q), 
\ {\rm Ext}^1_S (F_2,F_1) \simeq {\rm Ext}^1_\Sigma (F_2,F_1) \simeq T^{-a}(T^{-5}(1+T^3)+T),\ a=-5.$
\[
\CT_{[F]} =
T^6+T^3+2T+T^{-2}+T^{-5}+2T^{-8}+T^{-10}+T^{-13}.
\]

\noindent$(4.5)\ F_1 \simeq \CE_\Sigma,\ F_2 \simeq T^{a} \CO_\Sigma(2q), 
\ {\rm Ext}^1_S (F_2,F_1) \simeq {\rm Ext}^1_\Sigma (F_2,F_1) \simeq T^{3-a}(T^{-2}+T^{-4}),\ a=-1.$
\[
\CT_{[F]}=T^3+2T^2+T+T^{-2}+T^{-5}+T^{-8}+2T^{-9}+T^{-10}.
\]

\noindent{\bf Morse index 5} 

\noindent$(5.1)\ F_1 \simeq \CO_\Sigma(q),\ F_2 \simeq T^{a} \CO_\Sigma(2q),\ 
{\rm Ext}^1_S (F_2,F_1) \simeq {\rm Ext}^1_\Sigma (F_2,F_1)+{\rm Ext}^0_\Sigma (F_2,F_1(\Sigma)) \simeq T^{3-a}(1+T^{-2})+T^{-10-a},\ a=1$
\[
\CT_{[F]} = 
T^4+2T^3+T^2+T+T^{-8}+T^{-9}+2T^{-10}+T^{-11}.
\]

\noindent$(5.2)\ F_1 \simeq \CO_\Sigma(q),\ F_2 \simeq T^{a} \CO_\Sigma(2q),\ 
{\rm Ext}^1_S (F_2,F_1) \simeq {\rm Ext}^1_\Sigma (F_2,F_1)+{\rm Ext}^0_\Sigma (F_2,F_1(\Sigma)) \simeq T^{3-a}(1+T^{-2})+T^{-10-a},\ a=-10.$
\[
\CT_{[F]} = 
T^{13}+T^{11}+T^3+2T+2T^{-8}+T^{-10}+T^{-18}+T^{-20}.
\]

\medskip
{\bf Step 3}. Finally, one has to find all isomorphisms 
between stable sheaves $F$ with different extension 
presentations. Suppose 
\begin{align}
0\to F_1{\buildrel f_1\over \longto} F  {\buildrel f_2\over \longto} F_2 \to  0~,\cr
0\to G_1 {\buildrel g_1\over \longto}  G{\buildrel g_2\over \longto}  G_2 \to 0~, \label{eq:Gext}
\end{align}
are two extensions such that 
\begin{itemize}\setlength\itemsep{-.2em}
\item $F,G$ are stable with $\chi(F)=\chi(G)=1$, and 
\item $\chi(F_1)=\chi(G_1)=1$.
\end{itemize}
This covers all cases obtained at the previous step. 
Suppose
there is an isomorphism 
$f: F \to G$. Since $F_1, G_1$ are rank one torsion free sheaves the second condition implies 
${\rm Ext}^0(F_1,G_1)=0$. By exactness, it follows that $g_2\circ f \circ f_1 : F_1 \to G_2$ is a nonzero morphism 
of rank one torsion free sheaves. In particular it is injective. Let $Q= {\rm Coker}(f\circ f_1)$ and $Q_2={\rm Coker}(g_2\circ f \circ f_1)$. 
Then there is a commutative diagram 
$$
\xymatrix{ 
& & 0 \ar[d]& 0 \ar[d]\\
& & F_1\ar[r]^-{{\bf 1}} \ar[d]^-{f\circ f_1} & F_1\ar[d]^-{g_2\circ f\circ f_1} \\
0 \ar[r] & G_1\ar[d]^-{\bf 1} \ar[r]^-{g_1} & G \ar[d] \ar[r]^-{g_2} & G_2 \ar[r] \ar[d]& 0\\
0 \ar[r] & G_1 \ar[r] & Q \ar[r] \ar[d]& Q_2 \ar[r] \ar[d]& 0 \\
& & 0& 0 \\}
$$
with exact rows and columns. 
This implies that the extension 
class $\epsilon_G \in {\rm Ext}^1_S(G_2,G_1)$ is in the kernel of the map
\[ 
(g_2\circ f\circ f_1)^*: {\rm Ext}^1_S(G_2,G_1) \to 
{\rm Ext}^1_S(F_1,G_1)~.
\]

Conversely, suppose $G$ is a stable torsion free sheaf 
with $\chi(G)=1$ which fits in the extension \eqref{eq:Gext} 
with $G_1, G_2$ rank one torsion free sheaves on $\Sigma$ 
and $\chi(G_1)=1$. Let $F_1$ be a rank one torsion free 
sheaf on $\Sigma$ with $\chi(F_1)=0$ and suppose there is 
an injection $\iota: F_1 \hookrightarrow  G_2$ such that 
$\iota^*(\epsilon_G)=0$, where $\iota^*: {\rm Ext}^1_S(G_2,G_1) \to 
{\rm Ext}^1_S(F_1,G_1)$ is the natural pullback map. 
Then $\iota$ lifts to an injection $f_1:F_1 \to G$ and there is a commutative diagram 
\be\label{eq:extdiagB}
\xymatrix{ 
& & 0 \ar[d]& 0 \ar[d]\\
& & F_1\ar[r]^-{{\bf 1}} \ar[d]^-{f_1} & F_1\ar[d]^-{\iota} \\
0 \ar[r] & G_1\ar[d]^-{\bf 1} \ar[r]^-{g_1} & G \ar[d]^-{f_2} \ar[r]^-{g_2} & G_2 \ar[r] \ar[d]& 0\\
0 \ar[r] & G_1 \ar[r] & F_2 \ar[r] \ar[d]& G_2 \ar[r] \ar[d]& 0 \\
& & 0& 0 \\}
\ee
with exact rows and columns. Clearly, the cokernel 
$F_2 = {\rm Coker}(f_1)$ has numerical invariants $\ch_1(F_2)= \Sigma$, $\chi(F_2)=1$ and is set 
theoretically supported on $\Sigma$. 
 Since $G$ is stable, with $\chi(G)=1$, $F_2$ must be a rank one torsion free sheaf on $\Sigma$. Otherwise, $G$ would admit a destabilizing quotient. Therefore, for each such subsheaf $F_1$, one obtains an additional extension presentation of $G$.  Moreover, if $F_1$ is not isomorphic to 
$G_1$, the new extension presentation is clearly different from \eqref{eq:Gext}. 

There are however two cases below where one has to establish  such an equivalence although $F_1\simeq G_1$. These are 
the index 0 case and the last isomorphism of the index 3 case. Proving these identifications requires more work since it is not a priori obvious that the new extension presentation is different from the first one. 

Assuming $F_1\simeq G_1$, there is an exact sequence 
\be\label{eq:longextseqA} 
0\to {\rm Hom}_\Sigma(G_2,G_2)\to {\rm Hom}_\Sigma(G,G_2)
\to {\rm Hom}_\Sigma(G_1,G_2)  \xrightarrow{  \delta } {\rm Ext}^1_\Sigma(G_2, G_2)\to \cdots 
\ee
where the map $\delta :{\rm Hom}_\Sigma(G_1,G_2) \to {\rm Ext}^1_\Sigma(G_2, G_2)$ is given by the Yoneda 
product with the extension class $\epsilon_G \subset 
{\rm Ext}_\Sigma^1(G_2, G_1)$.  

Suppose ${\rm Ker}(\delta)=0$. Then ${\rm Hom}_\Sigma(G,G_2) \simeq {\rm Hom}_\Sigma(G_2,G_2) \simeq \IC$. 
Next suppose in the diagram \eqref{eq:extdiagB} one has an isomorphism $F_2 \simeq G_2$. The above isomorphism implies that  $f_2=\lambda g_2$
for some $\lambda \in \IC$. Since $G$ is stable, $\lambda \neq 0$. Then there is a commutative diagram 
\[ 
\xymatrix{ 
0\ar[r] & G_1 \ar[r]^-{g_1}  & G \ar[r]^-{g_2} \ar[d]^-{\bf 1} & G_2 
\ar[d]^-{\lambda {\bf 1}} \ar[r] & 0 \\
0\ar[r] & F_1 \ar[r]^-{f_1}  & G \ar[r]^-{f_2} & G_2 \ar[r] & 0 \\}
\]
This means that there is a morphism $G_1\to F_1$. Moreover, the snake lemma implies that this morphism has to be an isomorphism. However, this further implies that $g_2\circ f_1=0$, which leads to a contradiction as $g_2\circ f_1 = \iota \neq 0$ by construction. 

\medskip

\noindent{\bf Morse index 0}. $(0.1)$ is equivalent to $(0.2)$ 

Recall that 

\noindent $(0.1)\ F_1 \simeq \wCO_\Sigma(-q),\ F_2 \simeq T^{a} \wCO_\Sigma, 
\ {\rm Ext}^1_S (F_2,F_1) \simeq {\rm Ext}^1_\Sigma (F_2,F_1) \simeq T^{-a}T^{-5}(T^{-1}+T^{-2}+T^{-3}+T^{-4}),\ a=-2.$

\noindent $(0.2)\ G_1 \simeq \wCO_\Sigma(-q),\ G_2 \simeq T^{a} \CE_\Sigma(q), 
\ {\rm Ext}^1_S (F_2,F_1) \simeq {\rm Ext}^1_\Sigma (F_2,F_1) \simeq T^{-a}(T^{-5}+T^{-4}+T^{-1}),\ a=-1.$

There is a unique injective morphism $\iota: 
\wCO_\Sigma(-q) \to \CE_\Sigma(q)$ given locally by 
\[ 
1\mapsto t^2, \qquad s\mapsto s^{-1}. 
\]
Moreover, 
\[ 
{\rm Ext}^1_\Sigma(\CE_\Sigma(q), 
\wCO_\Sigma(-q)) = T^{-5}+T^{-4}+T^{-1}~,
\]
while the pull-back map 
$\iota^*: {\rm Ext}^1_\Sigma(\CE_\Sigma(q), 
\wCO_\Sigma(-q)) \to {\rm Ext}^1_\Sigma(\wCO_\Sigma(-q), 
\wCO_\Sigma(-q))$  is given by multiplication by $T^2$. This maps $\epsilon_G = T^{-1}$ to $0$, 
hence $\iota$ lifts to an injection $f_1:\wCO_\Sigma(-q) \hookrightarrow G$ as in the diagram \eqref{eq:extdiagB}. 
Furthermore, we have
\[ 
{\rm Hom}_\Sigma(\wCO_\Sigma(-q), \CE_\Sigma(q)) 
= \langle T^2\rangle ~,
\]
and 
\[
{\rm Ext}^1_\Sigma(\CE_\Sigma(q), \CE_\Sigma(q))= T^{-5}+T^{-2}+T~,
\]
in the exact sequence \eqref{eq:longextseqA}. The coboundary map $\delta$ is given by multiplication by $\epsilon_G = T^{-1}$, hence 
it is injective. As shown above, this implies
that the new extension presentation must be different from $(0.2)$. Thus, it must 
be isomorphic, as an extension, to $(0.1)$.

Proceeding by analogy, one finds the following 
identifications. 
\begin{align}\nonumber
\begin{array}{cc}
\textrm{{\bf Morse index 1}.} & \textrm{(1.1)  is equivalent to (1.2). }\\
\textrm{{\bf Morse index 2}.} &\textrm{(2.1) is equivalent to (2.3). }\\
\textrm{{\bf Morse index 3}.}& \textrm{(3.2) is equivalent to (3.3). }\\
&\textrm{(3.2) is equivalent to (3.4). }\\
&\textrm{(3.5) is equivalent to (3.6). }\\
&\textrm{(3.7) is equivalent to (3.8). }\\
\textrm{{\bf Morse index 4}.} &\textrm{(4.2) is equivalent to (4.5). }
\end{array}
\end{align}

Finally, one can verify that there are no further identifications left among pairs of fixed points with identical tangent spaces. For example, consider the fixed points  $(2.1)$ and $(2,2)$ which have identical equivariant tangent spaces. 
Note that there is a unique non-zero morphism 
$\iota: \CO_\Sigma(q)\to \CO_\Sigma(2q)$, given locally by $1\mapsto 1$, respectively $s^{-1}\mapsto s^{-1}$. For the extension 
$(2.2)$ one has $\epsilon_G = T^{-1}$ 
in 
\[ 
{\rm Ext}^1_\Sigma(\CO_\Sigma(2q), 
\wCO_\Sigma(-q)) = \langle T^{-2}, T^{-1}\rangle ~.
\]
At the same time, one has 
\[ 
{\rm Ext}^1_\Sigma(\CO_\Sigma(q), 
\wCO_\Sigma(-q)) = \langle T^{-1}\rangle,
\]
so that $\iota^*(\epsilon_G) = T^{-1}\neq 0$. Hence, these two extensions yield different fixed points. 

Collecting the results, the Poincar\'e polynomial of the moduli space is  
\[
P_{5,2,2}(v)=2v^{10} + 4v^{8}+4v^{6}+3v^{4}+v^{2}+1.
\]

\appendix

\section{Torsion-free sheaves on $\Sigma$}\label{torfreeapp} 

The goal of this section is to prove some basic results on torsion free sheaves on a singular plane curves $\Sigma$ as in \S\ref{gluingsect}. 
The notation and conventions will be as in that section. In particular, $\nu : \wSigma \to \Sigma$, 
$\wSigma=\IP^1$ is the normalization of $\Sigma$, and 
$\wq = \nu^{-1}(q)$ is the inverse image of the point at infinity on $\Sigma$.

\begin{lemm}\label{TFa}
Let $F$ be a rank $r$ torsion free coherent sheaf on $\Sigma$. Then there is a 
sequence of integers $d_1 \leq \cdots \leq d_r$ 
and a zero-dimensional sheaf
$T_F$ with set theoretic support $\{\sigma\}$, 
both determined by $F$,  such that $F$ fits in an exact sequence 
$$
0 \to F {\buildrel f \over \longto} \oplus_{i=1}^r \wCO_\Sigma(d_iq) \to T_F\to 0.
$$
\end{lemm} 

{\it Proof}. 
Let $G\subset \nu^*F$ be the maximal zero-dimensional subsheaf of $\nu^*F$ and let 
$\wF = \nu_*\big((\nu^*F)/G\big)$.  Thus, one has an exact sequence of $\CO_\Sigma$-modules 
\[ 
0 \to \nu_*G \to \nu_*\nu^*F {\buildrel g\over \longto} \wF\to 0,
\]
where $\nu_*G$ is zero-dimensional.
By construction there is an injective morphism $\iota:F\hookrightarrow \nu_*\nu^*F$. 
Let $f = g \circ \iota: F \to \wF$. Then the snake lemma implies that ${\rm Ker}(f) \subset 
\nu_*G$, hence $f$ must be injective since $F$ is torsion free. Let $T_F = {\rm Coker}(f)$. 
Then, it must be set theoretically supported at the singular point since $\nu$ is an isomorphism on the complement $\Sigma\setminus\{\sigma\}$. 
Moreover, the quotient $\nu^*F/G$ is a rank $r$ torsion free sheaf on $\wSigma =\IP^1$ so that there is a unique 
sequence of integers $d_1\leq \cdots \leq d_r$ 
such that 
\[
(\nu^*F)/G \simeq \oplus_{i=1}^{r}
\CO_{\wSigma}(d_i\wq).
\]
This implies the claim since $\nu_*\CO_{\wSigma}(d_i\wq)\simeq {\wCO}_\Sigma(d_iq)$. 

\hfill $\Box$

The next goal is to prove the existence of a maximal splitting for torsion free sheaves on $\Sigma$, 
generalizing a construction of Atiyah for smooth curves
\cite[\S  4]{VB_elliptic}.

\begin{lemm}\label{TFaa} 
In the setup of Lemma \ref{TFa}, there is a second exact sequence 
\be\label{eq:torfreeseqB} 
0 \to \oplus_{i=1}^r \CO_\Sigma(d_iq) \to F \to Q_F~,
\ee
where $Q_F$ is a zero-dimensional sheaf fitting into an exact sequence 
\be\label{eq:torseqA} 
0 \to Q_F \to \CT_\Sigma \to T_F \to 0.
\ee
\end{lemm} 

{\it Proof}. By construction the space of local sections $\Gamma_\CV(F)$ is isomorphic to an 
$R$-submodule of $\wR^{\oplus r}$. Moreover, 
one must have an isomorphism 
\[\Gamma_\CV(F) \otimes_R {\wR}\, /\, {\rm torsion} 
\simeq \wR^{\oplus r}.
\]
This implies that $\Gamma_\CV(F)$ must contain all the canonical generators of $\wR^{\oplus r}$. Thus, 
there is an injective morphism of $R$-modules 
$R^{\oplus r} \to \Gamma_\CV(F)$, which induces a local injection $\iota_\CV : \CO_\CV^{\oplus r} \to F|_\CV$. This easily extends to an injective morphism $\oplus_{i=1}^r \CO_\Sigma(d_iq)\to F$ 
since $F|_\CU$ is canonically identified with $\oplus_{i=1}^r \wCO_\Sigma(d_iq)$. This yields the  sequence \eqref{eq:torfreeseqB}. The second sequence 
\eqref{eq:torseqA} follows from the snake lemma. 

\hfill $\Box$

 \begin{lemm}\label{TFb}
 Let $F$ be a rank $r$ torsion free coherent sheaf on $\Sigma$. Then there exists a saturated rank one torsion free subsheaf $F_1\subset F$ such that $\chi(F_1)$ is maximal among all saturated rank one torsion free subsheaves. Such a subsheaf will be called a maximal rank one subsheaf in the following.
\end{lemm}

{\it Proof}. 
Using the notation in  Lemma \ref{TFa}, for each $1\leq i \leq r$ let 
\[
\wF_i = {\bigoplus_{\substack{j=1 \\  j\neq i}}^r}\, \wCO_\Sigma(d_jq)~,
\]
and let $p_i : \wF \to \wF_i$ denote the canonical projection. Let $G_i = {\rm Ker}(p_i\circ f)\subset F$. Then the snake lemma 
yields an exact sequence 
\[ 
0 \to G_i \to \wCO_\Sigma(d_iq) \to T_i\to 0~,
\]
where $T_i \subset T_F$. In particular, $G_i$ is a rank one torsion free subsheaf of $F$. It is also saturated since $F/G_i$ is isomorphic to a subsheaf of $\wF_i$, which is torsion free. 

Now let  $c ={\rm max}\{\chi(F_i)\, |\, 1\leq i \leq r\}$ and consider the set of isomorphism 
classes of saturated rank one torsion free subsheaves $G\subset F$ such that $\chi(G)\geq c$. This set is clearly non-empty since it contains at least one of the $G_i$. Grothendieck's lemma \cite[Lemma 1.7.9]{huylehn} 
implies that this set is bounded. Therefore, 
there exists a subsheaf $F_1\subset F$ as claimed above. 

\hfill $\Box$

Applying the above result recursively, one then constructs a filtration  
$$
0 =F_0\subset F_1 \subset F_2 \subset \cdots \subset F_r= F
$$
by saturated torsion free subsheaves such that  $F_i/F_{i-1}$ is a maximal rank one subsheaf 
for all $1\leq i \leq r$. Such a filtration will be called a maximal splitting of $F$ by analogy 
with \cite[\S 4]{VB_elliptic}.
The next goal is to prove a generalization of
\cite[Lemma 3]{VB_elliptic} in the present context. Recall that the dualizing sheaf of $\Sigma$ is isomorphic to a line bundle $\CO_\Sigma(2eq)$, with $e\in \IZ$. 
Moreover, under the current assumptions, $e\geq 0$. 

\begin{lemm}\label{TFc} 
Let $F$ be a rank one torsion free sheaf on $\Sigma$. Then there is an exact sequence 
\be\label{eq:rkoneseqA}
0\to F \to F \otimes \omega_\Sigma \to \CO_{2eq}(2eq)\to 0~.
\ee
Moreover, 
\be\label{eq:extrkoneA}
{\rm dim}\, {\rm Ext}^1_\Sigma(F,F)  = e +1. 
\ee
\end{lemm}

{\it Proof}. Since $\omega_\Sigma\simeq \CO_\Sigma(2eq)$ is an invertible sheaf, there is 
an exact sequence 
\[
0\to \CO_\Sigma\to \omega_\Sigma \to 
\CO_{2eq}(2eq)\to 0. 
\]
Since $F$ is locally free at $q$, taking a tensor product with $F$ one obtains the sequence \eqref{eq:rkoneseqA}. 
This yields 
\[ 
\chi(F, F \otimes \omega_\Sigma)=\chi(F,F) + 
\chi(F, \CO_{2eq}(2eq)). 
\]
Using Serre duality this implies 
\[ 
2 \chi(F,F) = - \chi(F, \CO_{2eq}(2eq))~.
\]
Moreover, since $\CO_{2eq}(2eq)$ is zero-dimensional, ${\rm Hom}_\Sigma(\CO_{2eq}(2eq), F\otimes \omega_\Sigma) =0$. Therefore, using Serre duality, 
\[
\chi(F, \CO_{2eq}(2eq)) = {\rm dim}\, 
 {\rm Hom}_\Sigma(F,\CO_{2eq}(2eq)) = 2e 
 \]
 since $F$ is locally free at $q$. Thus, $\chi(F,F) = -e$. Since $F$ is torsion free of rank one, ${\rm Ext}^0_\Sigma(F,F)\simeq \IC$, which implies the relation 
 \eqref{eq:extrkoneA}. 
 
\hfill $\Box$

\begin{lemm}\label{TFd} 
Let $F$ be a rank two torsion free sheaf on $\Sigma$ and let $F_1\subset F$ be a maximal rank one subsheaf as in Lemma \ref{TFb}. 
Let $F_2= F/F_1$. Then 
\be\label{eq:splitboundA} 
{\rm dim}\, {\rm Hom}_\Sigma(F_1, F_2) \leq e+2. 
\ee
\end{lemm}

{\it Proof}. 
If the extension 
\be\label{eq:maxsplitB}
0 \to F_1 \to F \to F_2 \to 0
\ee
is trivial, it follows that $\chi(F_2)\leq \chi(F_1)$ by construction. Hence ${\rm Hom}_\Sigma(F_1,F_2) =0$ since both $F_1, F_2$ 
are rank one torsion free sheaves. 

Suppose the extension \eqref{eq:maxsplitB} is nontrivial. Then there is a long exact sequence
\begin{align}
0 &\to {\rm Hom}_\Sigma(F_1,F_1) \to 
{\rm Hom}_\Sigma(F_1,F) \to  {\rm Hom}_\Sigma(F_1,F_2)\cr
&\xrightarrow{\delta} {\rm Ext}^1_\Sigma(F_1,F_1) \to 
{\rm Ext}^1_\Sigma(F_1,F) \to 
{\rm Ext}^1_\Sigma(F_1,F_2) \to 0~, \nonumber
\end{align}
which yields 
\[ 
\bal 
{\rm dim}\, {\rm Hom}_\Sigma(F_1,F) \geq 
{\rm dim}\, {\rm Hom}_\Sigma(F_1,F_1) + {\rm dim}\, {\rm Hom}_\Sigma(F_1,F_2)- {\rm dim}\, 
{\rm Ext}^1_\Sigma(F_1,F_1).
\eal
\]
Since ${\rm dim}\,{\rm Hom}_\Sigma(F_1,F_1) =1$ for a rank one torsion free sheaf, using Lemma \ref{TFc}, 
one obtains 
\[ 
{\rm dim}\, {\rm Hom}_\Sigma(F_1,F) \geq  
{\rm dim}\, {\rm Hom}_\Sigma(F_1,F_2)-e. 
\]

Now suppose ${\rm dim}\, {\rm Hom}_\Sigma(F_1,F_2)\geq e+3$. Then it will be shown below that this leads to a contradiction. Since $F_1$ is locally free at $q$, there is an exact sequence 
\be\label{eq:rkoneseqB}
0\to F_1 \to F_1(q) \to F_1\otimes \CO_q(q)\to 0 ~,
\ee
where $F_1(q) = F_1\otimes \CO_\Sigma(q)$. 
This yields an exact sequence
\[
0\to {\rm Hom}_\Sigma(F_1(q),F) \to {\rm Hom}_\Sigma(F_1,F) \to {\rm Ext}^1_\Sigma(\CO_q(q), F) \to \cdots 
\]
while Serre duality yields an isomorphism 
\[
{\rm Ext}^1_\Sigma(\CO_q(q), F) \simeq 
{\rm Ext}^0_\Sigma(F, 
\CO_q(q)\otimes \omega_\Sigma)^\vee.
\]
Since $F$ is locally free at $q$ of rank two, ${\rm Ext}^0_\Sigma(F, 
\CO_q(q)\otimes \omega_\Sigma)\simeq \IC^2$. Therefore, under the current assumptions it follows that ${\rm Hom}_\Sigma(F_1(q),F)$ is at least one-dimensional. However, any nonzero morphism $\psi: F_1(q)\hookrightarrow F$ must be injective since 
$F_1(q)$ is torsion free of rank one. Let ${\overline {{\rm Im}(\psi)}}$ be the saturation of the image of $\psi$ in $F$. By construction, this is a saturated 
rank one torsion free subsheaf of $F$ such that 
\[ 
\chi({\overline {{\rm Im}(\psi)}}) \geq \chi(F_1(q))~.
\]
However the exact sequence \eqref{eq:rkoneseqB}
yields the relation
\[
 \chi(F_1(q))=
\chi(F_1) +1~,
\]
since $F_1$ is locally free of rank one at $q$. 
This contradicts the assumption that $F_1\subset F$ is a maximal rank one subsheaf. Thus, the inequality 
\eqref{eq:splitboundA} must hold. 

\hfill $\Box$

An immediate corollary of Lemma \ref{TFd} is the following, which can be proved by induction.

\begin{coro}\label{TFe} 
Let $F$ be a rank $r\geq 2$ torsion free sheaf on $\Sigma$ and let $0=F_0 \subset F_1 \subset \cdots \subset F_r=F$ be a maximal splitting of $F$. Let 
$G_i= F_i/F_{i-1}$, $1\leq i \leq r$ be the successive quotients.
Then  
$$
{\rm dim}\, {\rm Hom}_\Sigma(G_{i}, G_{i+1}) \leq e+2 
$$
for all $1\leq i \leq r-1$. 
\end{coro} 

\section{Pure dimension one sheaves on $S$}\label{spsheaves} 

This section contains some basic facts on pure dimension one sheaves on $S$ with set theoretic support on $\Sigma$.

\subsection{Filtrations}\label{filtrsect} 
Let $F$ be a pure dimension one sheaf on $S$ with 
set theoretic support on $\Sigma$ such that $\ch_1(F) = r \Sigma$
for some $r\geq 1$. Then 
there is a canonical filtration 
\be\label{eq:filtrA} 
0 = K_0 \subseteq K_1 \subseteq \cdots \subseteq K_r = F ~,
\ee 
by saturated subsheaves such that all successive quotients $K_i/K_{i-1}$ are scheme theoretically supported on $\Sigma$. This filtration is defined by 
\[ 
K_i = {\rm Ker}( {\bf 1}_F \otimes \zeta^{\otimes i} :F{\buildrel {} \over \longto}
F\otimes \CO_S(i\Sigma) ) ~,
\]
where $\zeta \in H^0(S,\CO(\Sigma))$ is a defining 
section of $\Sigma$.

Furthermore, as shown in Appendix \ref{torfreeapp}, right after
Lemma \ref{TFb}, each successive quotient $K_i/K_{i-1}$ admits a specific filtration 
by rank one torsion free subsheaves called a maximal splitting. Therefore, the filtration 
\eqref{eq:filtrA} can be refined to a filtration 
$$
0 = F_0 \subset F_1 \subset \cdots \subset F_r=F ~,
$$
such that each successive quotient is a rank one 
torsion free sheaf on $\Sigma$. 

\subsection{Extensions}\label{extsect} 

Next suppose $F_1,F_2$ are torsion free sheaves on the 
curve $\Sigma$, embedded as a divisor in the surface $S$ as in \S\ref{spectralsect}. By extension by zero, they are canonically identified with pure dimension one sheaves on $S$. Then 
one has the following extension result. 

\begin{lemm}\label{extlemma} 
There is an isomorphism 
$$
{\rm Ext}^0_S(F_2,F_1) \simeq {\rm Ext}^0_\Sigma(F_2,F_1)
$$
and an exact sequence
$$ 
\bal
0 & \to {\rm Ext}^1_\Sigma (F_2,F_1) \to 
{\rm Ext}^1_S(F_2,F_1) \to  {\rm Ext}^0_\Sigma(F_2, F_1(\Sigma))\\
& \to {\rm Ext}^2_\Sigma(F_2,F_1) \to 
{\rm Ext}^2_S(F_2,F_1) 
 \to {\rm Ext}^1_\Sigma(F_2,F_1(\Sigma)) \\
 &\to {\rm Ext}^3_\Sigma(F_2,F_1)\to 0. 
 \eal
$$
\end{lemm}

{\it Proof.} 
Let $\iota: \Sigma \to S$ denote the canonical embedding. Then for any coherent sheaf $G$ on $S$ the derived adjunction formula yields a quasi-isomorphism 
\[
{\rm RHom}_S(R\iota_*F_2, G) \simeq {\rm RHom}_\Sigma(F_2, \iota^!G) 
\]
where $\iota^!G = L\iota^*G\otimes\omega_\Sigma[-1]$. 
Since $\Sigma$ is a divisor in $S$, its structure sheaf has a two term locally free resolution 
$\CO_S(-\Sigma) \to \CO_S$. Therefore, all local Tor 
sheaves ${\mathcal Tor}_k^S(\CO_\Sigma, G)$ vanish for 
$k \geq 2$. As a result, there is an exact triangle 
\[ 
{\mathcal Tor}_1^{S}(\CO_\Sigma, G)[1] \to L\iota^*G\to {\mathcal Tor}_0^{S}(\CO_\Sigma, G)
\]
in the derived category of $\Sigma$. For $G=\iota_*F_1$, one easily obtains 
\[
{\mathcal Tor}_1^{S}(\CO_\Sigma, \iota_*F_1)\simeq F_1(-\Sigma), \qquad {\mathcal Tor}_0^{S}(\CO_\Sigma, \iota_*F_1)\simeq F_1.
\]
Furthermore $\Omega_\Sigma\simeq \CO_\Sigma(\Sigma)$ is locally free. Thus, there is an exact triangle 
\[
\mathcal{RH}om_\Sigma(F_2,F_1)
\to \mathcal{RH}om_\Sigma(F_2, \iota^!\iota_*F_1)
\to \mathcal{RH}om_\Sigma(F_2,F_1(\Sigma))[-1].
\]
Using ${\rm RHom}=R\Gamma\circ\mathcal{RH}om$, 
the above lemma follows immediately from the 
this exact triangle. 

\hfill $\Box$

\begin{rema}\label{extremark}
$(i)$  In particular, there is an injection 
${\rm Ext}^1_\Sigma(F_2,F_1) \hookrightarrow 
{\rm Ext}^1_S(F_2,F_1)$. Naturally, given an extension 
\[
0\to F_1\to F \to F_2 \to 0
\]
parameterized by $\epsilon \in {\rm Ext}^1_S(F_2,F_1)$, the sheaf $F$ 
is scheme theoretically supported on $\Sigma$ if and only if $\epsilon \in {\rm Ext}^1_\Sigma(F_2,F_1)$. 

$(ii)$ Since $\Sigma$ is singular, the higher extension groups $Ext^k_\Sigma(F_2,F_1)$, $k \geq 2$ may be nonzero. However by Serre duality they vanish if at least one of $F_1,F_2$ is locally free on $\Sigma$. In that case, one obtains a short exact sequence 
$$ 
0  \to {\rm Ext}^1_\Sigma (F_2,F_1) \to 
{\rm Ext}^1_S(F_2,F_1) \to  {\rm Ext}^0_\Sigma(F_2, F_1(\Sigma))\to 0.
$$
\end{rema}

\subsection{A stability criterion for extensions}
Let 
\be\label{eq:shextA}
0\to F_1 \to F \to F_2\to 0
\ee
be an extension as in Lemma \ref{extlemma} with $F_1,F_2$ rank one torsion free sheaves  on $\Sigma$, 
and $\chi(F)=1$.  Let $\epsilon \in {\rm Ext}^1_S(F_2,F_1)$ be the corresponding extension class. Let $\iota:F_2'\hookrightarrow F_2$ be a subsheaf and let 
$\iota^*:{\rm Ext}^1_S(F_2,F_1) \to {\rm Ext}^1_S(F_2',F_1)$ be the induced map.
Then the following stability criterion is straightforward.

\begin{lemm}\label{stabext}
The sheaf $F$ in \eqref{eq:shextA} is stable if and only if 

$(i)$ $\chi(F_1) \leq 0$, $\chi(F_2)\geq 1$, and 

$(ii)$ for any nonzero subsheaf $\iota:F_2'\hookrightarrow F_2$ such that $\iota^*(\epsilon)=0$ one has $\chi(F_2')\leq 0$. 
\end{lemm} 

For applications, the condition 
$\iota^*(\epsilon)=0$ can be made more explicit 
if Serre duality holds for 
the pairs $(F_2,F_1)$ and $(F_2',F_1)$. 
Namely suppose $\epsilon \in {\rm Ext}^1_\Sigma
(F_2,F_1)$ and Serre duality holds for 
the pairs $(F_2,F_1)$ and $(F_2',F_1)$. 
Then there is a commutative diagram
\[
\xymatrix{
{\rm Ext}^0_\Sigma(F_1, F_2'\otimes\omega_\Sigma) \ar[rr]^-{\iota^\vee} \ar[dr]^-{\iota^*(\epsilon)}& & 
{\rm Ext}^0_\Sigma(F_1, F_2\otimes\omega_\Sigma)\ar[dl]^-{\epsilon}\\
& \IC & \\}
\]
For simplicity, $(\iota^*)^\vee$ has been denoted by 
$\iota^\vee$. 
This yields the relation $\iota^*(\epsilon) = \epsilon\circ \iota^\vee$. In particular, $\iota^*(\epsilon)=0$ if and only if $\epsilon \circ \iota^\vee=0$. 

Furthermore, suppose $\Sigma\subset S$ is a singular 
curve as in \S\ref{locsect}  preserved by the torus action. 
Let $F$ be a pure dimension one sheaf on $S$ with set-theoretic support on $\Sigma$ which is fixed by the torus action up to gauge transformations. This means 
that for each element $t \in \IC^\times$ there is an isomorphism of sheaves $\xi(t): F {\buildrel \sim \over \longto} \rho_t^*F$ where $\rho_t : S \to S$ is the 
corresponding automorphism of $S$. The assignment 
$t\mapsto \xi(t)$ must satisfy the cocycle condition 
\[
\xi(t't) = \left(\rho_{t'}^*\xi(t)\right) \circ \xi(t'). 
\]
Then, using basic properties of the Harder-Narasimhan filtration  one has:
\begin{lemm}\label{invHN}
Let $=F_0 \subset F_1 \subset F_2 \subset \cdots \subset F_h=F$ be the Harder-Narasimhan filtration 
of $F$. Then for any $t\in \IC^\times$, the morphism $\xi(t)$ maps each $F_i$ isomorphically to 
$\rho_t^*F_i$ for all $1\leq i \leq h$. 
\end{lemm} 

As a useful consequence of Lemma \ref{invHN}, note that 
in Lemma \ref{stabext} it suffices to test stability for 
subobjects $F_2'\subset F_2$ which are preserved by the torus action. 

\subsection{Extension groups on singular genus two curves}\label{extgenustwo} 

Let $k=3$ and $\ell=2$. Then the curve $\Sigma$ has a 
local singularity $v^2 = w^5$. In this case the normalization map reads locally $v=t^5$, $w=t^2$. Hence  
\[
R = \langle 1, t^2, t^4, t^5,t^6, \ldots \rangle 
\]
as a vector space over $\IC$, and $\wR/R \simeq \langle t,t^3\rangle$. Therefore, $\CT_\Sigma$ is isomorphic to $T \CO_Z$ as an equivariant sheaf, where 
$Z\subset \Sigma$ is the closed subscheme of $\Sigma$ defined by
the ideal $(v,w^2)$. Recall that $\CE$ denotes the rank one torsion free sheaf on $\Sigma$ with local sections 
\[
\Gamma(\CU, \CE) =\IC[s], \qquad \Gamma(\CV,\CE) = 
\langle 1,t^2,t^3, \ldots \rangle.
\]
In this subsection, we shall compute some extension groups on $\Sigma$ needed in the classification of stable fixed points. 

Note that $\wR$ has the following infinite equivariant free resolution 
\[
\cdots \to F_{k+1}  \xrightarrow{ f_{k+1} } F_{k} \to \cdots 
\]
with $k\geq 0$, 
where 
\[
F_k= T^{5k} R \oplus T^{5k+1} R 
\]
for $k \geq 0$. The differentials are 
\[
f_{k+1} = \left(\begin{array}{cc} -t^5 & -t^6 \\ t^4 & t^5
\end{array}\right)
\]
for $k\geq 0$ and the projection $F_0=R+TR \to \wR$ is $f_0=(1, t)$.

Let $E = \langle 1, t^3, t^4, \ldots \rangle$. Then $E$ has a similar equivariant free resolution 
\[
\cdots \to G_{k+1} \xrightarrow{ g_{k+1}} G_{k} \to \cdots 
\]
with $k\geq 0$, where $$G_k = T^{5k}R\oplus T^{5k+3}R$$ and 
\[
g_{k+1} = \left(\begin{array}{cc} -t^5 & -t^8 \\ t^2 & t^5
\end{array}\right)
\]
for $k\geq 0$. Note that the projection $G_0\to E$ is
$g_0=(1,t^3)$.

The free resolutions above tell us that
$$\begin{array}{lll}
 {\rm Hom}_R(\wR, \wR) \simeq \wR, &  
{\rm Ext}^k_R(\wR, \wR) \simeq T^{-5k}\langle 1,t,t^2,t^3\rangle, \
&k\geq 1~, \\
 {\rm Hom}_R(E, E) \simeq E, & 
{\rm Ext}^k_R(E, E) \simeq T^{-5k}\langle 1,t^3\rangle, \
&k\geq 1~, \\
 {\rm Hom}_R(\wR, E) \simeq t^2\wR, & 
{\rm Ext}^k_R(\wR, E) \simeq T^{-5k}\langle t^2,t^3\rangle, \
&k\geq 1~, \\
 {\rm Hom}_R(E, \wR) \simeq \wR, & 
{\rm Ext}^k_R(E, \wR) \simeq T^{-5k}\langle 1,t\rangle, \
&k\geq 1~. 
\end{array}$$
With the above local constructions, one obtains the following locally free resolutions 
\begin{align}
&\cdots \to \CF_{k+1} \xrightarrow{\phi_{k+1} } \CF_{k} \to \cdots \to \CF_0\xrightarrow{ \phi_0} \wCO_\Sigma~,\cr
&\cdots \to \CG_{k+1} \xrightarrow{ \psi_{k+1}} \CG_{k} \to \cdots \to \CG_0\xrightarrow{ \psi_0} \CE~,\nonumber
\end{align}
where 
\begin{align}
\CF_k &= T^{5k} \CO_\Sigma(-5k)\oplus T^{5k+1}\CO_\Sigma(-5k-1)~, \cr
\CG_k &= T^{5k} \CO_\Sigma(-5k)\oplus T^{5k+3}\CO_\Sigma(-5k-3)~. \nonumber
\end{align}
The differentials $\phi_k, \psi_k$ are naturally determined by $f_k, g_k$. This yields the following local ext sheaves 
$$\begin{array}{lll}
 {\mathcal Ext}^0_\Sigma(\wCO_\Sigma, \wCO_\Sigma) \simeq \wCO_\Sigma,   &
{\mathcal Ext}^k_\Sigma(\wCO_\Sigma, \wCO_\Sigma) \simeq T^{-5k}\langle 1,t,t^2,t^3\rangle^\sim, \
& k\geq 1~, \\
 {\mathcal Ext}^0_\Sigma(\CE, \CE) \simeq \CE,  &
{\mathcal Ext}^k_\Sigma(\CE, \CE) \simeq T^{-5k}\langle 1,t^3\rangle^\sim, \ &
k\geq 1~, \\
{\mathcal Ext}^0_\Sigma(\wCO_\Sigma, \CE) 
 \simeq \wCO_\Sigma(-2\delta),  &
{\mathcal Ext}^k_\Sigma(\wCO_\Sigma, \CE) 
\simeq T^{-5k}\langle t^2,t^3\rangle^\sim, \
&k\geq 1 ~,\\
 {\mathcal Ext}^0_\Sigma(\CE, \wCO_\Sigma) \simeq \wCO_\Sigma,   &
{\mathcal Ext}^k_\Sigma(\CE, \wCO_\Sigma) \simeq T^{-5k}\langle 1,t\rangle^\sim, \
&k\geq 1 ~.
\end{array}$$
In the above formulas $\delta$ is the divisor $t^2=0$
on $\Sigma$ and $V^\sim$ denotes the coherent sheaf associated to a length zero $R$-module $V$. 

Using the local to global spectral sequence one then 
finds 
\begin{align}\label{Ext-results}
 {\rm Ext}^0_\Sigma(\wCO, \wCO(dq)) &= 
\left\{\begin{array}{ll} 1+T+\cdots+T^{d}, & {\rm for}\ d\geq 0,\\
0, & {\rm for}\ d\leq -1.\\
\end{array}\right.
\cr 
 {\rm Ext}^1_\Sigma(\wCO, \wCO(dq)) &= 
\left\{\begin{array}{ll} T^{-5}(1+T+T^2+T^3), & {\rm for}\ d\geq -1,\\
T^{-5}(1+T+T^2+T^3)+ (T^{-1}+\cdots + T^{d+1}), & {\rm for}\ d\leq -2.\\
\end{array}\right.
\cr
{\rm Ext}^k_\Sigma(\wCO, \wCO(dq)) &= T^{-5k}(1+T+T^2+T^3), \  k\geq 2,\ {\rm for\ all}\ d\in \IZ~.
\cr  
 {\rm Ext}^0_\Sigma(\CE, \CE(dq)) &= 
\left\{\begin{array}{ll} 1+T^2+\cdots+T^{d}, & {\rm for}\ d\geq 2,\\
1, & {\rm for}\ 0\leq d\leq 1\\ 
0, & {\rm for}\ d\leq -1.\\
\end{array}\right.
\cr
 {\rm Ext}^1_\Sigma(\CE, \CE(dq)) &= 
\left\{\begin{array}{ll}
T^{-5}(1+T^3), & {\rm for}\ d\geq 1\\
T^{-5}(1+T^3)+T, & {\rm for}\  -1\leq d \leq 0\\
T^{-5}(1+T^3)+T+ T^{-1}+\cdots + T^{d+1}, & {\rm for}\ d\leq -2.\\
\end{array}\right.
\cr
{\rm Ext}^k_\Sigma(\CE, \CE(dq)) &= T^{-5k}(1+T^3),\  k\geq 2,\ {\rm for\ all}\ d\in \IZ~.
\cr  
 {\rm Ext}^0_\Sigma(\wCO_\Sigma, \CE(dq)) &= 
\left\{\begin{array}{ll} T^2+T^3+\cdots+T^d, & {\rm for}\ d\geq 2,\\
0, & {\rm for}\ d\leq 1.\\
\end{array}\right.
\cr  
 {\rm Ext}^1_\Sigma(\wCO_\Sigma, \CE(dq)) &= 
\left\{\begin{array}{ll} T^{-5}(T^2+T^3), & {\rm for}\ d\geq 1,\\
T^{-5}(T^2+T^3)+T(1+\cdots+T^{d}), & {\rm for}\ d\leq 0.\\
\end{array}\right.
\cr
{\rm Ext}^k_\Sigma(\wCO_\Sigma, \CE(dq)) &= 
T^{-5k}(T^2+T^3),\  k\geq 2,\ {\rm for\ all}\ d\in \IZ~.
\cr  
 {\rm Ext}^0_\Sigma(\CE, \wCO_\Sigma(dq)) &= 
\left\{\begin{array}{ll} 1+T+\cdots+T^{d}, & {\rm for}\ d\geq 0,\\
0, & {\rm for}\ d\leq -1.\\
\end{array}\right.
\cr  
 {\rm Ext}^1_\Sigma(\CE, \wCO_\Sigma(dq)) &= 
\left\{\begin{array}{ll} 
T^{-5}(1+T), & {\rm for}\ d\geq -1\\
T^{-5}(1+T) + T^{-1}+\cdots + T^{d+1}, & 
{\rm for}\ d\leq -2.\\
\end{array}\right.
\cr
{\rm Ext}^k_\Sigma(\CE, \wCO_\Sigma(dq)) &= 
T^{-5k}(1+T),\  k\geq 2,\ {\rm for\ all}\ d\in \IZ~.
\end{align}
 
\section{Examples}\label{examples}

\subsection{Refined genus zero examples}\label{refex}  

In all examples below, we set $g=0$ and $m=1$. 

$a)$ $n\geq 3$, $\ell=2$, $r=1$
\[
WP_{n,2,1}(u,v)  = \sum_{i=0}^{n-3} (uv)^{2i}.
\]

$b)$ $n\geq 3$, $\ell=3$, $r=1$
\[
WP_{n,3,1}(u,v)  = \sum_{j=0}^{n-3} \sum_{i=0}^{3j+1} 
u^{3k-3j+2i-3}v^{4k-4j+2i-4}.
\]

$c)$ $n=4$, $\ell=2$, $r=2$ 
\[
WP_{4,2,2}(u,v) = u^2v^2+1
\]

$d)$ $n=4$, $\ell=2$, $r=3$ 
\[
WP_{4,2,3}(u,v) = u^2v^2+1
\]

$e)$ $n=5$, $\ell=2$, $r=2$ 
\[
\bal 
WP_{5,2,2}(u,v) =\ & 
{u}^{10}{v}^{10}+{u}^{8}{v}^{10}+{u}^{8}{v}^{8}+{u}^{7}{v}^{8}+2\,{u}^{6}{v}^{8}+{u}^{6}{v}^{6}+{u}^{5}{v}^{6}+2\,{u}^{4}{v}^{6}\\
& +{u}^{4}{v}^{4}+{u}^{3}{v}^{4}+{u}^{2}{v}^{4}
+{u}^{2}{v}^{2}+1
\eal 
\]

$f)$ $n=6$, $\ell=2$, $r=2$ 
\[
\bal
WP_{6,2,2}(u,v) =\ & 
{u}^{18}{v}^{18}+{u}^{16}{v}^{18}+{u}^{16}{v}^{16}+{u}^{14}{v}^{18}+{u}^{15}{v}^{16}+2\,{u}^{14}{v}^{16}+{u}^{13}{v}^{16}\\
& +{u}^{14}{v}^{14}+2\,{u}^{12}{v}^{16}+{u}^{13}{v}^{14}+3\,{u}^{12}{v}^{14}+2\,{u}^{11}{v}^{14}+{u}^{12}{v}^{12}+3\,{u}^{10}{v}^{14}\\
& +{u}^{11}{v}^{12}+3\,{u}^{10}{v}^{12}+3\,{u}^{9}{v}^{12}+{u}^{10}{v}^{10}+3\,{u}^{8}{v}^{12}+{u}^{9}{v}^{10}+3\,{u}^{8}{v}^{10}+2\,{u}^{7}{v}^{10}\\
&+{u}^{8}{v}^{8}+2\,{u}^{6}{v}^{10}+{u}^{7}{v}^{8}+3\,{u}^{6}{v}^{8}+{u}^{5}{v}^{8}+{u}^{6}{v}^{6}+{u}^{4}{v}^{8}+{u}^{5}{v}^{6}+2\,{u}^{4}{v}^{6}\\
& +{u}^{4}{v}^{4}+{u}^{3}{v}^{4}+{u}^{2}{v}^{4}
+{u}^{2}{v}^{2}+1
\eal 
\]

$g)$ $n=4$, $\ell=3$, $r=2$ 
\[
\bal 
WP_{4,3,2}(u,v) =\ & 
{u}^{26}{v}^{26}+{u}^{24}{v}^{26}+{u}^{24}{v}^{24}+{u}^{22}{v}^{26}+{u}^{23}{v}^{24}+2\,{u}^{22}{v}^{24}+{u}^{20}{v}^{26}\\
&+2\,{u}^{21}{v}^{24}+{u}^{22}{v}^{22}+3\,{u}^{20}{v}^{24}+{u}^{21}{v}^{22}+2\,{u}^{19}{v}^{24}+3\,{u}^{20}{v}^{22}+3\,{u}^{18}{v}^{24}\\
&+3\,{u}^{19}{v}^{22}+{u}^{17}{v}^{24}+{u}^{20}{v}^{20}+6\,{u}^{18}{v}^{22}+{u}^{19}{v}^{20}+5\,{u}^{17}{v}^{22}+3\,{u}^{18}{v}^{20}\\
&+7\,{u}^{16}{v}^{22}+4\,{u}^{17}{v}^{20}+2\,{u}^{15}{v}^{22}+{u}^{18}{v}^{18}+7\,{u}^{16}{v}^{20}+{u}^{17}{v}^{18}+8\,{u}^{15}{v}^{20}\\
&+3\,{u}^{16}{v}^{18}+10\,{u}^{14}{v}^{20}+4\,{u}^{15}{v}^{18}+3\,{u}^{13}{v}^{20}+{u}^{16}{v}^{16}+8\,{u}^{14}{v}^{18}+{u}^{15}{v}^{16}\\
&+8\,{u}^{13}{v}^{18}+3\,{u}^{14}{v}^{16}+10\,{u}^{12}{v}^{18}+4\,{u}^{13}{v}^{16}+2\,{u}^{11}{v}^{18}+{u}^{14}{v}^{14}+8\,{u}^{12}{v}^{16}\\
&+{u}^{13}{v}^{14}+8\,{u}^{11}{v}^{16}+3\,{u}^{12}{v}^{14}+7\,{u}^{10}{v}^{16}+4\,{u}^{11}{v}^{14}+{u}^{9}{v}^{16}+{u}^{12}{v}^{12}\\
&+7\,{u}^{10}{v}^{14}+{u}^{11}{v}^{12}+5\,{u}^{9}{v}^{14}+3\,{u}^{10}{v}^{12}+3\,{u}^{8}{v}^{14}+4\,{u}^{9}{v}^{12}+{u}^{10}{v}^{10}+6\,{u}^{8}{v}^{12}\\
&+{u}^{9}{v}^{10}+2\,{u}^{7}{v}^{12}+3\,{u}^{8}{v}^{10}+{u}^{6}{v}^{12}+3\,{u}^{7}{v}^{10}+{u}^{8}{v}^{8}+3\,{u}^{6}{v}^{10}+{u}^{7}{v}^{8}+3\,{u}^{6}{v}^{8}\\ 
& +2\,{u}^{5}{v}^{8}+{u}^{6}{v}^{6}+{u}^{4}{v}^{8}+{u}^{5}{v}^{6}+2\,{u}^{4}{v}^{6}+{u}^{4}{v}^{4}+{u}^{3}{v}^{4}+{u}^{2}{v}^{4}+{u}^{2}{v}^{2}+1
\eal
\]

$h)$ $n=5$, $\ell=3$, $r=2$ 
\begin{align}
WP_{5,3,2}(u,v) =\ & 
{v}^{50}{u}^{50}+{v}^{50}{u}^{48}+{v}^{50}{u}^{46}+{v}^{50}{u}^{44}+{v}^{50}{u}^{42}+{v}^{50}{u}^{40}+{v}^{50}{u}^{38}\cr &
\mbox{}+{v}^{48}{u}^{48}+{v}^{48}{u}^{47}+2\,{v}^{48}{u}^{46}+2\,{v}^{48}{u}^{45}+3\,{v}^{48}{u}^{44}+2\,{v}^{48}{u}^{43}+3\,{v}^{48}{u}^{42}\cr &
\mbox{}+2\,{v}^{48}{u}^{41}+3\,{v}^{48}{u}^{40}+2\,{v}^{48}{u}^{39}+3\,{v}^{48}{u}^{38}+2\,{v}^{48}{u}^{37}+3\,{v}^{48}{u}^{36}+{v}^{48}{u}^{35}\cr &
\mbox{}+{v}^{46}{u}^{46}+{v}^{46}{u}^{45}+3\,{v}^{46}{u}^{44}+3\,{v}^{46}{u}^{43}+6\,{v}^{46}{u}^{42}+5\,{v}^{46}{u}^{41}+8\,{v}^{46}{u}^{40}\cr &
\mbox{}+6\,{v}^{46}{u}^{39}+9\,{v}^{46}{u}^{38}+6\,{v}^{46}{u}^{37}+9\,{v}^{46}{u}^{36}+6\,{v}^{46}{u}^{35}+8\,{v}^{46}{u}^{34}+3\,{v}^{46}{u}^{33}\cr &
\mbox{}+{v}^{46}{u}^{32}+{v}^{44}{u}^{44}+{v}^{44}{u}^{43}+3\,{v}^{44}{u}^{42}+4\,{v}^{44}{u}^{41}+7\,{v}^{44}{u}^{40}+8\,{v}^{44}{u}^{39}\cr &
\mbox{}+12\,{v}^{44}{u}^{38}+12\,{v}^{44}{u}^{37}+16\,{v}^{44}{u}^{36}+15\,{v}^{44}{u}^{35}+18\,{v}^{44}{u}^{34}+15\,{v}^{44}{u}^{33}\cr 
& \mbox{}+16\,{v}^{44}{u}^{32}+8\,{v}^{44}{u}^{31}+3\,{v}^{44}{u}^{30}+{v}^{42}{u}^{42}+{v}^{42}{u}^{41}+3\,{v}^{42}{u}^{40}+4\,{v}^{42}{u}^{39}\cr &
\mbox{}+8\,{v}^{42}{u}^{38}+9\,{v}^{42}{u}^{37}+15\,{v}^{42}{u}^{36}+16\,{v}^{42}{u}^{35}+23\,{v}^{42}{u}^{34}+23\,{v}^{42}{u}^{33}\cr &
\mbox{}+30\,{v}^{42}{u}^{32}+25\,{v}^{42}{u}^{31}+27\,{v}^{42}{u}^{30}+13\,{v}^{42}{u}^{29}+5\,{v}^{42}{u}^{28}+{v}^{40}{u}^{40}\cr &
\mbox{}+{v}^{40}{u}^{39}+3\,{v}^{40}{u}^{38}+4\,{v}^{40}{u}^{37}+8\,{v}^{40}{u}^{36}+10\,{v}^{40}{u}^{35}+16\,{v}^{40}{u}^{34}\cr &
\mbox{}+19\,{v}^{40}{u}^{33}+27\,{v}^{40}{u}^{32}+30\,{v}^{40}{u}^{31}+39\,{v}^{40}{u}^{30}+37\,{v}^{40}{u}^{29}+37\,{v}^{40}{u}^{28}\cr &
\mbox{}+19\,{v}^{40}{u}^{27}+6\,{v}^{40}{u}^{26}+{v}^{40}{u}^{25}+{v}^{38}{u}^{38}+{v}^{38}{u}^{37}+3\,{v}^{38}{u}^{36}+4\,{v}^{38}{u}^{35}\cr & 
\mbox{}+8\,{v}^{38}{u}^{34}+10\,{v}^{38}{u}^{33}+17\,{v}^{38}{u}^{32}+20\,{v}^{38}{u}^{31}+30\,{v}^{38}{u}^{30}+34\,{v}^{38}{u}^{29}\cr & 
\mbox{}+45\,{v}^{38}{u}^{28}+43\,{v}^{38}{u}^{27}+44\,{v}^{38}{u}^{26}+20\,{v}^{38}{u}^{25}+6\,{v}^{38}{u}^{24}+{v}^{36}{u}^{36}\cr 
& 
\mbox{}+{v}^{36}{u}^{35}+3\,{v}^{36}{u}^{34}+4\,{v}^{36}{u}^{33}+8\,{v}^{36}{u}^{32}+10\,{v}^{36}{u}^{31}+17\,{v}^{36}{u}^{30}\cr & 
\mbox{}+21\,{v}^{36}{u}^{29}+31\,{v}^{36}{u}^{28}+37\,{v}^{36}{u}^{27}+48\,{v}^{36}{u}^{26}+46\,{v}^{36}{u}^{25}+44\,{v}^{36}{u}^{24}\cr & 
\mbox{}+19\,{v}^{36}{u}^{23}+5\,{v}^{36}{u}^{22}+{v}^{34}{u}^{34}+{v}^{34}{u}^{33}+3\,{v}^{34}{u}^{32}+4\,{v}^{34}{u}^{31}+8\,{v}^{34}{u}^{30}\cr & 
\mbox{}+10\,{v}^{34}{u}^{29}+17\,{v}^{34}{u}^{28}+21\,{v}^{34}{u}^{27}+32\,{v}^{34}{u}^{26}+37\,{v}^{34}{u}^{25}+48\,{v}^{34}{u}^{24}\cr & 
\mbox{}+43\,{v}^{34}{u}^{23}+37\,{v}^{34}{u}^{22}+13\,{v}^{34}{u}^{21}+3\,{v}^{34}{u}^{20}+{v}^{32}{u}^{32}+{v}^{32}{u}^{31}\cr & 
\mbox{}+3\,{v}^{32}{u}^{30}+4\,{v}^{32}{u}^{29}+8\,{v}^{32}{u}^{28}+10\,{v}^{32}{u}^{27}+17\,{v}^{32}{u}^{26}+21\,{v}^{32}{u}^{25}\cr & 
\mbox{}+32\,{v}^{32}{u}^{24}+37\,{v}^{32}{u}^{23}+45\,{v}^{32}{u}^{22}+37\,{v}^{32}{u}^{21}+27\,{v}^{32}{u}^{20}+8\,{v}^{32}{u}^{19}\cr & 
\mbox{}+{v}^{32}{u}^{18}+{v}^{30}{u}^{30}+{v}^{30}{u}^{29}+3\,{v}^{30}{u}^{28}+4\,{v}^{30}{u}^{27}+8\,{v}^{30}{u}^{26}+10\,{v}^{30}{u}^{25}\cr &
\mbox{}+17\,{v}^{30}{u}^{24}+21\,{v}^{30}{u}^{23}+31\,{v}^{30}{u}^{22}+34\,{v}^{30}{u}^{21}+39\,{v}^{30}{u}^{20}+25\,{v}^{30}{u}^{19}\cr &
\mbox{}+16\,{v}^{30}{u}^{18}+3\,{v}^{30}{u}^{17}+{v}^{28}{u}^{28}+{v}^{28}{u}^{27}+3\,{v}^{28}{u}^{26}+4\,{v}^{28}{u}^{25}+8\,{v}^{28}{u}^{24}\cr &
\mbox{}+10\,{v}^{28}{u}^{23}+17\,{v}^{28}{u}^{22}+21\,{v}^{28}{u}^{21}+30\,{v}^{28}{u}^{20}+30\,{v}^{28}{u}^{19}+30\,{v}^{28}{u}^{18}\cr & 
\mbox{}+15\,{v}^{28}{u}^{17}+8\,{v}^{28}{u}^{16}+{v}^{28}{u}^{15}+{v}^{26}{u}^{26}+{v}^{26}{u}^{25}+3\,{v}^{26}{u}^{24}+4\,{v}^{26}{u}^{23}\cr &
\mbox{}+8\,{v}^{26}{u}^{22}+10\,{v}^{26}{u}^{21}+17\,{v}^{26}{u}^{20}+20\,{v}^{26}{u}^{19}+27\,{v}^{26}{u}^{18}+23\,{v}^{26}{u}^{17}\cr &
\mbox{}+18\,{v}^{26}{u}^{16}+6\,{v}^{26}{u}^{15}+3\,{v}^{26}{u}^{14}+{v}^{24}{u}^{24}+{v}^{24}{u}^{23}+3\,{v}^{24}{u}^{22}+4\,{v}^{24}{u}^{21}\cr &
\mbox{}+8\,{v}^{24}{u}^{20}+10\,{v}^{24}{u}^{19}+17\,{v}^{24}{u}^{18}+19\,{v}^{24}{u}^{17}+23\,{v}^{24}{u}^{16}+15\,{v}^{24}{u}^{15}\cr &
\mbox{}+9\,{v}^{24}{u}^{14}+2\,{v}^{24}{u}^{13}+{v}^{24}{u}^{12}+{v}^{22}{u}^{22}+{v}^{22}{u}^{21}+3\,{v}^{22}{u}^{20}+4\,{v}^{22}{u}^{19}\cr & 
\mbox{}+8\,{v}^{22}{u}^{18}+10\,{v}^{22}{u}^{17}+16\,{v}^{22}{u}^{16}+16\,{v}^{22}{u}^{15}+16\,{v}^{22}{u}^{14}+6\,{v}^{22}{u}^{13}\cr & 
\mbox{}+3\,{v}^{22}{u}^{12}+{v}^{20}{u}^{20}+{v}^{20}{u}^{19}+3\,{v}^{20}{u}^{18}+4\,{v}^{20}{u}^{17}+8\,{v}^{20}{u}^{16}+10\,{v}^{20}{u}^{15}\cr & 
\mbox{}+15\,{v}^{20}{u}^{14}+12\,{v}^{20}{u}^{13}+9\,{v}^{20}{u}^{12}+2\,{v}^{20}{u}^{11}+{v}^{20}{u}^{10}+{v}^{18}{u}^{18}+{v}^{18}{u}^{17}\cr &
\mbox{}+3\,{v}^{18}{u}^{16}+4\,{v}^{18}{u}^{15}+8\,{v}^{18}{u}^{14}+9\,{v}^{18}{u}^{13}+12\,{v}^{18}{u}^{12}+6\,{v}^{18}{u}^{11}\cr & 
\mbox{}+3\,{v}^{18}{u}^{10}+{v}^{16}{u}^{16}+{v}^{16}{u}^{15}+3\,{v}^{16}{u}^{14}+4\,{v}^{16}{u}^{13}+8\,{v}^{16}{u}^{12}+8\,{v}^{16}{u}^{11}\cr &
\mbox{}+8\,{v}^{16}{u}^{10}+2\,{v}^{16}{u}^{9}+{v}^{16}{u}^{8}+{v}^{14}{u}^{14}+{v}^{14}{u}^{13}+3\,{v}^{14}{u}^{12}+4\,{v}^{14}{u}^{11}\cr & 
\mbox{}+7\,{v}^{14}{u}^{10}+5\,{v}^{14}{u}^{9}+3\,{v}^{14}{u}^{8}+{v}^{12}{u}^{12}+{v}^{12}{u}^{11}+3\,{v}^{12}{u}^{10}+4\,{v}^{12}{u}^{9}\cr & 
\mbox{}+6\,{v}^{12}{u}^{8}+2\,{v}^{12}{u}^{7}+{v}^{12}{u}^{6}+{v}^{10}{u}^{10}+{v}^{10}{u}^{9}+3\,{v}^{10}{u}^{8}+3\,{v}^{10}{u}^{7}+3\,{v}^{10}{u}^{6}\cr & 
\mbox{}+{v}^{8}{u}^{8}+{v}^{8}{u}^{7}+3\,{v}^{8}{u}^{6}+2\,{v}^{8}{u}^{5}+{v}^{8}{u}^{4}+{v}^{6}{u}^{6}+{v}^{6}{u}^{5}+2\,{v}^{6}{u}^{4}+{v}^{4}{u}^{4}\cr
& \mbox{}+{v}^{4}{u}^{3}+{v}^{4}{u}^{2}+{v}^{2}{u}^{2}
+1\nonumber
\end{align}

\subsection{Unrefined higher genus examples}\label{unrefex} 

 Some reduced $E$-polynomials defined as 
${\widetilde E}_{n,\ell,r}(u) = (1-u)^{-2g} E_{n,\ell,r}(u)$ are listed below. 

$a)$ $g=2$, $m=1$, $n=4$, $\ell=2$, $r=2$ 
\[
\bal 
{\widetilde E}_{4,2,2}(u) = \ & {u}^{62}-2\,{u}^{60}-3\,{u}^{59}-2\,{u}^{58}+6\,{u}^{57}+9\,{u}^{56}+6\,{u}^{55}-6\,{u}^{54}-19\,{u}^{53}-12\,{u}^{52}+2\,{u}^{51}\\
& +20\,{u}^{50}
+20\,{u}^{49}+2\,{u}^{48}-12\,{u}^{47}-19\,{u}^{46}-6\,{u}^{45}+3\,{u}^{44}-{u}^{43}+16\,{u}^{42}+30\,{u}^{41}\\
& -11\,{u}^{40}-44\,{u}^{39}
-17\,{u}^{38}+43\,{u}^{37}+33\,{u}^{36}-26\,{u}^{35}-4\,{u}^{34}-42\,{u}^{33}-11\,{u}^{32}\\
&+92\,{u}^{31}
 -11\,{u}^{30}-42\,{u}^{29}
-4\,{u}^{28}-26\,{u}^{27}+33\,{u}^{26}+43\,{u}^{25}-17\,{u}^{24}-44\,{u}^{23}\\ 
& -11\,{u}^{22}
+30\,{u}^{21}+16\,{u}^{20}-{u}^{19}+3\,{u}^{18}
-6\,{u}^{17}-19\,{u}^{16}-12\,{u}^{15}+2\,{u}^{14}+20\,{u}^{13}\\ 
& +20\,{u}^{12}+2\,{u}^{11}
-12\,{u}^{10}-19\,{u}^{9}-6\,{u}^{8}+6\,{u}^{7}+9\,{u}^{6}
+6\,{u}^{5}-2\,{u}^{4}-3\,{u}^{3}-2\,{u}^{2}+1
\eal 
\]

$b)$ $g=1$, $m=3$, $n=4$, $\ell=2$, $r=2$ 
\[
\bal 
{\widetilde E}_{4,2,2}(u) = \ & {u}^{96}-3\,{u}^{93}-3\,{u}^{92}+3\,{u}^{90}+9\,{u}^{89}+3\,{u}^{88}-{u}^{87}-9\,{u}^{86}-9\,{u}^{85}-{u}^{84}+3\,{u}^{83}+9\,{u}^{82}\\
& +3\,{u}^{81}
 -3\,{u}^{79}-3\,{u}^{78}+{u}^{75}-3\,{u}^{69}-3\,{u}^{68}+9\,{u}^{67}+12\,{u}^{66}-9\,{u}^{65}-18\,{u}^{64}+2\,{u}^{63}\\
& +12\,{u}^{62}+6\,{u}^{61}
-3\,{u}^{60}-12\,{u}^{59}-9\,{u}^{58}+9\,{u}^{57}+30\,{u}^{56}+3\,{u}^{55}-26\,{u}^{54}-21\,{u}^{53}\\
&-12\,{u}^{52}+37\,{u}^{51}+27\,{u}^{50}
-21\,{u}^{49}-20\,{u}^{48}-21\,{u}^{47}+27\,{u}^{46}+37\,{u}^{45}-12\,{u}^{44}\\
&-21\,{u}^{43}-26\,{u}^{42}+3\,{u}^{41}+30\,{u}^{40}
+9\,{u}^{39}-9\,{u}^{38}-12\,{u}^{37}-3\,{u}^{36}+6\,{u}^{35}+12\,{u}^{34}+2\,{u}^{33}\\
& -18\,{u}^{32}-9\,{u}^{31}+12\,{u}^{30}+9\,{u}^{29}-3\,{u}^{28}
-3\,{u}^{27}+{u}^{21}-3\,{u}^{18}-3\,{u}^{17}+3\,{u}^{15}+9\,{u}^{14}\\ & +3\,{u}^{13}-{u}^{12}
-9\,{u}^{11}-9\,{u}^{10}-{u}^{9}+3\,{u}^{8}+9\,{u}^{7}+3\,{u}^{6}
-3\,{u}^{4}-3\,{u}^{3}+1
\eal 
\]

$c)$ $g=1$, $m=3$, $n=5$, $\ell=2$, $r=2$ 
\[
\bal 
{\widetilde E}_{5,2,2}(u) = \ & 
{u}^{120}-3\,{u}^{117}-3\,{u}^{116}+3\,{u}^{114}+9\,{u}^{113}+3\,{u}^{112}-{u}^{111}-9\,{u}^{110}-9\,{u}^{109}-{u}^{108}+3\,{u}^{107}\\
&+9\,{u}^{106}+3\,{u}^{105}-3\,{u}^{103}-3\,{u}^{102}+{u}^{99}-3\,{u}^{87}-3\,{u}^{86}+9\,{u}^{85}+12\,{u}^{84}-9\,{u}^{83}-18\,{u}^{82}\\
&+2\,{u}^{81}+12\,{u}^{80}+3\,{u}^{79}-3\,{u}^{78}-3\,{u}^{77}+4\,{u}^{75}-9\,{u}^{73}-9\,{u}^{72}+9\,{u}^{71}+27\,{u}^{70}+5\,{u}^{69}\\
& -24\,{u}^{68}
-30\,{u}^{67}+7\,{u}^{66}+33\,{u}^{65}-9\,{u}^{64}+{u}^{63}+6\,{u}^{62}-12\,{u}^{61}+4\,{u}^{60}-12\,{u}^{59}+6\,{u}^{58}\\ 
&+{u}^{57}-9\,{u}^{56}
+33\,{u}^{55}+7\,{u}^{54}-30\,{u}^{53}-24\,{u}^{52}+5\,{u}^{51}+27\,{u}^{50}+9\,{u}^{49}-9\,{u}^{48}-9\,{u}^{47}\\
&+4\,{u}^{45}-3\,{u}^{43}-3\,{u}^{42}
+3\,{u}^{41}+12\,{u}^{40}+2\,{u}^{39}-18\,{u}^{38}-9\,{u}^{37}+12\,{u}^{36}+9\,{u}^{35}-3\,{u}^{34}\\
& -3\,{u}^{33}+{u}^{21}-3\,{u}^{18}-3\,{u}^{17}
+3\,{u}^{15}+9\,{u}^{14}+3\,{u}^{13}-{u}^{12}-9\,{u}^{11}-9\,{u}^{10}-{u}^{9}\\
& +3\,{u}^{8}+9\,{u}^{7}+3\,{u}^{6}-3\,{u}^{4}-3\,{u}^{3}+1\\
\eal
\]

\bibliography{Twisted_ref}

\newcommand{\etalchar}[1]{$^{#1}$}
\begin{thebibliography}{DBMM{\etalchar{+}}13}

\bibitem[AS15]{refCS}
M.~Aganagic and S.~Shakirov.
\newblock {Knot Homology and Refined Chern-Simons Index}.
\newblock {\em Commun. Math. Phys.}, 333(1):187--228, 2015,
  \href{http://arxiv.org/abs/1105.5117}{{\tt arXiv:1105.5117}}.

\bibitem[Ati57]{VB_elliptic}
M.~F. Atiyah.
\newblock Vector bundles over an elliptic curve.
\newblock {\em Proc. London Math. Soc. (3)}, 7:414--452, 1957.

\bibitem[BB04]{Wild_curves}
O.~Biquard and P.~Boalch.
\newblock Wild non-abelian {H}odge theory on curves.
\newblock {\em Compos. Math.}, 140(1):179--204, 2004,
  \href{http://arxiv.org/abs/math/0111098}{{\tt arXiv:math/0111098}}.

\bibitem[Boa07]{Quasi_hamiltonian}
P.~Boalch.
\newblock Quasi-{H}amiltonian geometry of meromorphic connections.
\newblock {\em Duke Math. J.}, 139(2):369--405, 2007,
  \href{http://arxiv.org/abs/math/0203161}{{\tt arXiv:math/0203161}}.

\bibitem[Boa14]{Braiding_stokes}
P.~P. Boalch.
\newblock Geometry and braiding of {S}tokes data; fission and wild character
  varieties.
\newblock {\em Ann. of Math. (2)}, 179(1):301--365, 2014,
  \href{http://arxiv.org/abs/1111.6228}{{\tt arXiv:1111.6228}}.

\bibitem[BY15]{Twisted_wild}
P.~Boalch and D.~Yamakawa.
\newblock {Twisted wild character varieties}.
\newblock 2015, \href{http://arxiv.org/abs/1512.08091}{{\tt arXiv:1512.08091}}.

\bibitem[CDDP15]{Par_ref}
W.-y. Chuang, D.-E. Diaconescu, R.~Donagi, and T.~Pantev.
\newblock {Parabolic refined invariants and Macdonald polynomials}.
\newblock {\em Commun. Math. Phys.}, 335(3):1323--1379, 2015,
  \href{http://arxiv.org/abs/1311.3624}{{\tt arXiv:1311.3624}}.

\bibitem[CDP11]{wallpairs}
W.-y. Chuang, D.-E. Diaconescu, and G.~Pan.
\newblock {Wallcrossing and Cohomology of The Moduli Space of Hitchin Pairs}.
\newblock {\em Commun.Num.Theor.Phys.}, 5:1--56, 2011,
  \href{http://arxiv.org/abs/1004.4195}{{\tt arXiv:1004.4195}}.

\bibitem[CDP14]{BPSPW}
W.~Y. Chuang, D.~E. Diaconescu, and G.~Pan.
\newblock {BPS states and the P=W conjecture}.
\newblock In {\em {Proceedings, School on Moduli Spaces: Cambridge, UK, 5-14
  Jan 2011}}, volume 411, pages 132--150. 2014,
  \href{http://arxiv.org/abs/1202.2039}{{\tt arXiv:1202.2039}}.

\bibitem[Che95]{cherednik1995macdonald}
I.~Cherednik.
\newblock Double affine {H}ecke algebras and {M}acdonald's conjectures.
\newblock {\em Ann. of Math. (2)}, 141(1):191--216, 1995.

\bibitem[Che13]{Cherednik:2011nr}
I.~Cherednik.
\newblock {Jones polynomials of torus knots via DAHA}.
\newblock {\em Int.Math.Res.Not.}, 23:5366--5425, 2013,
  \href{http://arxiv.org/abs/1111.6195}{{\tt arXiv:1111.6195}}.

\bibitem[Che16]{cherednik2016daha}
I.~Cherednik.
\newblock {DAHA-Jones polynomials of torus knots}.
\newblock {\em Selecta Mathematica}, 22(2):1013--1053, 2016,
  \href{http://arxiv.org/abs/1406.3959}{{\tt arXiv:1406.3959}}.

\bibitem[DBMM{\etalchar{+}}13]{Super_evol}
P.~Dunin-Barkowski, A.~Mironov, A.~Morozov, A.~Sleptsov, and A.~Smirnov.
\newblock {Superpolynomials for toric knots from evolution induced by
  cut-and-join operators}.
\newblock {\em JHEP}, 03:021, 2013, \href{http://arxiv.org/abs/1106.4305}{{\tt
  arXiv:1106.4305}}.

\bibitem[dCHM12]{hodgechar}
M.~A.~A. d.~Cataldo, T.~Hausel, and L.~Migliorini.
\newblock Topology of {H}itchin systems and {H}odge theory of character
  varieties: the case {$A_1$}.
\newblock {\em Ann. of Math. (2)}, 175(3):1329--1407, 2012,
  \href{http://arxiv.org/abs/1004.1420}{{\tt arXiv:1004.1420}}.

\bibitem[DDP18]{BPS_wild}
D.-E. Diaconescu, R.~Donagi, and T.~Pantev.
\newblock {BPS states, torus links and wild character varieties}.
\newblock {\em Comm. Math. Phys.}, 357, 2018,
  \href{http://arxiv.org/abs/1704.07412}{{\tt arXiv:1704.07412}}.

\bibitem[DGT16]{Ind_obj}
G.~{Dobrovolska}, V.~{Ginzburg}, and R.~{Travkin}.
\newblock {Moduli spaces, indecomposable objects and potentials over a finite
  field}.
\newblock 2016, \href{http://arxiv.org/abs/1612.01733}{{\tt arXiv:1612.01733}}.

\bibitem[DHS12]{DHS}
D.-E. Diaconescu, Z.~Hua, and Y.~Soibelman.
\newblock {HOMFLY polynomials, stable pairs and motivic Donaldson-Thomas
  invariants}.
\newblock {\em Commun. Num. Theor. Phys.}, 6:517--600, 2012,
  \href{http://arxiv.org/abs/1202.4651}{{\tt arXiv:1202.4651}}.

\bibitem[Dia17]{Wild_deg}
D.-E. Diaconescu.
\newblock {Local curves, wild character varieties, and degenerations}.
\newblock 2017, \href{http://arxiv.org/abs/1705.05707}{{\tt arXiv:1705.05707}}.

\bibitem[DSV13]{DSV}
D.~E. Diaconescu, V.~Shende, and C.~Vafa.
\newblock {Large N duality, lagrangian cycles, and algebraic knots}.
\newblock {\em Commun. Math. Phys.}, 319:813--863, 2013,
  \href{http://arxiv.org/abs/1111.6533}{{\tt arXiv:1111.6533}}.

\bibitem[EK96]{kirillov1996inner}
P.~I. Etingof and A.~A. Kirillov, Jr.
\newblock Representation-theoretic proof of the inner product and symmetry
  identities for {M}acdonald's polynomials.
\newblock {\em Compositio Math.}, 102(2):179--202, 1996,
  \href{http://arxiv.org/abs/math/9411232}{{\tt arXiv:math/9411232}}.

\bibitem[ES98]{etingof1996representation}
P.~Etingof and K.~Styrkas.
\newblock Algebraic integrability of {M}acdonald operators and representations
  of quantum groups.
\newblock {\em Compositio Math.}, 114(2):125--152, 1998,
  \href{http://arxiv.org/abs/q-alg/9603022}{{\tt arXiv:q-alg/9603022}}.

\bibitem[FSS17]{Mot_conn_higgs}
R.~Fedorov, A.~Soibelman, and Y.~Soibelman.
\newblock {Motivic classes of moduli of Higgs bundles and moduli of bundles
  with connections}.
\newblock 2017, \href{http://arxiv.org/abs/1705.04890}{{\tt arXiv:1705.04890}}.

\bibitem[GGS13]{Gorsky:2013jxa}
E.~Gorsky, S.~Gukov, and M.~Stosic.
\newblock {Quadruply-graded colored homology of knots}.
\newblock 2013, \href{http://arxiv.org/abs/1304.3481}{{\tt arXiv:1304.3481}}.

\bibitem[GH93]{graded-rep}
A.~M. Garsia and M.~Haiman.
\newblock A graded representation model for {M}acdonald's polynomials.
\newblock {\em Proc. Nat. Acad. Sci. U.S.A.}, 90(8):3607--3610, 1993.

\bibitem[GN15]{Ref_knots_Hilb}
E.~Gorsky and A.~Negut.
\newblock {Refined knot invariants and Hilbert schemes}.
\newblock {\em J. Math. Pure. Appl.}, 104:403--435, 2015,
  \href{http://arxiv.org/abs/1304.3328}{{\tt arXiv:1304.3328}}.

\bibitem[GPH13]{y_genus_higgs}
O.~Garc\'ia-Prada and J.~Heinloth.
\newblock The {$y$}-genus of the moduli space of {${\rm PGL}_n$}-{H}iggs
  bundles on a curve (for degree coprime to {$n$}).
\newblock {\em Duke Math. J.}, 162(14):2731--2749, 2013,
  \href{http://arxiv.org/abs/1207.5614}{{\tt arXiv:1207.5614}}.

\bibitem[GPHS14]{Mot_chains}
O.~Garc\'ia-Prada, J.~Heinloth, and A.~Schmitt.
\newblock On the motives of moduli of chains and {H}iggs bundles.
\newblock {\em J. Eur. Math. Soc. (JEMS)}, 16(12):2617--2668, 2014,
  \href{http://arxiv.org/abs/1104.5558}{{\tt arXiv:1104.5558}}.

\bibitem[Hai99]{MDgeom}
M.~Haiman.
\newblock Macdonald polynomials and geometry.
\newblock In {\em New perspectives in algebraic combinatorics ({B}erkeley,
  {CA}, 1996--97)}, volume~38 of {\em Math. Sci. Res. Inst. Publ.}, pages
  207--254. Cambridge Univ. Press, Cambridge, 1999.

\bibitem[Haua]{Hausel_slides}
T.~Hausel.
\newblock {Arithmetic of wild character varieties}.
\newblock \newline
  \href{https://ist.ac.at/fileadmin/user_upload/group_pages/hausel/Singapore2014.pdf}{Talk
  slides at Geometry, Topology and Physics of Moduli of Higgs Bundles, National
  University of Singapore}, \newline
  \href{https://ist.ac.at/fileadmin/user_upload/group_pages/hausel/Utah15.pdf}{Talk
  slides at AMS summer institute in algebraic geometry, University of Utah,
  Salt Lake City}.

\bibitem[Haub]{Hausel_private}
T.~Hausel.
\newblock {Private communication}.

\bibitem[HL97]{huylehn}
D.~Huybrechts and M.~Lehn.
\newblock {\em The geometry of moduli spaces of sheaves}.
\newblock Aspects of Mathematics, E31. Friedr. Vieweg \& Sohn, Braunschweig,
  1997.

\bibitem[HLRV11]{HLRV}
T.~Hausel, E.~Letellier, and F.~Rodriguez-Villegas.
\newblock Arithmetic harmonic analysis on character and quiver varieties.
\newblock {\em Duke Math. J.}, 160(2):323--400, 2011,
  \href{http://arxiv.org/abs/0810.2076}{{\tt arXiv:0810.2076}}.

\bibitem[HLRV13]{HLRVII}
T.~Hausel, E.~Letellier, and F.~Rodriguez-Villegas.
\newblock Arithmetic harmonic analysis on character and quiver varieties {II}.
\newblock {\em Adv. Math.}, 234:85--128, 2013,
  \href{http://arxiv.org/abs/1109.5202}{{\tt arXiv:1109.5202}}.

\bibitem[HMW16]{Arithmetic_wild}
T.~Hausel, M.~Mereb, and M.~L. Wong.
\newblock {Arithmetic and representation theory of wild character varieties}.
\newblock 2016, \href{http://arxiv.org/abs/1604.03382}{{\tt arXiv:1604.03382}}.

\bibitem[HRV08]{HRV}
T.~Hausel and F.~Rodriguez-Villegas.
\newblock Mixed {H}odge polynomials of character varieties.
\newblock {\em Invent. Math.}, 174(3):555--624, 2008,
  \href{http://arxiv.org/abs/math/0612668}{{\tt arXiv:math/0612668}}.
\newblock With an appendix by Nicholas M. Katz.

\bibitem[HST01]{HST}
S.~Hosono, M.-H. Saito, and A.~Takahashi.
\newblock Relative {L}efschetz action and {BPS} state counting.
\newblock {\em Internat. Math. Res. Notices}, (15):783--816, 2001,
  \href{http://arxiv.org/abs/math/0105148}{{\tt arXiv:math/0105148}}.

\bibitem[{Jia}17]{Non_arch_sheaves}
Y.~{Jiang}.
\newblock {The moduli space of stable coherent sheaves via non-archimedean
  geometry}.
\newblock 2017, \href{http://arxiv.org/abs/1703.00497}{{\tt arXiv:1703.00497}}.

\bibitem[KS08]{wallcrossing}
M.~Kontsevich and Y.~Soibelman.
\newblock Stability structures, {D}onaldson-{T}homas invariants and cluster
  transformations.
\newblock 2008, \href{http://arxiv.org/abs/0811.2435}{{\tt arXiv:0811.2435}}.

\bibitem[KS14]{structures}
M.~Kontsevich and Y.~Soibelman.
\newblock Wall-crossing structures in {D}onaldson--{T}homas invariants,
  integrable systems and mirror symmetry.
\newblock In {\em Homological mirror symmetry and tropical geometry}, pages
  197--308. Springer, 2014, \href{http://arxiv.org/abs/1303.3253}{{\tt
  arXiv:1303.3253}}.

\bibitem[Let15]{Zariski_closures}
E.~Letellier.
\newblock Character varieties with {Z}ariski closures of {${GL}_n$}-conjugacy
  classes at punctures.
\newblock {\em Selecta Math. (N.S.)}, 21(1):293--344, 2015,
  \href{http://arxiv.org/abs/1309.7662}{{\tt arXiv:1309.7662}}.

\bibitem[{Let}16]{Higgs_ind_proj_line}
E.~{Letellier}.
\newblock {Higgs bundles and indecomposable parabolic bundles over the
  projective line}.
\newblock 2016, \href{http://arxiv.org/abs/1609.04875}{{\tt arXiv:1609.04875}}.

\bibitem[Mar94]{spectral_int}
E.~Markman.
\newblock Spectral curves and integrable systems.
\newblock {\em Compositio Math.}, 93(3):255--290, 1994.

\bibitem[{Mau}]{HRV_proof}
D.~{Maulik}.
\newblock {Refined stable pair invariants for local curves}.
\newblock to appear.

\bibitem[Mau16]{Homfly_pairs}
D.~Maulik.
\newblock Stable pairs and the {HOMFLY} polynomial.
\newblock {\em Invent. Math.}, 204(3):787--831, 2016,
  \href{http://arxiv.org/abs/1210.6323}{{\tt arXiv:1210.6323}}.

\bibitem[Moc11]{Wild_harmonic}
T.~Mochizuki.
\newblock Wild harmonic bundles and wild pure twistor {$D$}-modules.
\newblock {\em Ast\'erisque}, (340):x+607, 2011,
  \href{http://arxiv.org/abs/0803.1344}{{\tt arXiv:0803.1344}}.

\bibitem[MS14]{Counting_Higgs}
S.~{Mozgovoy} and O.~{Schiffmann}.
\newblock {Counting Higgs bundles}.
\newblock 2014, \href{http://arxiv.org/abs/1411.2101}{{\tt arXiv:1411.2101}}.

\bibitem[NO14]{Membranes_Sheaves}
N.~Nekrasov and A.~Okounkov.
\newblock {Membranes and Sheaves}.
\newblock 2014, \href{http://arxiv.org/abs/1404.2323}{{\tt arXiv:1404.2323}}.

\bibitem[ORS18]{ORS}
A.~{Oblomkov}, J.~{Rasmussen}, and V.~{Shende}.
\newblock {The Hilbert scheme of a plane curve singularity and the HOMFLY
  homology of its link}.
\newblock {\em Geometry \& Topology}, 22(2):645--691, 2018,
  \href{http://arxiv.org/abs/1201.2115}{{\tt arXiv:1201.2115}}.

\bibitem[OS12]{OS}
A.~Oblomkov and V.~Shende.
\newblock {The Hilbert scheme of a plane curve singularity and the HOMFLY
  polynomial of its link}.
\newblock {\em Duke Math. J.}, 161:1277--1303, 2012,
  \href{http://arxiv.org/abs/1003.1568}{{\tt arXiv:1003.1568}}.

\bibitem[PT09]{stabpairsI}
R.~Pandharipande and R.~P. Thomas.
\newblock Curve counting via stable pairs in the derived category.
\newblock {\em Invent. Math.}, 178(2):407--447, 2009,
  \href{http://arxiv.org/abs/0707.2348}{{\tt arXiv:0707.2348}}.

\bibitem[RMGP09]{Sheaves_genus_one}
D.~H. Ruip{\'e}rez, A.~C.~L. Mart{\'\i}n, D.~S. G{\'o}mez, and C.~T. Prieto.
\newblock Moduli spaces of semistable sheaves on singular genus 1 curves.
\newblock {\em International Mathematics Research Notices},
  2009(23):4428--4462, 2009, \href{http://arxiv.org/abs/0806.2034}{{\tt
  arXiv:0806.2034}}.

\bibitem[Sab99]{Harmonic_metrics}
C.~Sabbah.
\newblock Harmonic metrics and connections with irregular singularities.
\newblock {\em Ann. Inst. Fourier}, 49(4):1265--1291, 1999,
  \href{http://arxiv.org/abs/math/9905039}{{\tt arXiv:math/9905039}}.

\bibitem[Sch16]{Ind_vb_higgs}
O.~Schiffmann.
\newblock Indecomposable vector bundles and stable {H}iggs bundles over smooth
  projective curves.
\newblock {\em Ann. of Math. (2)}, 183(1):297--362, 2016,
  \href{http://arxiv.org/abs/1406.3839}{{\tt arXiv:1406.3839}}.

\bibitem[Sha13]{Colored_HL}
S.~Shakirov.
\newblock {Colored knot amplitudes and Hall-Littlewood polynomials}.
\newblock 2013, \href{http://arxiv.org/abs/1308.3838}{{\tt arXiv:1308.3838}}.

\bibitem[STWZ15]{Cluster_legendrian}
V.~Shende, D.~Treumann, H.~Williams, and E.~Zaslow.
\newblock {Cluster varieties from Legendrian knots}.
\newblock 2015, \href{http://arxiv.org/abs/1512.08942}{{\tt arXiv:1512.08942}}.

\bibitem[STZ17]{Fukaya_knots}
V.~Shende, D.~Treumann, and E.~Zaslow.
\newblock Legendrian knots and constructible sheaves.
\newblock {\em Inventiones mathematicae}, 207(3):1031--1133, 2017,
  \href{http://arxiv.org/abs/1402.0490}{{\tt arXiv:1402.0490}}.

\end{thebibliography}
 \bibliographystyle{halpha}

\bigskip

\noindent
Wu-Yen Chuang, {\sf Department of Mathematics, National Taiwan University, No.1 Sec.4 Roosevelt Road Taipei 10617, Taiwan} wychuang@gmail.com

 \

\noindent 
Duiliu-Emanuel 
Diaconescu, {\sf
 NHETC, Rutgers University, 
 126 Frelinghuysen Road, Piscataway NJ 08854, USA},
 duiliu@physics.rutgers.edu

 \

\noindent
Ron Donagi, {\sf Department of Mathematics, University of
  Pennsylvania, David Rittenhouse 
Laboratory, 209 South 33rd Street,
  Philadelphia, PA 19104, USA}, donagi@math.upenn.edu

\

\noindent
Satoshi Nawata, {\sf Department of Physics and Center for Field Theory and Particle Physics, Fudan University, 220
Handan Road, 200433 Shanghai, China}, snawata@gmail.com

\

\noindent
Tony Pantev, {\sf Department of Mathematics, University of
  Pennsylvania, David Rittenhouse 
Laboratory, 209 South 33rd Street,
  Philadelphia, PA 19104, USA}, tpantev@math.upenn.edu

\end{document}